\documentclass[
draftclsnofoot,
onecolumn,
]{IEEEtran}

\usepackage{algorithm,algorithmic}
\usepackage{multicol}
\usepackage{subfigure} 
\usepackage{cite}
\usepackage{amsmath,amssymb,amsfonts}
\usepackage{algorithmic}
\usepackage{textcomp}
\usepackage[usenames]{color}
\usepackage{psfrag}
\usepackage{url}
\usepackage{soul}
\usepackage{comment}
\usepackage{graphicx,color}
\usepackage{epstopdf}
\usepackage[nolist]{acronym}
\usepackage{algorithm,algorithmic} %algorithm2e
\usepackage{pgfplots}
\usepackage{pgfplotstable}
\pgfplotsset{compat=1.16}

\usepackage{caption}

\title{Green Tethered UAVs for EMF-Aware Cellular Networks}
\author{Zhengying Lou, \textit{Student Member, IEEE}, Ahmed Elzanaty,\textit{ Member, IEEE}, and Mohamed-Slim Alouini, \textit{Fellow, IEEE} 
	\thanks{The authors are with the Computer, Electrical, and Mathematical Science
		and Engineering (CEMSE) Division, King Abdullah University of Science
		and Technology (KAUST), Thuwal, Makkah Province, Saudi Arabia. (e-mail:
		{zhengying.lou,ahmed.elzanaty,slim.alouini}@kaust.edu.sa).}}

\begin{document}
	\pgfplotsset{every axis/.append style={
			line width=1pt,
			%	label style={font=\Huge},
			%	tick label style={font=\normalsize},
			legend style={font=\large, at={(0.97,0.85)}}},
		%	legend style={font=\small, at={(0.1,0.95)}},legend cell align=left},
		%legend style={font=\footnotesize, at={(0.99,0.75)}},
	} %
	\pgfplotsset{compat=1.16}
	\begin{acronym}

\acro{RV}{random variable}

\acro{5G-NR}{5G New Radio}
\acro{3GPP}{3rd Generation Partnership Project}
\acro{ABS}{aerial base station}
\acro{AC}{address coding}
\acro{ACF}{autocorrelation function}
\acro{ACR}{autocorrelation receiver}
\acro{ADC}{analog-to-digital converter}
\acrodef{aic}[AIC]{Analog-to-Information Converter}     
\acro{AIC}[AIC]{Akaike information criterion}
\acro{aric}[ARIC]{asymmetric restricted isometry constant}
\acro{arip}[ARIP]{asymmetric restricted isometry property}

\acro{ARQ}{Automatic Repeat Request}
\acro{AUB}{asymptotic union bound}
\acrodef{awgn}[AWGN]{Additive White Gaussian Noise}     
\acro{AWGN}{additive white Gaussian noise}

\acro{APSK}[PSK]{asymmetric PSK} 

\acro{waric}[AWRICs]{asymmetric weak restricted isometry constants}
\acro{warip}[AWRIP]{asymmetric weak restricted isometry property}
\acro{BCH}{Bose, Chaudhuri, and Hocquenghem}        
\acro{BCHC}[BCHSC]{BCH based source coding}
\acro{BEP}{bit error probability}
\acro{BFC}{block fading channel}
\acro{BG}[BG]{Bernoulli-Gaussian}
\acro{BGG}{Bernoulli-Generalized Gaussian}
\acro{BPAM}{binary pulse amplitude modulation}
\acro{BPDN}{Basis Pursuit Denoising}
\acro{BPPM}{binary pulse position modulation}
\acro{BPSK}{Binary Phase Shift Keying}
\acro{BPZF}{bandpass zonal filter}
\acro{BSC}{binary symmetric channels}              
\acro{BU}[BU]{Bernoulli-uniform}
\acro{BER}{bit error rate}
\acro{BS}{base station}
\acro{BW}{BandWidth}
\acro{BLLL}{ binary log-linear learning }

\acro{CP}{Cyclic Prefix}
\acrodef{cdf}[CDF]{cumulative distribution function}   
\acro{CDF}{Cumulative Distribution Function}
\acrodef{c.d.f.}[CDF]{cumulative distribution function}
\acro{CCDF}{complementary cumulative distribution function}
\acrodef{ccdf}[CCDF]{complementary CDF}               
\acrodef{c.c.d.f.}[CCDF]{complementary cumulative distribution function}
\acro{CD}{cooperative diversity}

\acro{CDMA}{Code Division Multiple Access}
\acro{ch.f.}{characteristic function}
\acro{CIR}{channel impulse response}
\acro{cosamp}[CoSaMP]{compressive sampling matching pursuit}
\acro{CR}{cognitive radio}
\acro{cs}[CS]{compressed sensing}                   
\acrodef{cscapital}[CS]{Compressed sensing} %will not include it in the list
\acrodef{CS}[CS]{compressed sensing}
\acro{CSI}{channel state information}
\acro{CCSDS}{consultative committee for space data systems}
\acro{CC}{convolutional coding}
\acro{Covid19}[COVID-19]{Coronavirus disease}

\acro{DAA}{detect and avoid}
\acro{DAB}{digital audio broadcasting}
\acro{DCT}{discrete cosine transform}
\acro{dft}[DFT]{discrete Fourier transform}
\acro{DR}{distortion-rate}
\acro{DS}{direct sequence}
\acro{DS-SS}{direct-sequence spread-spectrum}
\acro{DTR}{differential transmitted-reference}
\acro{DVB-H}{digital video broadcasting\,--\,handheld}
\acro{DVB-T}{digital video broadcasting\,--\,terrestrial}
\acro{DL}{downlink}
\acro{DSSS}{Direct Sequence Spread Spectrum}
\acro{DFT-s-OFDM}{Discrete Fourier Transform-spread-Orthogonal Frequency Division Multiplexing}
\acro{DAS}{Distributed Antenna System}
\acro{DNA}{DeoxyriboNucleic Acid}

\acro{EC}{European Commission}
\acro{EED}[EED]{exact eigenvalues distribution}
\acro{EIRP}{Equivalent Isotropically Radiated Power}
\acro{ELP}{equivalent low-pass}
\acro{eMBB}{Enhanced Mobile Broadband}
\acro{EMF}{electromagnetic field}
\acro{EU}{European union}
\acro{EI}{exposure index}
\acro{eICIC}{enhanced Inter-Cell Interference Coordination}

\acro{FC}[FC]{fusion center}
\acro{FCC}{Federal Communications Commission}
\acro{FEC}{forward error correction}
\acro{FFT}{fast Fourier transform}
\acro{FH}{frequency-hopping}
\acro{FH-SS}{frequency-hopping spread-spectrum}
\acrodef{FS}{Frame synchronization}
\acro{FSsmall}[FS]{frame synchronization}  
\acro{FDMA}{Frequency Division Multiple Access}

\acro{GA}{Gaussian approximation}
\acro{GF}{Galois field }
\acro{GG}{Generalized-Gaussian}
\acro{GIC}[GIC]{generalized information criterion}
\acro{GLRT}{generalized likelihood ratio test}
\acro{GPS}{global positioning system}
\acro{GMSK}{Gaussian Minimum Shift Keying}
\acro{GSMA}{Global System for Mobile communications Association}
\acro{GS}{ground station}
\acro{GMG}{ Grid-connected MicroGeneration}

\acro{HAP}{high altitude platform}
\acro{HetNet}{Heterogeneous network}

\acro{IDR}{information distortion-rate}
\acro{IFFT}{inverse fast Fourier transform}
\acro{iht}[IHT]{iterative hard thresholding}
\acro{i.i.d.}{independent, identically distributed}
\acro{IoT}{Internet of Things}                      
\acro{IR}{impulse radio}
\acro{lric}[LRIC]{lower restricted isometry constant}
\acro{lrict}[LRICt]{lower restricted isometry constant threshold}
\acro{ISI}{intersymbol interference}
\acro{ITU}{International Telecommunication Union}
\acro{ICNIRP}{International Commission on Non-Ionizing Radiation Protection}
\acro{IEEE}{Institute of Electrical and Electronics Engineers}
\acro{ICES}{IEEE international committee on electromagnetic safety}
\acro{IEC}{International Electrotechnical Commission}
\acro{IARC}{International Agency on Research on Cancer}
\acro{IS-95}{Interim Standard 95}

\acro{KPI}{Key Performance Indicator}

\acro{LEO}{low earth orbit}
\acro{LF}{likelihood function}
\acro{LLF}{log-likelihood function}
\acro{LLR}{log-likelihood ratio}
\acro{LLRT}{log-likelihood ratio test}
\acro{LoS}{Line-of-Sight}
\acro{LRT}{likelihood ratio test}
\acro{wlric}[LWRIC]{lower weak restricted isometry constant}
\acro{wlrict}[LWRICt]{LWRIC threshold}
\acro{LPWAN}{Low Power Wide Area Network}
\acro{LoRaWAN}{Low power long Range Wide Area Network}
\acro{LiFi}[Li-Fi]{light-fidelity}
 \acro{LED}{light emitting diode}
 \acro{LABS}{LoS transmission with each ABS}
 \acro{NLABS}{NLoS transmission with each ABS}

\acro{MB}{multiband}
\acro{MC}{macro cell}
\acro{MDS}{mixed distributed source}
\acro{MF}{matched filter}
\acro{m.g.f.}{moment generating function}
\acro{MI}{mutual information}
\acro{MIMO}{multiple-input multiple-output}
\acro{MISO}{multiple-input single-output}
\acrodef{maxs}[MJSO]{maximum joint support cardinality}                       
\acro{ML}[ML]{maximum likelihood}
\acro{MMSE}{minimum mean-square error}
\acro{MMV}{multiple measurement vectors}
\acrodef{MOS}{model order selection}
\acro{M-PSK}[${M}$-PSK]{$M$-ary phase shift keying}                       
\acro{M-APSK}[${M}$-PSK]{$M$-ary asymmetric PSK} 
\acro{MP}{ multi-period}
\acro{MINLP}{mixed integer non-linear programming}

\acro{M-QAM}[$M$-QAM]{$M$-ary quadrature amplitude modulation}
\acro{MRC}{maximal ratio combiner}                  
\acro{maxs}[MSO]{maximum sparsity order}                                      
\acro{M2M}{Machine-to-Machine}                                                
\acro{MUI}{multi-user interference}
\acro{mMTC}{massive Machine Type Communications}      
\acro{mm-Wave}{millimeter-wave}
\acro{MP}{mobile phone}
\acro{MPE}{maximum permissible exposure}
\acro{MAC}{media access control}
\acro{NB}{narrowband}
\acro{NBI}{narrowband interference}
\acro{NB-IoT}{narrowband – Internet of things}
\acro{NLA}{nonlinear sparse approximation}
\acro{NLOS}[NLoS]{non-line of sight}
\acro{NTIA}{National Telecommunications and Information Administration}
\acro{NTP}{National Toxicology Program}
\acro{NHS}{National Health Service}

\acro{NP}{non-deterministic polynomial-time}

\acro{LOS}[LoS]{line of sight}

\acro{OC}{optimum combining}                             
\acro{OC}{optimum combining}
\acro{ODE}{operational distortion-energy}
\acro{ODR}{operational distortion-rate}
\acro{OFDM}{orthogonal frequency-division multiplexing}
\acro{omp}[OMP]{orthogonal matching pursuit}
\acro{OSMP}[OSMP]{orthogonal subspace matching pursuit}
\acro{OQAM}{offset quadrature amplitude modulation}
\acro{OQPSK}{offset QPSK}
\acro{OFDMA}{Orthogonal Frequency-division Multiple Access}
\acro{OPEX}{Operating Expenditures}
\acro{OQPSK/PM}{OQPSK with phase modulation}

\acro{PAM}{pulse amplitude modulation}
\acro{PAR}{peak-to-average ratio}
\acrodef{pdf}[PDF]{probability density function}                      
\acro{PDF}{probability density function}
\acrodef{p.d.f.}[PDF]{probability distribution function}
\acro{PDP}{power dispersion profile}
\acro{PMF}{probability mass function}                             
\acrodef{p.m.f.}[PMF]{probability mass function}
\acro{PN}{pseudo-noise}
\acro{PPM}{pulse position modulation}
\acro{PRake}{Partial Rake}
\acro{PSD}{power spectral density}
\acro{PSEP}{pairwise synchronization error probability}
\acro{PSK}{phase shift keying}
\acro{PD}{power density}
\acro{8-PSK}[$8$-PSK]{$8$-phase shift keying}
\acro{PPP}{Poisson point process}
\acro{PCP}{Poisson cluster process}
 
\acro{FSK}{Frequency Shift Keying}

\acro{QAM}{Quadrature Amplitude Modulation}
\acro{QPSK}{Quadrature Phase Shift Keying}
\acro{OQPSK/PM}{OQPSK with phase modulator }

\acro{RD}[RD]{raw data}
\acro{RDL}{"random data limit"}
\acro{ric}[RIC]{restricted isometry constant}
\acro{rict}[RICt]{restricted isometry constant threshold}
\acro{rip}[RIP]{restricted isometry property}
\acro{ROC}{receiver operating characteristic}
\acro{rq}[RQ]{Raleigh quotient}
\acro{RS}[RS]{Reed-Solomon}
\acro{RSC}[RSSC]{RS based source coding}
\acro{r.v.}{random variable}                               
\acro{R.V.}{random vector}
\acro{RMS}{root mean square}
\acro{RFR}{radiofrequency radiation}
\acro{RIS}{reconfigurable intelligent surface}
\acro{RNA}{RiboNucleic Acid}
\acro{RRM}{Radio Resource Management}
\acro{RUE}{reference user equipments}
\acro{RAT}{radio access technology}
\acro{RB}{resource block}

\acro{SCgNB}{small cell gNB}
\acro{MgNB}{Macro gNB}

\acro{SA}[SA-Music]{subspace-augmented MUSIC with OSMP}
\acro{SC}{small cell}
\acro{SCBSES}[SCBSES]{Source Compression Based Syndrome Encoding Scheme}
\acro{SCM}{sample covariance matrix}
\acro{SEP}{symbol error probability}
\acro{SG}[SG]{sparse-land Gaussian model}
\acro{SIMO}{single-input multiple-output}
\acro{SINR}{signal-to-interference plus noise ratio}
\acro{SIR}{signal-to-interference ratio}
\acro{SISO}{Single-Input Single-Output}
\acro{SMV}{single measurement vector}
\acro{SNR}[\textrm{SNR}]{signal-to-noise ratio} 
\acro{sp}[SP]{subspace pursuit}
\acro{SS}{spread spectrum}
\acro{SW}{sync word}
\acro{SAR}{specific absorption rate}
\acro{SSB}{synchronization signal block}
\acro{SR}{shrink and realign}
\acro{SAR}{specific absorption rate}
\acro{tUAV}[TUAV]{tethered unmanned aerial vehicle}
\acro{TBS}{terrestrial base station}

\acro{uUAV}{untethered Unmanned Aerial Vehicle}
\acro{PDF}{probability density functions}

\acro{PL}{path-loss}

\acro{TH}{time-hopping}
\acro{ToA}{time-of-arrival}
\acro{TR}{transmitted-reference}
\acro{TW}{Tracy-Widom}
\acro{TWDT}{TW Distribution Tail}
\acro{TCM}{trellis coded modulation}
\acro{TDD}{Time-Division Duplexing}
\acro{TDMA}{time division multiple access}
\acro{Tx}{average transmit}

\acro{UAV}{unmanned aerial vehicle}
\acro{uric}[URIC]{upper restricted isometry constant}
\acro{urict}[URICt]{upper restricted isometry constant threshold}
\acro{UWB}{ultrawide band}
\acro{UWBcap}[UWB]{Ultrawide band}   
\acro{URLLC}{Ultra Reliable Low Latency Communications}
         
\acro{wuric}[UWRIC]{upper weak restricted isometry constant}
\acro{wurict}[UWRICt]{UWRIC threshold}                
\acro{UE}{user equipment}
\acro{UL}{uplink}

\acro{WiM}[WiM]{weigh-in-motion}
\acro{WLAN}{wireless local area network}
\acro{wm}[WM]{Wishart matrix}                               
\acroplural{wm}[WM]{Wishart matrices}
\acro{WMAN}{wireless metropolitan area network}
\acro{WPAN}{wireless personal area network}
\acro{wric}[WRIC]{weak restricted isometry constant}
\acro{wrict}[WRICt]{weak restricted isometry constant thresholds}
\acro{wrip}[WRIP]{weak restricted isometry property}
\acro{WSN}{wireless sensor network}                        
\acro{WSS}{Wide-Sense Stationary}
\acro{WHO}{World Health Organization}
\acro{Wi-Fi}{Wireless Fidelity}

\acro{sss}[SpaSoSEnc]{sparse source syndrome encoding}

\acro{VLC}{Visible Light Communication}
\acro{VPN}{Virtual Private Network} 
\acro{RF}{radio frequency}
\acro{FSO}{Free Space Optics}
\acro{IoST}{Internet of Space Things}

\acro{GSM}{global System for Mobile Communications}
\acro{2G}{second-generation cellular network}
\acro{3G}{third-generation cellular network}
\acro{4G}{fourth-generation cellular network}
\acro{5G}{fifth-generation cellular network}	
\acro{gNB}{next-generation node-B base station}
\acro{NR}{New Radio}
\acro{UMTS}{Universal Mobile Telecommunications Service}
\acro{LTE}{long term evolution}

\acro{QoS}{quality of Service}
\end{acronym}
	
	%% EMF definitions
\newcommand{\SAR} {\mathrm{SAR}}
\newcommand{\WBSAR} {\mathrm{SAR}_{\mathsf{WB}}}
\newcommand{\gSAR} {\mathrm{SAR}_{10\si{\gram}}}
\newcommand{\Sab} {S_{\mathsf{ab}}}
\newcommand{\Eavg} {E_{\mathsf{avg}}}
\newcommand{\ft}{f_{\textsf{th}}}
\newcommand{\alphatf}{\alpha_{24}}

\def\SAR{\mathrm{SAR}}

\def\PN{P_k({\boldsymbol{\gamma}})}

	\maketitle

%\tableofcontents

\section{Abstract}
A prevalent theory circulating among the non-scientific community is that the intensive deployment of base stations over the territory significantly increases the level of \ac{EMF} exposure and affects population health. To alleviate this concern, in this work, we propose a network architecture that introduces \acp{tUAV} carrying green antennas to minimize the \ac{EMF} exposure while guaranteeing a high data rate for users. In particular, each \ac{tUAV} can attach itself to one of the possible ground stations at the top of some buildings.  The location of the \acp{tUAV}, transmit power of \acl{UE} and association policy are optimized to minimize the \ac{EMF} exposure. Unfortunately, the problem turns out to be a \ac{MINLP}, which is \ac{NP} hard. We propose an efficient low-complexity algorithm composed of three submodules. Firstly, we propose an algorithm based on the greedy principle to determine the optimal association matrix between the users and base stations. Then, we offer two approaches, modified $K$-mean and \ac{SR} process, to associate each \ac{tUAV} with a ground station. Finally, we put forward two algorithms based on the golden search and \ac{SR} process to adjust the \ac{tUAV}’s position within the hovering area over the building. After that, we consider the dual problem that maximizes the sum rate while keeping the exposure below a predefined value, such as the level enforced by the regulation. Next, we perform extensive simulations to show the effectiveness of the proposed \acp{tUAV} to reduce the exposure compared to various architectures. Eventually, we show that \acp{tUAV} with green antennas can effectively mitigate the \ac{EMF} exposure by more than  $20\%$  compared to fixed green \aclp{SC} while achieving a higher data rate.

\textbf{Index Terms}: Electromagnetic radiation; electromagnetic fields; mobile communication; \acs{EMF}-aware design; green communications; \acs{EMF} exposure; cellular systems; Internet of things; enhanced mobile broadband; airborne small cells;  \acsp{UAV}; resource allocation.

\acresetall % resent all acronym after abstract
\section{Introduction}

Unlike previous generations of cellular systems, the ongoing deployment of the \ac{5G} has attracted wide attention and controversy from researchers and non-specialists. Attitudes toward \ac{5G} and beyond are quite different that some even regard these new technologies as threats. Although the improvements from the communications perspective are obvious and recognized by the public, the dense deployment of \acp{gNB} over the territory generally generates a sentiment of suspect and fear \cite{warnemf}. Recently, media myths and false reports about the health effects caused by \ac{EMF} exposure have intensified negative feelings towards this technology \cite{healthrisk}. 

In this context, movements against installing new \acp{gNB} have been very active in recent years. A widely accepted fallacy among laymen is that the level of \ac{EMF} exposure is positively correlated with the amount of deployed antennas \cite{chiaraviglio2020health}. Although the fault of this argument can be easily detected by scholars, such as the power emitted by \acp{gNB} is not a fixed parameter. The claims about the adverse impacts of 5G densification spread throughout society leading to various sabotages to destroy towers hosting cellular equipment in several cities \cite{BBCderby5gfire}. 
%
% 	In this scenario, another controversial area of concern is the adoption of \ac{mm-Wave} in 5G and beyond. Although \ac{mm-Wave} has not been used in previously cellular communications, the biological effects of \ac{mm-Wave} have been well studied in the past few years \cite{erwin1981assessment, gandhi1986absorption, walters2000heating}. For humans, over 90\% of the transmitted power is absorbed by the epidermis and dermis. Moreover, due to this shallow penetration of \ac{mm-Wave}, the eyes and skin are organs of primary concern \cite{wu2015body}. 
	%For instance, \ac{mm-Wave} irradiation can cause lots of biological reactions in the skin because shallow depths lead to faster heating and less dissipation \cite{NelnelWal:20}. 
	In fact, there is no scientifically proven causality relation between the exposure to \ac{RF} waves and  adverse thermal effects for \ac{EMF} levels below the limits prescribed by law \cite{ICNIRPGuidelines:20}.  Therefore, many countries in the world adopt the \ac{EMF} limits established by the \ac{ICNIRP} to restrict the \ac{EMF} level caused by different \ac{EMF} sources \cite{chiaraviglio2020health}. Nevertheless, some experiments have found  adverse non-thermal impacts on animals for long-term \ac{EMF} exposure \cite{NTP:18a,FalBua:18,chiaraviglio2020health}. Based on these results, the \ac{RF} radiation was classified as ”Possibly carcinogenic to humans” by \ac{IARC}  \cite{Ciaula:18}. 
	%Moreover, to further reduce threats perceived by the public, the \ac{WHO} is conducting a health risk assessment of exposure over the entire range of \ac{RF} which also includes \ac{mm-Wave} \cite{5Gnetwork}.

    % Given this confusing scenario, the general public suspects that \ac{5G} and beyond communication systems will increase \ac{EMF} exposure, putting population health at risk. 
    % %Moreover, there is a debate about possible health consequences due to long term exposure to \ac{EMF}, even with a low level  \cite{NTP:18a,FalBua:18,chiaraviglio2020health}. 
    % Therefore, as a precautionary measure, we propose to design \ac{EMF}-aware system to reduce the exposure while achieving the target \ac{QoS}. In this paper, we propose an \ac{EMF}-aware architecture, where  \acp{tUAV} are deployed in cellular networks to reduce \ac{EMF} exposure and achieve high data rate. The related works about \ac{EMF}-aware design is described below. 
    % %and \ac{UAV} in communications are
    
    Given this confusing scenario, the general public suspects that \ac{5G} and beyond communication systems will increase \ac{EMF} exposure, putting population health at risk.   Therefore, as a precautionary measure, we propose to design \ac{EMF}-aware system to reduce the exposure while achieving the target \ac{QoS}. In this paper, {\color{black} we} propose an \ac{EMF}-aware architecture, where  \acp{tUAV} are deployed in cellular networks to reduce \ac{EMF} exposure and achieve a high data rate. The related works about \ac{EMF}-aware design are described below.
    
\subsection{Related Work}

In \ac{EMF}-aware design, cellular systems are designed with the aim of reducing the health risks due to the \ac{RF} radiations from cellular networks. This can be done either by reducing the exposure itself or by taking into consideration the constraint on the \ac{EMF} while designing the networks. 
In this regard, several effective radio resource allocation schemes and communication protocols have been proposed to mitigate EMF exposure. For instance, the authors of \cite{SamHelImr:14} design a user-scheduling approach based on their total transmit power in the past frames to reduce transmit power, thus lessen the \ac{UL} exposure in \ac{TDMA} systems. In \cite{SamImaHelImr:17}, two \ac{OFDM} based systems have been proposed to minimize \ac{UL} exposure while guaranteeing predefined throughput of users. Focusing then on the \ac{DL} exposure,  an algorithm for the exposure-aware association of user to \acp{gNB} is given in \cite{MatDerTanJoseph:18}. It is shown that although the number of \acp{gNB} deployed in massive \ac{MIMO} \ac{5G} networks is almost twice as much as that needed in the \ac{LTE} network, the exposure in the former is almost an order of magnitude lower than that of the later with the same network coverage. However, in \ac{5G} and beyond networks, the adopted \acp{mm-Wave} are subject to high blockage of the \ac{LOS} signal in unfavorable channel conditions, and it suffers from high path loss. Recently, \acp{RIS}, which can reflect the incident electromagnetic wave toward the specified direction, have been proposed to reduce exposure \cite{IprElzanaty:21}. However, in order to operate on a large scale, several \acp{RIS} need to be installed and optimized jointly, which can be a challenging task. Moreover, the signal processing capabilities of the \ac{RIS} are extremely limited. 

Another candidate for widespread deployment of 5G and beyond communication systems is small cell offloading, which has proven to improve the performance and enhance the energy efficiency  \cite{SidAlt:16, de2014emf}. The main idea of this method is to unload the user traffic from the macro cell to \acp{SC}. Thoroughly, a significant reduction in the \ac{UL} radiated power can be achieved \cite{de2014emf}. The radiation can be further reduced by  considering the decoupling between \ac{UL} and \ac{DL} \cite{BocDohPop:16}, where green antennas, i.e., receiving only radio, are used to assist the \ac{UL} \cite{green2009ezri}. %Specifically, the green antenna, which is the receiving equipment of the transceiver macro cell, can reduce the original radiation without generating additional radiation at all.

	Although the use of fixed deployed \acp{SC} and \acp{RIS} can effectively reduce \ac{EMF} exposure,  the network cannot adapt to user distribution changes \cite{green2009ezri,IprElzanaty:21}. Therefore, a more flexible deployment with airborne-\acp{SC} can assist the existing cellular infrastructure to provide users with better service quality, e.g., coverage and capacity \cite{GiorgettiChianiWin:11,mozaffari2015drone,dai2018energy,HamHamShiAloSha:20,10754/667536}. {\color{black}In \cite{SHARMA20181}, the authors proposed an efficient cell-based allocation method, which provides an optimized \ac{UAV} positioning to enhance the performance of communication systems. Moreover, \acp{UAV} can hover over geographical regions that experience heavy traffic conditions due to natural disasters or mass events to assist the existing  infrastructure in providing users with better service quality \cite{s20216140}.} But the limited onboard energy and flight time impose a critical challenge for the deployment of \acp{UAV} \cite{kishk2019capacity}. In this regard, \acp{tUAV} can be a viable alternative, as they can be powered and backhauled through cables connecting them to \acp{GS} usually located on the rooftop. \acp{tUAV} can outperform regular \acp{UAV} in terms of coverage and capacity \cite{kishk2020tuav,Bushnaq2020Cellular}. Nevertheless, the exposure in the \ac{DL} direction can increase. Also, the position of the \acp{tUAV} and the resource allocation problem on a large scale have not been investigated.

% Although the use of fixed deployed \acp{SC} and \acp{RIS} can effectively reduce \ac{EMF} exposure,  the network cannot adapt to user distribution changes \cite{green2009ezri,IprElzanaty:21}. Therefore, a more flexible deployment with airborne-\acp{SC} can assist the existing cellular infrastructure to provide users with better service quality, e.g., coverage and capacity \cite{GiorgettiChianiWin:11,mozaffari2015drone,dai2018energy,HamHamShiAloSha:20}.  The limited on-board energy and flight time impose a critical challenge for the deployment of \acp{UAV} \cite{kishk2019capacity}. In this regard, \acp{tUAV} can be a viable alternative as they can be powered and backhauled through cables connecting them to \acp{GS} usually located on the rooftop. \acp{tUAV} can outperform regular \acp{UAV} in terms of coverage and capacity \cite{kishk2020tuav,Bushnaq2020Cellular}. Nevertheless, the exposure in the \ac{DL} direction can increase. Also, the position of the \acp{UAV} and the resource allocation problem on a large scale have not been investigated.

In contrast to existing works that mainly consider analyzing the coverage, our goal is to design \ac{EMF}-aware cellular networks with low complexity algorithms to minimize the exposure and guarantee a target \ac{QoS}. To the best of our knowledge, \ac{EMF}-efficient architectures and association algorithms involving \ac{tUAV} carrying green antennas have not been discussed in the literature.

\subsection{Contributions}
%This paper studies the resource allocation to reduce human exposure to \ac{EMF} using airborne and terrestrial network integration. The objective of the proposed framework is to decrease \ac{EMF} exposure while respecting to resource limitations. To the best of our knowledge, using green  \ac{tUAV} to reduce \ac{UL} radiation has not been discussed in the literature to date. Hence, the main contribution of the paper can be summarized as follows:

%In order to alleviate some of the issues in the literature schemes, e.g., the reduced mobility, limited power capacity,  excessive radiation in \ac{DL}, and backhauling, We propose a system with green \ac{tUAV} to minimize exposure by optimizing the transmit power, users' allocation, and the location of \acp{UAV}. More precisely, we propose an architecture where there are several \acp{GS} that \acp{tUAV} can attach to, also each \ac{tUAV} has a limited number of \acp{RB}. This problem tends to be a \ac{MINLP}; hence, we design several efficient algorithms to solve the optimization problem. 

\begin{table}[t]
	\caption{Summary of notations}
	\label{table_notation}
	\centering
	\begin{tabular}{|m{8em}|m{35em}|}
		\hline
		Notation & Description \\ \hline
		$K$     &  Number of users in the considered area \\ \hline
		$\bar{K}$     & Number of   residents including users and non-users in the considered area \\ \hline
		$M$    &    Number of \acp{tUAV}   \\ \hline
		$N$    &    Number of \acp{GS}    \\ \hline
		$J$        &  Number of \acp{gNB}  \\ \hline
		$W_{j, \max}$      &    Number of available \acp{RB} that \ac{gNB} $j$ has    \\ \hline
{\color{black}${\mathsf{L}_{kj}^{\rm{LoS}}}$,  ${\mathsf{L}_{kj}^{\rm{NLoS}}}$   }   &  {\color{black}Path loss of \ac{LOS}, \ac{NLOS} between user $k$ and \acp{gNB} $j$, respectively }        \\ \hline
{\color{black}${r_{kj}}$ }     & {\color{black} Distance between the \ac{gNB}  $j$ and the user  $k$ } \\ \hline
{\color{black}${d_{kj}}$   }   & {\color{black} Distance from the projected position of \ac{gNB} $j$ to user $k$ } \\ \hline
		$p_{kj}^{{\rm{LoS}}}$,  $p_{kj}^{{\rm{NLoS}}}$     &  Probability of \ac{LOS}, \ac{NLOS} between user $k$ and \acp{gNB} $j$, respectively          \\ \hline
		${L_{kj}}$         & Path loss between user $k$ and \ac{gNB} $j$     \\ \hline
		$\epsilon _{kj}$       &   A binary variable indicates the association between user $k$ and \acp{gNB} $j$           \\ \hline
		$\vartheta _{mn}$       &   A binary variable indicates the deployment between \ac{tUAV} $m$ and \ac{GS} $n$       \\ \hline
{\color{black}$\mathrm{SAR}_k^{\text{UL}}$, $\mathrm{SAR}_k^{\text{DL}}$} &  {\color{black} Whole body \ac{SAR} or localized \ac{SAR} normalized to unit transmit power or reference received power density    }   \\ \hline
		$\mathrm{EI}$,  $\mathrm{EI}^\text{UL}$, $\mathrm{EI}^\text{DL}$  &   \Ac{EI} for total, \ac{UL} and \ac{DL}, respectively        \\ \hline
		$R_{kj}^\text{UL}$, $R_{kj}^\text{DL}$    &  Transmit rate between user $k$ and \acp{gNB} $j$ for \ac{UL} and \ac{DL}, respectively           \\ \hline
		$P_{kj}^\text{UL}$, $P_{kj}^\text{DL}$      &   Transmit power between user $k$ and \acp{gNB} $j$ for \ac{UL} and \ac{DL}, respectively          \\ \hline
	{\color{black}	$P_{\max}$  }    &  {\color{black}Maximum transmit power }         \\ \hline
$B$      &  Bandwidth of each \ac{RB}          \\ \hline
		$ \sigma^2 $     &    Thermal noise power         \\ \hline
       {\color{black} $T _{m}$,$T _{\max}$}   &   {\color{black} Tether length of \ac{tUAV} $m$, and the maximum tether length  }  \\ \hline
{\color{black}$\theta _{m}$,$\theta _{\min}$   } &  {\color{black}  Elevation angle of \ac{tUAV} $m$, and the minimum elevation angle   }\\ \hline
{\color{black}$\varphi _{m}$ } &   {\color{black} Azimuth angle of \ac{tUAV} $m$ }  \\ \hline
{\color{black}$\gamma$}  &  {\color{black}  A set which contains all optimization variables  } \\ \hline
{\color{black}$e_{kj}$  }&  {\color{black}  \ac{EMF} exposure generated by user $k$ when he is assigned to \ac{gNB} $j$  } \\ \hline
{\color{black} $\mathrm{SAR}_\mathrm{limit}$ } & {\color{black} Whole body or local \ac{SAR} threshold }\\ \hline
		$\mathbb{I}\left \{ \cdot \right \}$ & Indicator function, which one's output is 1 if conditions are satisfied and 0, otherwise \\ \hline
	\end{tabular}
\end{table}

{\color{black} In order to alleviate some of the issues in the common adopted architectures with \acp{UAV}, e.g., the reduced mobility, limited power capacity,  excessive radiation in \ac{DL}, and backhauling, {\color{black}we} propose a system with green \acp{tUAV} to minimize the \ac{EMF} exposure by optimizing the transmit power, users' allocation, and the location of the \acp{tUAV}. More precisely, we propose an architecture where there are several \acp{GS} that \acp{tUAV} can attach to. Moreover, each \ac{tUAV} has a limited number of \acp{RB}. This problem tends to be a \ac{MINLP}; hence, we design several efficient algorithms to solve the optimization problem.
The main contributions of the paper can be summarized as follows:
\begin{itemize}
    \item We propose a novel architecture for cellular networks that exploit green \acp{tUAV} as an efficient tool to minimize the \ac{EMF} exposure.
    \item We formulate the optimization problem aiming to minimize the users' exposure taking into consideration the \ac{RB} constraint of each \ac{tUAV}, the maximum transmit power of users, and the finite number of \acp{GS}.
   % \item We propose an alternate optimization algorithms for the problem.
    \item We design alternate optimization algorithms to optimize the users' association,  deployment of \acp{tUAV}, and fine positioning of the \acp{tUAV} within their hovering area. The complexities of the proposed algorithms are analyzed.
    \item We consider the dual problem, i.e., maximizing the average \ac{UL} rate with a constraint on the exposure, e.g., the \ac{ICNIRP} limit. Then, we present algorithms for the efficient deployment of the \acp{tUAV} for the dual problem.
%Furthermore, this problem can also be solved by the proposed algorithms, after some modifications.
    \item We assess the performance of the proposed scheme against various architectures, e.g., different schemes with \acp{tUAV} and fixed \acp{SC} that assist the \ac{UL} and/or \ac{DL}.
    %Formulate an optimization problem aiming to minimize the users’ exposure in a populous urban environment taking into consideration the \ac{RB} constraint of each \ac{tUAV}.
%    \item Due to the non-convexity of the formulated problem, we propose a valid three-step approach. Firstly, we use adopted greedy algorithm to optimally determine the users’ association. Then, we solve the deployment of \acp{tUAV} by using modified $K$-mean algorithm and an algorithm based on 2D \ac{SR} process. In the last step, we adjust tether length and inclination angle for each \ac{tUAV} based on 3D \ac{SR} process and golden search. 
%    \item The complexity and effect of different algorithms are analyzed. Also, the performance of various types of \ac{tUAV} which densify \ac{UL} and/or \ac{DL} on reducing radiation is compared. Moreover, we compare fixed \ac{SC} and \ac{tUAV} in terms of \ac{EMF} exposure and users’ satisfaction.
\end{itemize}}

\subsection{Notations and Organization}

The main notations used in the paper are listed in Table~\ref{table_notation}. 
The remainder of the paper is organized as follows {\color{black}:} Section \ref{sec:sys} explains the considered system model. The problem formulation for minimizing the exposure with achieving a target \ac{QoS} is given in Section \ref{sec:pro}. Section \ref{sec:alg} details three submodularity of the objective functions and heuristic algorithms. Next, we present the dual problem, where we propose a novel design of cellular networks with an \ac{EMF} constraint in Section \ref{sec:dual}. Then, simulation results are provided in Section \ref{sec:num}. Lastly, Section \ref{sec:con} concludes the paper with a few remarks. 

%\subsection{Notations}
%

\section{System Model} \label{sec:sys}

\begin{figure}[t!]
	\centering
	\includegraphics[width=1\linewidth]{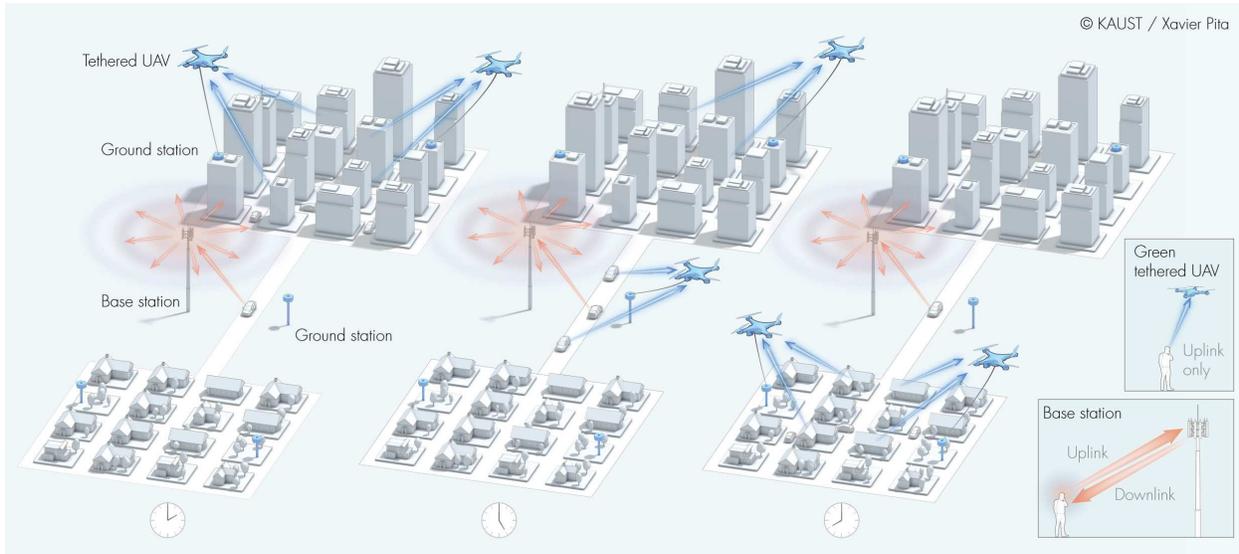}
	\caption{An illustrative use case for the proposed architecture with green \acp{tUAV} and a central \ac{BS}. From left to right, the locations of \acp{tUAV} in the morning, afternoon, and evening are adapted to the users' spatial distribution. In the lower right corner, the model of green \ac{tUAV} and \ac{BS} are shown separately.}
	\label{fig:system model_1}
\end{figure}

We consider a \acp{tUAV}-enabled system consisting of one macro \ac{BS}, $K$ users and $M$ \acp{tUAV} with green antennas, i.e., receiving only antennas \cite{green2009ezri}, as show in Fig.~\ref{fig:system model_1}. We also consider $N$ \ac{GS} placed in fixed locations at the top of some buddings and connected to links to provide backhaul and power to the \acp{tUAV}, where each \ac{tUAV} can choose one \ac{GS} to connect to. We consider that the \ac{BS} is located in the center of the area and is responsible for resource management.\footnote{Although distributed algorithms have been popular over the last decade \cite{UAV2015chmaj,UAV2018TRO}, we consider centralized schemes where the \ac{BS} orchestrates the \acp{tUAV} and users' resources.}

\newpage
\subsection{Channel Model}  \label{sec:channel}
%\begin{enumerate}
%\item Channel Model
{\color{black} 
	 In the considered scenario, each user can receive two main signals, i.e., \ac{LOS} signal and strongly reflected \ac{NLOS} signal from \acp{gNB} \cite{Al-Hourani2014air-to-ground},  the corresponding path loss for the \ac{LOS} and \ac{NLOS} propagation can be written as
\begin{equation}
{\mathsf{L}_{kj}^{\rm{LoS}}}=\left (  \frac{4\pi {f_c}}{c} \right )^2 {r_{kj}}^{\alpha_{\rm{l}}}F _{\rm{l}}\,G_{\rm{l}},
\end{equation}
\begin{equation}
{\mathsf{L}_{kj}^{\rm{NLoS}}}=\left (  \frac{4\pi {f_c}}{c} \right )^2 {r_{kj}}^{\alpha_{\rm{n}}}F _{\rm{n}}\,G_{\rm{n}},
\end{equation}
respectively, where $f_c$ is the carrier frequency, $c$ is the speed of light, $r_{kj}$ is the distance between the \ac{gNB}  $j$ and the user  $k$,  $\alpha_{\rm{l}} $ and $\alpha_{\rm{n}}$ are the path loss exponent for the \ac{LOS} and \ac{NLOS} links, respectively. The symbols $F_{\rm{l}}$ and $F_{\rm{n}}$ denotes the \acp{RV}  representing the power loss due to large scale fading and shadowing  with mean ${{\eta  _{\rm{l}}}}$ and ${{\eta  _{\rm{n}}}}$ \cite{LAP2014Lardner,UAV2017Alzenad,UAV2018Alzenad}.
%and they are typically modeled as lognormal \acp{RV}
Moreover, $G_{\rm{l}}$ and $G_{\rm{n}}$ are  \acp{RV} accounting for the power loss due to small scale fading with mean unity \cite{qin2021influence}.

%, as they typically follow gamma distributions with shape and scale parameters $(a_{\rm{l}},\frac{1}{a_{\rm{l}}})$ and  $(a_{\rm{n}},\frac{1}{a_{\rm{n}}})$, for Nakagami fading with shape factor $a_{\rm{l}}$ and $a_{\rm{l}}$, induced in the \ac{LOS} and \ac{NLOS} links,  respectively. %The mean of the fading gains is the product of the shape and scale parameters, which is one . 

% As we cannot get an exact \ac{LOS} and \ac{NLOS} map or the exact value of addition random losses, w
We adopt the channel model based on the probabilistic \ac{LOS} model, averaged over the large and small scale fading \cite{LAP2014Lardner,UAV2017Alzenad,UAV2018Alzenad}. The \ac{LOS} and \ac{NLOS} links  have different probabilities of occurrence, where the probability of \ac{LOS} can be written as
\begin{equation}\label{eq:los}
	p_{kj}^{{\rm{LoS}}}({r_{kj}},{d_{kj}}) = \frac{1}{{1 + a \, \exp \left( { - b\left( {\frac{{180}}{\pi }\arctan \frac{{\sqrt {r_{kj}^2 - d_{kj}^2} }}{{{d_{kj}}}} - a} \right)} \right)}},
\end{equation}
where $d_{kj}$ is the distance from the projected position of \ac{gNB} $j$ to user $k$, and $a$ and $b$ are constants that depend on the environment. For the probability of the \ac{NLOS} signal, we have $p_{kj}^{{\rm{NLoS}}}=1 - p_{kj}^{{\rm{LoS}}}$. Therefore,  the average path loss can be writte as 
\begin{equation}\label{eq:avgpathloss}
	{L_{kj}}({r_{kj}},{d_{kj}}) = {\left( {\frac{{4\pi {f_c}}}{c}} \right)^{2} }\left( {{\eta  _{\rm{l}}}{r_{kj}}^{\alpha_{\rm{l}}}p_{kj}^{{\rm{LoS}}}({r_{kj}},{d_{kj}}) + {\eta  _{\rm{n}}}{r_{kj}}^{\alpha_{\rm{n}}}p_{kj}^{{\rm{NLoS}}}({r_{kj}},{d_{kj}}))} \right).
\end{equation}
}

% We adopt a common channel model  that the user can receive two main signals, i.e., \ac{LOS} signal and strongly reflected \ac{NLOS} signal, from \acp{gNB} \cite{Al-Hourani2014air-to-ground}.\footnote{In this paper, a \ac{gNB} can be either the macro \ac{BS} or a \ac{tUAV}, where the total number of \acp{gNB} is $J \triangleq  M+1$.} These two links have different probabilities of occurrence, where the probability of \ac{LOS} can be written as
% \begin{equation}
% 	p_{kj}^{{\rm{LoS}}}({r_{kj}},{d_{kj}}) = \frac{1}{{1 + a \, \exp \left( { - b\left( {\frac{{180}}{\pi }\arctan \frac{{\sqrt {r_{kj}^2 - d_{kj}^2} }}{{{d_{kj}}}} - a} \right)} \right)}},
% \end{equation}
% where $d_{kj}$ is the distance from the projected position of \ac{gNB} $j$ to user $k$, $r_{kj}$ is the distance between the \ac{gNB} and the user, and $a$ and $b$ are constants that depend on the environment. The path loss can be given as
% \begin{equation}
% 	{L_{kj}}({r_{kj}},{d_{kj}}) = {\left( {\frac{{4\pi {f_c}{r_{kj}}}}{c}} \right)^\alpha }\left( {{\xi _{{\rm{LoS}}}}p_{kj}^{{\rm{LoS}}}({r_{kj}},{d_{kj}}) + {\xi _{{\rm{NLoS}}}}(1 - p_{kj}^{\rm{LoS}}({r_{kj}},{d_{kj}}))} \right),
% \end{equation}
% where $p_{kj}^{{\rm{NLoS}}}=1 - p_{kj}^{{\rm{LoS}}}$ is the probability of \ac{NLOS}, $f_c$ is the carrier frequency, $c$ is the speed of light, $\alpha $ is the path loss exponent, and the parameters ${{\xi _{{\rm{LoS}}}}}$ and ${{\xi _{{\rm{NLoS}}}}}$ are the losses due to shadow effect and reflection.

\subsection{\acs{UL} Association } \label{sys:ULass}
%\item \ac{UL} Associations \label{sys:ULass}

%We consider two type of associations: the \ac{UL} association between \acp{UAV}(or \ac{BS}) and users, and deployment association between \acp{UAV} and \acp{GS}.

  In the system, \acp{tUAV} with green antennas are designed to reduce exposure to \ac{EMF}. In the \ac{DL}, all users are connected to the \ac{BS}. For \ac{UL}, some users are connected to \acp{tUAV}, while the rest can be associated with the \ac{BS}. Furthermore, we introduce a binary variable $\epsilon_{kj}$, which is equal to $1$ if  user $k$ is associated with \ac{gNB} $j$ and $0$ otherwise, i.e.,
\begin{equation}
	\epsilon _{kj}= \begin{cases}
		1 & \text{if user } k \text{ is associated with \ac{gNB} }  j\\ 
		0 & \text{otherwise} 
	\end{cases}.
\end{equation}
We also consider that the number of available spectrum resources at \ac{gNB} $j$ is $W_ {j,\max}$ \acp{RB}, and each \ac{RB} has a bandwidth of $B$ that can be allocated to a single user for either \ac{UL} or \ac{DL} communications, i.e., the maximum number of users that can be associated with that \ac{tUAV} is $W_ {j,\max}$. Hence, the following conditions should be satisfied
\begin{equation}
	\label{eq.user}
\sum_{k=1}^{K}\epsilon _{kj}\leq W_ {j,\max}, \;  \forall j \in \mathcal{J} \triangleq \{0,1,\cdots,M\}, \text{ }\sum_{j=0}^{J-1}\epsilon _{kj} = 1 , \; \forall k \in \mathcal{K} \triangleq \{1,2,\cdots,K\}.
\end{equation}
%where $W_ {j,\max}$ is the number of resource blocks in \ac{tUAV} $j$ which corresponds to . 
For convenience, we consider that the index $j=0$ refers to the \ac{BS}, while the other elements represent the \acp{tUAV}.

%Moreover, we using vector $\boldsymbol{\epsilon}_{k} = \left ( {\epsilon}_{k1}, {\epsilon}_{k2},  \cdots,  {\epsilon}_{kJ}   \right ) $ to  represent the association between user $k$ and all \acp{gNB}, and elements in the row vector satisfy \eqref{eq.user}.

\subsection{Deployment of \acs{tUAV}}
%\item Deployment of \ac{UAV}

 For the deployment of \ac{tUAV}, each \ac{tUAV} can only be connected with one \ac{GS} at a certain time. More precisely, we have
\begin{equation}
\sum_{n=1}^{N}\vartheta _{mn} = 1 , \; \forall m \in \mathcal{M} \triangleq \{1,2,\cdots,M\},
\end{equation}
%\end{enumerate}
where $\vartheta _{mn}$ is a binary variable that equals $1$ only if the $m^{\text{th}}$ \ac{tUAV}  is attached to the $n^{\text{th}}$ \ac{GS}.
%%%%%%%%%%%%%%%%%%%%%%%%%%%%%%%%%%%%%%%%
\section{Problem Formulation} \label{sec:pro}   
%In this work, we are interested in whether it is possible to reduce the \ac{EMF} exposure due to \ac{UL} transmissions while guaranteeing a certain \ac{QoS} in terms of the users' \ac{UL} rate requirement.
In this section, we  describe a metric usually adopted to quantify the \ac{EMF} exposure and propose a power allocation strategy for the devices.  Then, we formalize the optimization problem such that the \ac{EMF} exposure due to \ac{UL} transmissions is minimized while guaranteeing a certain \ac{QoS} in terms of the users' \ac{UL} rate requirement.

\subsection{EMF Assessment}
%	\item EMF Assessment
	To evaluate the EMF radiation in a cellular network, we adopt an exposure metric called \ac{EI} \cite{Tes2014LEXNET}, which is able to model the population exposure to \ac{EMF}, accounting for various elements such as different technologies, environments, and usages. The $\mathrm{EI}$ can be written as
%	\text{EI}=\sum _{i}\sum _{j}\sum _{k}\sum _{l}\sum _{m}f\left ( \text{SAR}^\text{UL},\bar{P}_\text{TX},t^\text{UL}, \text{SAR}^\text{DL},\bar{S}_\text{RX},t^\text{DL}\right )	
	\begin{equation}
		\label{eq.ei}		
		\mathrm{EI}=\sum _{l}\sum _{e}\sum _{r}\sum _{g}f\left ( \mathrm{SAR}^\text{UL},\bar{P}_\text{TX}, \mathrm{SAR}^\text{DL},\bar{S}_\text{RX}\right ),	
    \end{equation}
	where the exposure in each case can be represented as a function $f$ of the mean  emitted  power  $\bar{P}_\text{TX}$ from the device, and mean received power density $\bar{S}_\text{RX}$, and the \ac{SAR} reference levels $\mathrm{SAR}^\text{UL}$ and $\mathrm{SAR}^\text{DL}$. 
	The \ac{EI} depends on the population age group $l$ (e.g., adults, seniors, and children), environments $e$ (e.g., indoor and outdoor), \acp{RAT} and layers $r$ (e.g., WiFi, 5G, \ac{NB-IoT}, and macro, micro), usage types $g$ (e.g., voice, data, and machine-type communication). 
	
In this paper, we consider one \ac{RAT} which is next-generation cellular network and two layers: \textit{i)} macro \ac{BS}; \textit{ii)} \ac{tUAV} carrying a \ac{SC} with green antennas. Then for simplicity, we consider adults as the population, while data and voice are the considered usage types. More precisely, now the \ac{EI} for both \ac{UL} and \ac{DL} can be given by
	
	\begin{equation}\label{eq:EIboth}
			\mathrm{EI}=\mathrm{EI}^\text{{UL}}+\mathrm{EI}^\text{{DL}},
	\end{equation}
with $\mathrm{EI}^\text{{UL}}$  representing the \ac{UL} exposure given as
	%then, \ac{UL} exposure $\mathrm{EI}^\text{{UL}}$ which related to users' association matrix $\bf{\epsilon} $ and \acp{tUAV}' deployment $\vartheta $ is given as
	\begin{equation}
	\label{eq:ul}
	\mathrm{EI}^{\text{UL}}=\sum_{k=1}^{K}\mathrm{SAR}_{k}^{\text{UL}}\, P_{{k}}^{\text{UL}},
	\end{equation}
	where $\mathrm{SAR}_{k}^{\text{UL}}$  is the whole body \ac{SAR} or localized \ac{SAR} normalized to unit transmit power, which depend on the required service  and the posture (how the device is hold), $P_{{k}}^{\text{UL}}$ is the transmit power of users $k$.
	For the \ac{DL} exposure $\mathrm{EI}^\text{{DL}}$, it can be obtained as
	\begin{equation}
		\label{eq.dl}
	\mathrm{EI}^{\text{DL}} =\sum_{k=1}^{\bar{K}}\sum_{j=0}^{J-1} \mathrm{SAR}^{\text{DL}}\, P_{jk}^{\text{DL}},
	\end{equation}	
	where $\mathrm{SAR}^{\text{DL}}$ is the \ac{DL} \ac{SAR} value which is normalized to a
	reference received power density of $1$~W/m$^2$ and $\bar{K}$ is the number residents in the region (including users and non-users), i.e., $\mathcal{K} \subseteq \mathcal{\bar{K}}$, 
	$ P_{jk}^{\text{DL}} $ is the received power density at resident $k$ from \ac{gNB} $j$.

%where $\bf{\epsilon} $ and $\bf{\vartheta} $ are \ac{UL} association matrix and deployment association matrix respectively, ${SAR}_{g}^{\text{UL}}$ can be whole body \ac{SAR} in the usage $g$,  $P_{kg}^{\text{UL}}$ is the \ac{Tx} power of users $k$ in usage $g$, and $t_{kg}^{\text{UL}}$ is the time of user $k$ spent in the usage $g$.
	
%	It can be seen that the exposure index is the summation of functions about \ac{SAR} value $\text{SAR}^\text{UL},\text{SAR}^\text{DL} $, 
%	mean  emitted  power  $\bar{P}_\text{TX}$, mean received density of power $\bar{S}_\text{RX}$, and corresponding duration time $t^\text{UL}, t^\text{DL}$, over diffentent times $i$, population $j$, environments $k$, \acp{RAT} $l$ and usage types $m$. Those parameters obtained from numerical dosimetry or network simulations. 
%However, due to the proximity of the device to the body, we only focus on the \ac{UL} exposure which is more important than \ac{DL} exposure. Then, we only consider mobile phone as devices, and adults as users to simplify analysis. So the \ac{UL} exposure in the network given by:
	
%	\begin{equation}
%		\mathrm{EI}^{\text{UL}}\left ( \bf{\epsilon ,\vartheta } \right )=\sum_{k=1}^{K}\sum_{g=1}^{G} {SAR}_{g}^{\text{UL}}\, P_{kg}^{\text{UL}}\, t_{kg}^{\text{UL}}
%	\end{equation}	
	\subsection {Transmit Rate and Power}
% 	{\color{black} 	Each user should be associated with one of the \acp{gNB} through a distinct \ac{RB}. therefore there  i.e., all the access transmissions between the \acp{gNB} and users operate sparsely. Hence, there is no interference between users, and only thermal noise needs to be considered.}
% 	Each user is allocated a distinct \ac{RB}, hence, the interference between users can be ignored, and only thermal noise needs to be considered.
	Let $P_{kj}$ be the transmit power of user $k$ when connected to the $j^{\text{th}}$ \ac{gNB}, his  \ac{UL} rate can be expressed as
	%The  rate $R_{kj}$ between \ac{gNB}  $j$ and user $k$ can be expressed as,
	\begin{equation} \label{eq:rate}
		R_{kj}^\text{UL}=B\,\text{log}\left ( 1+\frac{P_{kj}^\text{UL}}{\sigma^2L_{kj}} \right ),
	\end{equation}
	where $\sigma^2=k_{B}\,T\,B$ is the noise variance with $k_B$ being the Boltzmann constant and $T$ being the temperature in Kelvin.\footnote{\color{black}  Since each user is allocated a distinct \ac{RB}, there is no interference between users.}
	%$k_B$ is the Boltzmann constant, $T$ is the temperature and the product of them is the noise power  $\sigma^2=k_{B}TB$. 
	%For \ac{UL}, using \eqref{eq:rate} to calculate the transmit power $P_{kj}^\text{UL}$ is sufficient for calculating the total \ac{UL} exposure. But for \ac{DL},  we also need to calculate the total transmit power for \ac{gNB}. 
		Next, we focus on \ac{DL} and assume that \acp{gNB} are powerful enough to meet \ac{DL} rate requirement of the connected user, and total transmit power of \ac{gNB} $j$ can be given as
	
	\begin{equation}\label{eq:downlnikrefpower}
		{P_j^\text{DL}} =  \sum_{k=1}^{K_j} {\sigma^2L_{kj}} \left ( 2^{\frac{R_{kj}^\text{DL}}{B}}-1 \right ),
	\end{equation}
where  the subset $K_j$ represents users connected to \ac{gNB} $j$. The required \ac{DL} power of each user is a function of the \ac{DL} rate $R_{kj}^{\text{DL}}$ and the path loss $L_{kj}$.

\subsection{Complete Optimization Model}
In \eqref{eq:EIboth}, we consider a {\color{black}general} model where  both \ac{UL} and \ac{DL} exposure are accounted for. 
{\color{black} The total \ac{EMF} exposure is mainly dominated by \ac{UL} due to the proximity of the device to the body \cite{de2014emf,chiaraviglio2020health}, as shown also in Section~\ref{sec:num}.}
%{\color{black} The total \ac{EMF} exposure is mainly dominated by \ac{UL} due to the proximity of the device to the body \cite{de2014emf,chiaraviglio2020health}, as shown also in Section~\ref{sec:num}. By \eqref{eq:ul}, we know that minimizing the \ac{UL} radiation is equal to minimizing the weighted sum of transmit power. }
%the total \ac{EMF} exposure is mainly dominated by \ac{UL} due to the proximity of the device to the body \cite{de2014emf,chiaraviglio2020health}, as shown also in Section~\ref{sec:num}. 
{\color{black} Therefore, we only focus on the \ac{UL} exposure, and our objective is to optimally deploy the \acp{tUAV} and associate the users in order to minimize the \ac{UL} \ac{EI}.  Moreover, there should be a minimal power received by \ac{gNB} to satisfy the required rate, which depends on the path loss and transmit power.  In fact, the transmit power should compensate for the path loss, which depends on the association between \acp{tUAV} and \acp{GS} $\vartheta _{mn}$, the association between users and \acp{gNB} $\epsilon _{kj}$, as indicated by \eqref{eq:rate}. Also, the path loss depends the distance between the \ac{tUAV} and the users, $r_{kj}$, as in \eqref{eq:avgpathloss}, which, by its turn on, depends on the tether length $T _{m}$,  elevation angle $\theta _{m}$, and  azimuth angle $\varphi _{m}$ of the tether, as shown in Fig.~\ref{fig:system model_2}.
%We set up the rectangular coordinates with a \ac{GS} as the origin, a user is on the ground, and a \ac{tUAV} is in the hovering area.
\begin{figure}[t!]
	\centering
	\includegraphics[width=0.6\linewidth]{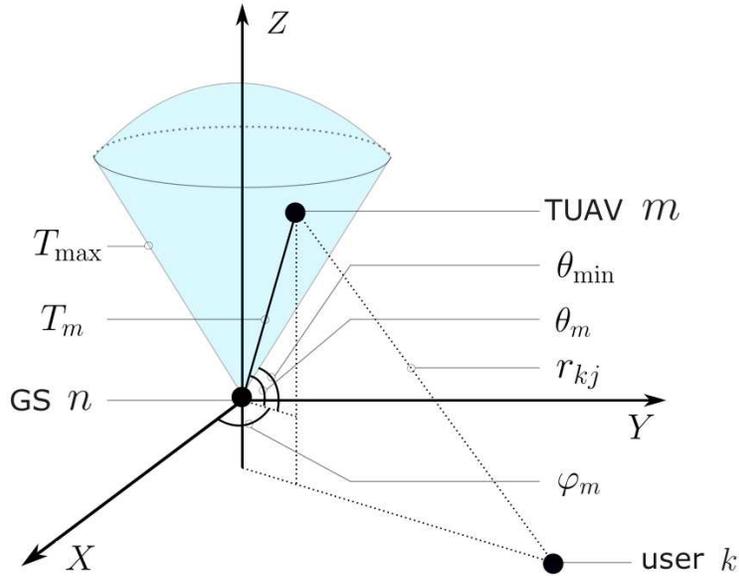}
	\caption{\color{black}A schematic diagram of variables.}
	\label{fig:system model_2}
\end{figure}
		
		Let $\boldsymbol{\gamma} \triangleq \left\{ \vartheta _{mn},\epsilon _{kj},T _{m},\theta _{m},\varphi _{m} \right\}, j \in \mathcal{J}, k \in \mathcal{K}, m \in \mathcal{M}, n \in \mathcal{N}$ represents all optimization variables, i.e., the association between users and \acp{gNB},  association between \acp{tUAV} and \acp{GS}, tether length, and inclination angles.} Then, our optimization problem can be formulated as
%Therefore, we only focus on \ac{UL} exposure and our objective is to optimally deploy the \acp{tUAV} and associate the users  in order to minimize the \ac{UL} exposure index. Let $\boldsymbol{\gamma} \triangleq \left\{ \vartheta _{mn},\epsilon _{kj},\theta _{m},T _{m},\varphi _{m} \right\}, j \in \mathcal{J}, k \in \mathcal{K}, m \in \mathcal{M}, n \in \mathcal{N}$ represents all optimization variables, i.e., the association between users and \acp{gNB},  association between \acp{tUAV} and \acp{GS}, tether length, and inclination angles.  Then, our optimization problem can be formulated as

% \epsilon _{k} is a vector, a 

%$\boldsymbol{\epsilon}_{k} = {\epsilon _{k,j}, \text{for} j \in \mathcal{J}}$

\begin{subequations} 
	\begin{alignat}{2}
		\text{(P0): }&\underset{ \boldsymbol{\gamma}}{\text{minimize}}        &\qquad& \sum_{k=1}^{K} \mathrm{SAR}_{k}^{\text{UL}}\, P_{{k}}^{\text{UL}} \label{eq:optProb}\\
%		\text{(P0): }&\underset{ \vartheta _{mn},\epsilon _{kj},\theta _{k},T _{k},\varphi _{k}}{\text{minimize}}        &\qquad& \sum_{k=1}^{K} \mathrm{SAR}_{k}^{\text{UL}}\, P_{{k}}^{\text{UL}} \label{eq:optProb}\\
		&\textrm{subject to:}    &      & R_{k}^\text{UL}(\boldsymbol{\gamma})\geq R_{k,\min}, \; \forall k \in \mathcal{K}, \label{eq:constraint6} \\
		&                  &      &   {P_{k}^\text{UL}(\boldsymbol{\gamma})\leq P_{\max}, \; \forall k \in \mathcal{K} ,} \label{eq:constraint7} \\
		&                  &      &\sum_{k=1}^{K}\epsilon _{kj}\leq W_{j,\max} , \;  \forall j \in \mathcal{J} , \label{eq:constraint1}\\
		&                  &      & \sum_{j=0}^{J-1}\epsilon _{kj}= 1 , \; \forall k \in \mathcal{K} , \label{eq:constraint2}\\
		&                  &      & \sum_{n=1}^{N}\vartheta _{mn}= 1 \text{ }, \; \forall m \in \mathcal{M}, \label{eq:constraint3}\\
		&                  &      & \theta _{\min }\leq \theta _{m}\leq \frac{\pi}{2}, \; \forall m \in \mathcal{M}, \label{eq:constraint4}\\
		&                  &      & 0\leq T _{m}\leq T_{\max}, \; \forall m \in \mathcal{M}, \label{eq:constraint5}\\
		&                  &      & \varphi _{m}\in \left [ 0,2\pi \right ],\; \forall m \in \mathcal{M}, \label{eq:constraint8}
	\end{alignat}
\end{subequations}
{\color{black}
where the objective function is the exposure index that can be seen as a weighted sum of the users' transmit power such that the weight of each user is his constant reference \ac{SAR}, $\mathrm{SAR}_{k}^{\text{UL}}$.} %which is a constant for each user. % depending on user $k$.
For the constraints, (\ref{eq:constraint6}) guarantees achieving a minimum required  \ac{UL} data rate,   \eqref{eq:constraint7} limits the transmit power of users due to the hardware restrictions, and 
%Next, constraints (\ref{eq:constraint1})-(\ref{eq:constraint2}) guarantees the \ac{UL} access association of each user, constraint (\ref{eq:constraint3}) indicate \ac{tUAV}'s deployment, and 
%
 (\ref{eq:constraint4})-(\ref{eq:constraint8}) ensure that \acp{tUAV} are within their hovering area with $\theta_{m}> \theta _{\min}$ for safety reasons. The problem \textbf{P0} is mathematically challenging because it includes integer and continuous variables, as well as non-convex objective function and non-convex constraints, which makes it a \ac{MINLP}. 

%\begin{subequations} 
%	\begin{alignat}{2}
%		\text{(P1): }&\underset{P_{k}^\text{UL}, \vartheta _{mn},\epsilon _{kj},\theta _{k},T _{k},\varphi _{k}}{\text{maximize}}        &\qquad& \sum_{k=1}^{K}  R_{kg}^{\text{UL}} \label{eq:optProb}\\
%		&\textrm{subject to:}    &      & R_{k}^\text{UL}\geq R_{k,\min}^\text{UL}, \text{ } \forall \text{ } k, \label{eq:constraint6} \\
%		&                  &      & P_{k}^\text{UL}\leq P_{\max}^\text{UL}, \text{ } \forall \text{ } k, \label{eq:constraint7} \\
%		&                  &      & \mathrm{SAR}_{g}^{\text{UL}}\, P_{kg}^{\text{UL}} \leq \mathrm{SAR}_{limit}, \text{ } \forall \text{ } k, \label{eq:constraint9} \\
%		&                  &      &\sum_{k=1}^{K}\epsilon _{kj}\leq W_j , \text{ } \forall \text{ } j, \label{eq:constraint1}\\
%		&                  &      & \sum_{j=1}^{J}\epsilon _{kj}= 1 , \text{ } \forall \text{ } k, \label{eq:constraint2}\\
%		&                  &      & \sum_{n=1}^{N}\vartheta _{mn}= 1 \text{ }, \text{ }\forall m, \label{eq:constraint3}\\
%		&                  &      & \theta _{\min }\leq \theta _{k}\leq \frac{\pi}{2}, \text{ } \forall \text{ } k, \label{eq:constraint4}\\
%		&                  &      & 0\leq T _{k}\leq T_{\max}, \text{ } \forall \text{ } k, \label{eq:constraint5}\\
%		&                  &      & \varphi _{k}\in \left [ 0,2\pi \right ],\text{ } \forall \text{ } k. \label{eq:constraint8}
%	\end{alignat}
%\end{subequations}
%
%	\begin{equation} 
%	P_{k}^\text{UL}=\text{min}\left \{ P_{\text{max}}^\text{UL},   \frac{\mathrm{SAR}_{limit}}{\mathrm{SAR}_{g}} \right \}	
%\end{equation}

%\end{enumerate}

\section{Optimization Algorithms}\label{sec:alg}
In order to solve the target optimization \textbf{P0}, we propose an alternate optimization algorithm, where we split the problem \textbf{P0} into three sub-optimization programs, \textit{(i)} finding the association matrix, \textit{(ii)} determining the deployment of \acp{tUAV}, and \textit{(iii)} adjusting the tether length and angles for each \ac{tUAV}. However, it is still complicated to solve each step due to the numerous combinations or non-convex continuous problem. Thus, we propose low complexity heuristic algorithms to find the optimal 3D placement of \acp{tUAV} and users association to minimize the \ac{EMF} exposure.
 \subsection{Power Allocation}
 The actual rate should be larger than or equal to the required rate according to constraint \eqref{eq:constraint6}. Since the \ac{EI} monotonically increases with the users' transmit power, the optimal power is the one that achieves the required rate with equality. However, there is also a limitation in power due to hardware limitations, as seen in constraint \eqref{eq:constraint7}. 
%Notice constraint \eqref{eq:constraint6}, the actual rate should be larger than or equal to the required rate, so we need to make it equal to minimize the EMF exposure. However, there is also a limitation in power, as seen in constraint \eqref{eq:constraint7}. Then, these two constraints may not be satisfied simultaneously. In numeric results, we will show how the satisfaction rate varies with the required rate. 
%If the power required to achieve the target rate is less than the maximum power, then the actual rate would equal the target rate. However, in some cases, the user is far away from the \ac{gNB}, and the path loss between them is too large. 
Therefore,  the  transmit power $\PN$ can be allocated as
	\begin{equation} \label{equ:rate}
	\PN=\text{min}\left \{ P_{\text{max}}, {\sigma^2}{L_{{{j}_{k}^{*}}}} \left ( 2^{\frac{R_{k,\min}}{B}}-1 \right )\right \},	
	\end{equation}
	where $P_\text{max}$ is the maximum transmit power of the users, and ${{j}_{k}^{*}}$ is the index of serving \ac{gNB} for user $k$, i.e., $\epsilon_{k{{j}_{k}^{*}}}=1$. It can be seen that power allocation scheme consider the minimum between required  and maximum transmit power, where the power constraint is more dominant than that of rate, because it is a hardware constraint that the mobile equipment cannot exceed. If the required transmit power to achieve the rate is larger than the maximum transmit power, the problem will be infeasible. 
	%Then, these two constraints may not be satisfied simultaneously, and the problem can be infeasible. 
	Therefore, the achieved rate will be less than the required one.
	%In this regard, we define the \emph{satisfied-users ratio} as the ratio of the number of users who satisfy their rate constraint \eqref{eq:constraint6} to that of total active users. 
	
	%And combined with \eqref{eq:rate}, the rate would be $R_{kj}^\text{UL}=B\text{log}\left ( 1+\frac{P_{kj}^\text{UL}L_{kj}}{k_{B}TB} \right )$.

%So now, we can remove the rate constraint (\ref{eq:constraint6}) and power constraint \eqref{eq:constraint7} from the program. 

%\subsection{Associations Optimization for Given \acp{tUAV}' Locations} 
\subsection{Optimize the Association between Users and \acsp{gNB}}

	For arbitrary given locations of the \acp{tUAV}, we focus on how to associate each user with the appropriate \ac{gNB} for the \ac{UL}. Under this conditions, the association problem can be formulated as
	%ignoring the limits of accessibility, 
	
	%First, given the initial location of \acp{tUAV}, we purpose to find the optimal association matrix for users, and now \textbf{P0} can be simplified as,
	
	\begin{subequations} 
		\begin{alignat}{2}
			\text{(P1): }&\underset{ \epsilon _{kj}}{\text{minimize}}        &\qquad& \sum_{k=1}^{K}  \mathrm{SAR}_{k}^{\text{UL}}\, \PN \\
			&\textrm{subject to:}& & {\color{black}\eqref{eq:constraint6},
			\eqref{eq:constraint7}},
			\eqref{eq:constraint1}, \eqref{eq:constraint2},
		\end{alignat}
	\end{subequations}
where we consider the power allocation strategy in \eqref{equ:rate}.	
	The problem is also \ac{NP}-hard, as we have to go through all the combinations between users and \acp{gNB}.
	{\color{black}
In order to solve the problem, we propose an algorithm, which starts with each user connected to the best \ac{gNB},  i.e., the one corresponding to the minimum path loss. Then, for the \acp{gNB} with an excessive number of connected users, we will change the connection of some users. That is, some users will be served by other \acp{gNB} which are not fully occupied, rather than the best one. Moreover, the users to be moved are selected such that the induced increase in radiation, which results from the non-optimal selection of the \ac{gNB}, is minimal. 
	
	More precisely, the increase in exposure when user $k$ is connected to the $j^\text{th}$ \ac{gNB} instead of the best choice $j^*$ can be quantified as $\bar{g}_{kj} \triangleq e_{kj}-e_{kj^*}$, where $e_{kj} \triangleq \SAR_{k}^{\text{UL}} \, P_{kj}$ is the \ac{EMF} exposure generated by user $k$ when assigned to \ac{gNB} $j$. Let us define
	%$\mathcal{K}_j$ as the set of users connected to \ac{gNB} $j$,
	$\bar{\mathcal{K}}$ as the set of users connected to excessive occupied \acp{gNB}, and  $\bar{\mathcal{J}}$ as the set of \acp{gNB} that are still not fully occupied. If the number of connected users exceeds the resource limit at some \acp{gNB}, we search for the user $k_{\mathrm c} \in \bar{\mathcal{K}}$ with the smallest induced increase in radiation (i.e., smallest $\bar{g}_{kj}$ for all $k\in \bar{\mathcal{K}}$ and $j \in \bar{\mathcal{J}}$) to be moved first. 
This user $k_{\mathrm c}$ will be moved to the \ac{gNB} that minimizes the induced radiation when the user moves.  This process is repeated until each \ac{gNB} is associated with a number of users that do not exceed the available resources. 

For the next stage, where we optimize the location of the \acp{tUAV}, we need to quantify the role that each \ac{tUAV} contributes for reducing the \ac{EMF}. This contribution of the $j^\text{th}$ \ac{tUAV} to reduce the radiation can be defined as $ g_j \triangleq \sum \limits_{k \in \mathcal{K}_j } g_{kj}, \forall j \in \mathcal{M} $, where $g_{kj} \triangleq e_{k0} - e_{kj}$ is the gain, when user $k$ is connected to the $j^\text{th}$ \ac{tUAV} instead of the \ac{BS}, as described in \textbf{Algorithm} \ref{alg.limited_1}.
% \begin{equation}
%  [k_{c},   j_{k}^{*}]= \arg \min_{k\in \bar{\mathcal{K}},j \in \bar{\mathcal{J}}} \bar{g}_{kj}
% \end{equation}
}

 Next, we analyze the complexity of \textbf{Algorithm} \ref{alg.limited_1}, which is mainly composed of two parts: \textit{i)} the initial association, \textit{ii)} users' re-association due to the limited resources at \acp{tUAV}. The algorithm's complexity mainly comes from the first stage, that is in the order of $J \times K$. In the second stage, from the users' perspective, each user only needs to be changing his association once at most, and the number of available \acp{gNB} is also less than $J$. Hence, the complexity of \textbf{Algorithm}~\ref{alg.limited_1} is $\mathcal{O}(J \times K)$.

%It can be seen that the most challenging step is to calculate the exposure $e_{kj}$ generated by each user in the different association. Therefore, in this paper, the computational complexity of exposure $e_{kj}$ or gain $g_{kj}$ is considered as one unit when analyzing complexity.

%, where $J$ is the number of \acp{gNB}, $K$ is the number of users. 

 \begin{algorithm}[t!] 
	\caption{Association matrix}
	\label{alg.limited_1} 
	\begin{algorithmic} [1]
		
			\STATE \textbf{Input} $W_{j, \max}, j \in \mathcal{J} $, location of \ac{gNB} $j \in \mathcal{J}$, location of users $k \in \mathcal{K}$ 
			\STATE $W_j= 0,\forall j \in \mathcal{J}$, generate initial number of associated users to \ac{gNB} $j$
			\STATE initial association matrix $\epsilon _{kj}=0$ for each $k \in \mathcal{K}, j \in \mathcal{J} $
			\STATE compute the quantified \ac{EMF} exposure and gain,  $e_{kj}$ and $g_{kj}$ for each $k \in \mathcal{K}, j \in \mathcal{J} $, respectively
		
%		\STATE \textbf{Initialization:} 
%		\STATE $W_j= 0,\forall j \in \mathcal{J}$, generate initial number of associated users to \ac{gNB} $j$, and $W_{j, \max}$ is the maximum number
%		\STATE $e_{kj}$ and $g_{kj}$, the quantified \ac{EMF} exposure and gain between user $k$ and \ac{gNB} $j$, respectively
%		
		\FOR[Initial assignment of users ]{$k=1:K$}
		\STATE find $j_{k}^{*}$ s.t.  ${{j}_{k}^{*}} = \arg {\min\limits _{j\in \mathcal{J}}}({e_{kj}})$ \label{alg:ass_ini} (i.e., find the best serving \ac{gNB} for user $k$, regardless of resource constraint)
		\STATE $W_{{{j}_{k}^{*}}} = W_{{{j}_{k}^{*}}} + 1$, $\epsilon _{k{{{j}_{k}^{*}}}}=1$, ${{j}_{k}^{*}}$ is the index of selected \ac{gNB}
	%	\STATE $e_{k} = {e_{kj}}$
		\ENDFOR
		
		\FOR[Adjust user-\ac{gNB} association ]{$j=1:M$}  \label{alg:ass_st}
		\WHILE{$W_j> W_{j,\max}$}
		\STATE find $({k_{\mathrm c}}$,$j_{\mathrm c})$ s.t.  $({k_{\mathrm c}}$,$j_{\mathrm c}) = \arg {\min\limits _{{k\in \mathcal{K}_j},j\in{\bar{\mathcal{J}}}}}({\bar{g}_{kj}})$ (i.e., move one user to a non-fully occupied \ac{gNB}, the corresponding increment of this alteration is minimal.)
% 		\STATE find ${k_{\mathrm c}}$ s.t.  ${k_{\mathrm c}} = \arg {\min\limits _{k\in \mathcal{K}_j}}({g_{kj}})$ (i.e., choose one user to move, which lead to lowest increase in radiation)
% 		\STATE find ${j_{\mathrm c}}$ s.t.  ${j_{\mathrm c}} = \arg {\max\limits _{j\in {\bar{\mathcal{J}}}}}({g_{kj}}) $ (i.e., choose a non-fully occupied \ac{gNB} which could provide greatest gain to associate with user ${k_{\mathrm c}}$) \label{alg.ass.com}

		\STATE $W_{j_{\mathrm c}} = W_{j_{\mathrm c}} + 1$, and $\epsilon _{k_{\mathrm c}j_{\mathrm c}}=1$
		\STATE ${W_j} = {W_j} - 1$, and $\epsilon _{k_{\mathrm c}j}=0$
%		\STATE $e_{{k_{\mathrm c}} }= {e_{{k_{\mathrm c}}{j_{\mathrm c}}}}$
		\ENDWHILE
		\ENDFOR \label{alg:ass_end}
		%\FOR{$j=2: J$}
		
		\STATE calculate \ac{tUAV}'s contribute to reducing radiation $g_{j}=\sum\limits_{k\in \mathcal{K}_j}{g_{kj}},  \forall j \in \mathcal{M}$
		\STATE \textbf{Output} $e_{kj},\epsilon _{kj},g_j, \forall k \in \mathcal{K}, \forall j \in \mathcal{J}$
	%	\STATE$g_{j}=\sum\limits_{k\in \mathcal{K}_j}{g_{kj}}$ (i.e., each \ac{gNB}'s contribute to reducing radiation)
	%	\ENDFOR

	\end{algorithmic}
\end{algorithm}

%\subsection{\ac{tUAV}'s Deployment Optimization}

\subsection{Optimize the Deployment of \acsp{tUAV} to \acsp{GS}}

Unlike regular \acp{UAV}, the freedom of \acp{tUAV} is limited by the predefined locations of the ground stations, tether length and inclination angles. In order to reduce the exposure, a proper location for the \acp{tUAV} should be determined, i.e., \acp{tUAV}' association with \acp{GS}.  The deployment problem is a \ac{MINLP}, as we need to consider all possible configurations between \acp{tUAV} and the \ac{GS} as well as the inclination angle and tether length of each \ac{tUAV}.  For given tethered length and inclination angles, e.g., \acp{tUAV} are located in the center of their hovering areas above the \ac{GS}, the \ac{tUAV} association to \acp{GS} problem can be written as
\begin{subequations} 
	\begin{alignat}{2}
		\text{(P2): }&\underset{\epsilon _{kj}, \vartheta _{mn}}{\text{minimize}}        &\qquad&\sum_{k=1}^{K}  \mathrm{SAR}_{k}^{\text{UL}}\, \PN\\
		&\textrm{subject to:} &      &\theta _{m}= \frac{\pi}{2}\\
		&                  &      &T _{m} = \frac{T_{\max}}{2} \\
		&                  &      & {\color{black}\eqref{eq:constraint6},
			\eqref{eq:constraint7}}, \eqref{eq:constraint1}, \eqref{eq:constraint2}, \eqref{eq:constraint3}.
	\end{alignat}
\end{subequations}
% to assume that \ac{tUAV} is located in the center of \ac{GS}'s  hovering area,

The Problem \textbf{P2} is a binary integer program for variables $\vartheta _{mn}$ and $\epsilon _{kj}$. One approach to avoid exhaustive search  over all possible configurations is to select the locations that are most likely to provide satisfactory performance, such as the barycenter of users’ concentrations weighted by their exposure.\footnote{\color{black}Note that the users' position is needed to find their barycenter, which can be found through wireless localization or \acp{GPS} \cite{ElzanatyGueGuiSlim:20,ElzanatyGuerraDardari:21}.} Another method is  a divide-and-conquer like algorithm, where we iteratively prune the search space on the premise of ensuring that each iteration is moving towards a better result. For this purpose, we propose two heuristic algorithms, based on \ac{SAR}-aware $K$-mean and \ac{SR} process \cite{GCAlsharoa2020}, respectively. Both of them are composed of three steps: $\mathit{(i)}$ deploying randomly the \acp{tUAV}, $\mathit{(ii)}$ changing the position of \acp{tUAV}, $\mathit{(iii)}$ searching for the nearest \ac{GS}. In the following, we  describe the two proposed algorithms in detail. 

%an adapted $K$-mean algorithm based on the amount of exposure produced by each user.
\subsubsection{Modified $K$-mean Algorithm}

Initially, \acp{tUAV} are randomly scattered  over the entire area, while the location of the \ac{BS} remained unchanged in the middle of the area. Then, the users are associated to the \acp{gNB} according to \textbf{Algorithm} \ref{alg.limited_1} and the gain of \acp{tUAV} at the initial iteration, $g_j^{i}, j \in \mathcal{M}$, of their deployment is also computed.
%i.e., the number of clusters equal to that of \acp{gNB}, and the users associated with the same \ac{gNB} belong to the same cluster. 
%Moreover, we can also quantify the \ac{EMF} exposure $e_{kj}$ for users and gain $g_j$ for \acp{tUAV}. 
Let us define $x_j^{\rm{mid}}, y_j^{\rm{mid}}$ as the  \ac{SAR}-weighted barycenter of users associated with the $j^{\text{th}}$ \ac{tUAV}. We move the \acp{UAV} to the barycenters and run \textbf{Algorithm} \ref{alg.limited_1} again to obtain the gains of \acp{tUAV} if they are moved to the new positions, $g_j^{\mathrm{mid}},j \in \mathcal{M}$. Then, we pick the coordinates based on the value of gains, i.e., select either the current location of the \ac{tUAV} or the user's weighted barycenter depending on the location that can reduce more the exposure. This process is repeated until 
the difference between the $i^{\mathrm{th}}$ and the ${(i+1)}^{\mathrm{th}}$ coordinates are less than the tolerance $\delta$, or reaching the maximum number of iteration $I_{\max}$. 
Finally, each \ac{tUAV} is connected to nearest \ac{GS} from unoccupied \acp{GS} $\mathcal{\bar{N}}$, while the location of the \ac{BS} remains the same, as described in \textbf{Algorithm} \ref{alg.kmean}.

 \begin{algorithm}[h] 
	\caption{Determine \acp{tUAV}' deployment by $K$-mean }
	\label{alg.kmean} 
	\begin{algorithmic} [1]
		
		\STATE \textbf{Input}  $W_{j, \max}, j \in \mathcal{J} $, location of users $k \in \mathcal{K}$, $I_{\max}$
		\STATE $i=1$
		\STATE initial connection matrix $\vartheta _{mn}=0$ for each $m \in \mathcal{M}, n \in \mathcal{N} $
     	\STATE generate  coordinates $x_j^i, y_j^i$ that	\acp{tUAV} are randomly scattered in the 2D space, and the \ac{BS} is fixed in the center.
  
     	%	\label{alg.kmean.ini} 
%		\STATE \textbf{Initialization:} 
%		
%		\STATE $i=1$, set the maximum number of iteration 
%		
		\REPEAT
		\STATE Run \textbf{Algorithm \ref{alg.limited_1}} with \acp{gNB}' coordinates $x_j^i, y_j^i, j \in \mathcal{J}$ to get \ac{EMF} exposure $e_{kj}$ and association matrix $\epsilon _{kj}$ for users and gain $g_{j}^i$ for \acp{tUAV}
		\STATE ${x_j^{\rm{mid}}} =  \frac{\sum\limits_{k \in \mathcal{K}_j}x_{kj}e_{kj}}{\sum\limits_{k \in \mathcal{K}_j} e_{kj}}, j\in \mathcal{M}$
		\STATE ${y_j^{\rm{mid}}} =  \frac{\sum\limits_{k \in \mathcal{K}_j}y_{kj}e_{kj}}{\sum\limits_{k \in \mathcal{K}_j} e_{kj}}, j\in \mathcal{M}$
		\STATE Run \textbf{Algorithm \ref{alg.limited_1}} with the coordinates of \acp{gNB} $x_j^{\rm{mid}}, y_j^{\rm{mid}}, j \in \mathcal{J}$ to get gain $g_{j}^{\rm{mid}}$ for \acp{tUAV}   
		\STATE $x_j^{i+1}=x_j^{i}\mathbb{I}\left \{ g_j^{i}\geq g_j^{\text{mid}}\right \}+x_j^{\text{mid}}\mathbb{I}\left \{ g_j^{i}< g_j^{\text{mid}}\right \}, j\in \mathcal{M}$
		\STATE $y_j^{i+1}=y_j^{i}\mathbb{I}\left \{ g_j^{i}\geq g_j^{\text{mid}}\right \}+y_j^{\text{mid}}\mathbb{I}\left \{ g_j^{i}< g_j^{\text{mid}}\right \}, j\in \mathcal{M}$
			\STATE $i=i+1$
		\UNTIL{$ \left( \left | x_j^i- x_j^{i+1} \right |<\delta\right)\,\text{and}\,\left( \left | y_j^i - y_j^{i+1}\right |<\delta\right),\forall j \in \mathcal{M}$ or $i>I_{\max}$}
		
		\FOR{$m =1: M$}
		\STATE find $n_0$ s.t. $n_0 = \mathop {\arg \min }\limits_{n \in \mathcal{\bar{N}}} {({x_m} - {x_n})^2} + {({y_m} - {y_n})^2}$, (i.e., $n_0$ indicates the index of the nearest \ac{GS})
		\STATE $\vartheta _{m{n_0}}=1$
		%\STATE $x_m=x_{n_0}$, $y_m=y_{n_0}$, $\vartheta _{m{n_0}}=1$
		\ENDFOR
		\STATE \textbf{Output} $\epsilon _{kj},j \in \mathcal{J}, k \in \mathcal{K}$, $\vartheta _{m{n}}, m \in \mathcal{M}, n \in \mathcal{N}$
		%$x_m, y_m, m \in \mathcal{J}$
	
	\end{algorithmic}
\end{algorithm}	

\subsubsection{Algorithm Based on 2D \acs{SR} Process} 

Initially, \acp{tUAV}  are uniformly scattered over the entire area, while the location of the \ac{BS} remained unchanged in the middle of the area, $q_j^i, j \in \mathcal{J}$. Then, we generate initial next position candidates $q_{j,t}^{i}, \forall t \in \mathcal{T} \triangleq \{1,2,\cdots T\}$ as a circle with radius $r^{i}$ around all current locations of the \acp{tUAV} to form initial  positions. Next, we will select one \ac{tUAV} and iterate over the candidate locations of it, solve the associations by \textbf{Algorithm} \ref{alg.limited_1}, and calculate the total \ac{EMF} exposure $\mathrm{EI}_{\mathrm{SR}} (q_{j,t}^{i}) \triangleq \sum\limits_{k \in \mathcal{K}} \sum\limits_{j \in \mathcal{J}} e_{kj} \, \epsilon_{kj}$, 
while keeping all the other \acp{tUAV} fixed. Then, we find the best next \ac{tUAV} candidate and compare it to where the \ac{tUAV} is located to decide whether to move it or not, i.e., choose the location where there is less radiation over the area when this \ac{tUAV} is in that location. Furthermore, we will move \acp{tUAV} around one by one in this way, then generate new candidates on a circle of radius $r^{i+1}=r^{i}/{2}$
around each local solution. We repeat this process until the size of sample space decrease below a certain threshold $r_{\min}$. Eventually, we move the \acp{tUAV} directly over the corresponding nearest \ac{GS} $\text{gs}_n$, as described in \textbf{Algorithm} \ref{alg.SR_d}.

%We purpose a heuristic algorithm based on 2D \ac{SR} process to find the optimal deployment of the \ac{tUAV}. The set $\mathcal{J}$ contains all \acp{gNB}, where the $j=1$ represents the fixed macro cell and the remaining are \acp{tUAV}. As described in \textbf{Algorithm} \ref{alg.SR_d}, each time current location $q_j^i$ would be the center of a circle, with candidates $q_{jt}^{i}$ evenly distributed around a circle with a radius of $r(i)$. Next, we find the best next candidate and compare it to where the \ac{tUAV} is located to decide whether to move or not. Furthermore, we will move \acp{tUAV} around one by one in this way, then reduce the radius to half of the previous iteration. The minimum search radius is $r_{\min}$, and the iteration will not stop until the radius $r(i)$ is less than $r_{\min}$. Eventually, we move the \acp{tUAV} directly over the corresponding nearest \ac{GS} $ \text{gs}_n$.

\begin{algorithm}[t!] 
	\caption{Optimize the deployment of \ac{tUAV} through 2D \ac{SR} process}
	\label{alg.SR_d} 
	\begin{algorithmic} [1]
     	\STATE \textbf{Input} $W_{j, \max}, j \in \mathcal{J} $, location of users $k \in \mathcal{K}$, $r_{\min}$
     	\STATE  $i=1$
     	\STATE initial connection matrix $\vartheta _{mn}=0$ for each $m \in \mathcal{M}, n \in \mathcal{N} $
     	\STATE generate initial location of \acp{gNB} $q_{j}^{i}, j \in \mathcal{J}$ that \acp{UAV} uniformly scattered and the \ac{BS} fixed in the center
     	 \STATE generate initial candidates in a circle of radius $r^{i}$ around each \ac{tUAV} $q_{j,t}^{i}, t\in \mathcal{T}$

%		\STATE \textbf{Initialization:} 
%		\STATE  $i=1$, and $r_{\min}$ is the minimum radius
%		\STATE generate initial location of \acp{gNB} $q_{j}^{i}, j \in \mathcal{J}$ that \acp{UAV} uniformly scattered and the \ac{BS} fixed in the center
%       \STATE generate initial candidates in a circle of radius $r(i)$ around each \ac{tUAV} $q_{j,t}^{i}, t\in \mathcal{T}$
		
	%	arg {\min\limits _{j\in \mathcal{J}}}({e_{kj}})
		\WHILE{$r^{i}>r_{\min}$}
		\FOR{$j=1:M$} 
		\STATE find $q_{j,t_0}^{i}$ s.t.  $q_{j,t_0}^{i}=\arg {\min\limits _{t\in \mathcal{T}}}(\mathrm{EI}_{\mathrm{SR}}(q_{j,t}^{i}))$ (i.e., run \textbf{Algorithm} \ref{alg.limited_1} to find association matrix $\epsilon _{kj},j \in \mathcal{J}, k \in \mathcal{K}$ and calculate the total exposure $\mathrm{EI}_{\mathrm{SR}}$)
		
		\STATE $q_{j}^{i+1}=\arg\min{\left \{ \mathrm{EI}_{\mathrm{SR}} (q_{j}^{i}),\mathrm{EI}_{\mathrm{SR}}(q_{j,t_0}^i)\right \}}$
		\ENDFOR 
		\STATE $r^{i+1}=r^{i}/2$	
		\STATE $i=i+1$
		\ENDWHILE
		\FOR{$m =1:M$}
		\STATE find $n_0$ s.t. $n_0 = \mathop {\arg \min }\limits_{n \in \mathcal{\bar{N}}} \left \| q_j^i-\text{gs}_n \right \|$, (i.e., $n_0$ indicates the index of the nearest \ac{GS})
	    \STATE $\vartheta _{m{n_0}}=1$ %$q_j^i=\text{gs}_n$
		\ENDFOR
		\STATE \textbf{Output} $\epsilon _{kj},j \in \mathcal{J}, k \in \mathcal{K}$, $\vartheta _{m{n}}, m \in \mathcal{M}, n \in \mathcal{N}$
		%$q_j^i, j \in \mathcal{J}$

	\end{algorithmic}
\end{algorithm}

\subsubsection{Complexity Analysis}
%As we assumed before that the computational complexity of exposure $e_{kj}$ or gain $g_{kj}$ is considered as one unit.
We analyze the complexity of \textbf{Algorithm} \ref{alg.kmean} and \textbf{Algorithm} \ref{alg.SR_d}. 
Firstly, for \textbf{Algorithm} \ref{alg.kmean}, the process of finding the \ac{SAR}-weighted barycenter and moving \acp{tUAV} to new location is the most complex step. In each iteration, its complexity mainly comes from the operation of \textbf{Algorithm} \ref{alg.limited_1}. After that, the number of iterations is finite, its value is much smaller than the number of users. Hence, the complexity is $\mathcal{O}(J \times K)$. Moreover, for \textbf{Algorithm \ref{alg.SR_d}}, the process of gradually narrowing the search radius is the most complex. More specifically, before each reduction in the search radius, we need to search over all candidate locations of each \ac{tUAV}, during which we run \textbf{Algorithm} \ref{alg.limited_1} constantly to calculate the total radiation. So the complexity is $\mathcal{O}(J^2 \times K)$.  

%The second stage of \textbf{Algorithm} \ref{alg.kmean} is the most complex, and for each iteration, its complexity mainly comes from the operation of \textbf{Algorithm} \ref{alg.limited_1}.And the number of iterations is finite, its value is much smaller than the number of users. So the complexity is $\mathcal{O}(J \times K)$. 

%For \textbf{Algorithm \ref{alg.SR_d}}, the process of gradually narrowing the search radius is the most complex. Before each reduction, all the candidate locations of the \acp{tUAV} need to search and the corresponding EI calculated based on \textbf{Algorithm \ref{alg.limited_1}}. So the complexity is $\mathcal{O}(J^2 \times K)$.  
%

\subsection{Optimize \acs{tUAV}'s Position within the Hovering Areas} 
Through the above two-step iteration, we can fix the deployment of \acp{tUAV} and association matrix. Now, we optimize the location of \acp{tUAV} on a small scale in their hovering zone. The goal is to find optimal inclination angels and tether length of each \ac{tUAV}, and this sub-optimization can be written as
	\begin{subequations} 
	\begin{alignat}{2}
		\text{(P3): }&\underset{\theta _{m},T _{m},\varphi _{m}}{\text{minimize}}        &\qquad& \sum_{k=1}^{K}  \mathrm{SAR}_{k}^{\text{UL}}\, \PN  \\
		&\textrm{subject to:}& &   {\color{black}\eqref{eq:constraint6},
			\eqref{eq:constraint7}},
			\eqref{eq:constraint4}, \eqref{eq:constraint5}, \eqref{eq:constraint8}. 
	\end{alignat}
\end{subequations}
In order to avoid entanglement with surrounding buildings and ensure safety, the inclination angel $\theta $ of the tether has a minimum value $\theta_{\min} $, which is related to environments  \cite{kishk2020tuav}. Then, the distance between \ac{tUAV} and \ac{GS}, $T_{m}$, should be smaller than the maximum tether length $T_{\max}$, and there are no additional restrictions on azimuth $\varphi _{k}$. 

It can be seen that problem \textbf{P3} includes continuous variables, i.e., $\theta _{m}, \varphi _{m}$  and $T_{m}$, and it involves a non-convex objective function. 
%by considering only the users associated with it, ignoring the other users and \acp{gNB}. 
We propose to solve this sub-optimization via two iterative algorithms, i.e., golden section search and \ac{SR} process. For both algorithms, we consider alternate optimization, where each \ac{tUAV} is individually optimized while keeping the users' association unchanged, as computed in preceding algorithms.

\subsubsection{Algorithm Based on Golden Search}

%The proposed algorithm involves two main steps: \textit{(i)} selecting altitudes, \textit{(ii)}  picking points on the horizontal planes. 
%The proposed algorithm is based on the golden search method.

At first, we consider a random tether length $T_m$ for the $m^{\text{th}}$ \ac{tUAV}. Then, we find the \ac{SAR}-weighted barycenter of its associated users, i.e., the projection in the horizontal $x$-$y$ plan. If that location is not in the hovering area, we will choose the nearest point in the flyable zone. Next, we compute the \ac{EI} at this height when the \ac{tUAV} is moved to the selected point $\mathbf{O}_m \triangleq (x_m(h_m),y_m(h_m),h_m)$, i.e., 
\begin{equation}
   \mathrm{EI}_{\textrm{GS}} \left(\mathbf{O}_m \right) \triangleq \sum\limits_{k\in {{\mathcal{K}_m}}}e_{km}, \qquad h_m \in \left[ h_{\mathrm{G}}, h_{\mathrm{G}}+T_{\max}    \right]
\end{equation}
 where $h_{\mathrm{G}}$ is the height of \ac{GS} at the top of the building, and $T_{\max}$ is the maximum tether length.  After that, we could narrow down the radius based on the golden search, where $v= \frac{\sqrt{5}-1}{2}$ is the golden ratio. The initial heights are  the upper and lower bounds,  i.e., $h_{\mathrm{u}}= h_{\mathrm{G}} +T_{\max} $ and $h_{\mathrm{l}}=h_{\mathrm{G}}$, respectively. Also, we will pick two golden section altitudes $h_{\mathrm{u}'} = h_{\mathrm{l}} + v\,(h_{\mathrm{u}}-h_{\mathrm{l}}) $ and $h_{\mathrm{l}'} = h_{\mathrm{u} }-v\,(h_{\mathrm{u}}-h_{\mathrm{l}})$. Then, we  calculate exposure at this two altitude $\mathrm{EI}(\mathbf{O}_{\mathrm{u}'})$ and $\mathrm{EI}(\mathbf{O}_{\mathrm{l}'})$. Based on the radiation of this two points, we will narrow down the search area. 
 %For example, if $\mathrm{EI}(h_{\mathrm{l}'}) < \mathrm{EI}(h_{\mathrm{u}'})$, then $h_{\mathrm{u}}= h_{\mathrm{u}'}$, and we will pick two golden section altitudes and calculate radiation again to change upper or lower limit. 
 Then, this process will iterate over and over to narrow the limits until the interval length is smaller than the tolerance height $h_{\min}$, as described in \textbf{Algorithm} \ref{alg.gs}. {\color{black} We opt for the golden search method, as it converges fast and is less computationally demanding than an exhaustive search over a large number of position candidates \cite{10.2307/2032161}. }
 %In this method, the shrinkage rate of the search interval in each iteration is the golden ratio. This method only needs to calculate the function value once per iteration.

%We will find the possible upper $h_1$ or lower $h_2$ limit, and calculate the corresponding $\mathrm{EI} \left(h \right) $ at the particular altitude. Then iterate over and over to narrow the limits until the gap between them is smaller than the accuracy set $error$. 

%As stated before, we optimize each \ac{tUAV} individually and will only consider the users associated with one \ac{tUAV}, ignoring the other user and \ac{gNB}. 

%We slice the whole space according to altitude, and f
%For a particular height, we place the \ac{tUAV} at the user's weighted barycenter of \ac{EMF} exposure. Moreover, if that location is not in the flyable zone, we will choose the nearest point in the flyable zone.

%As described in \textbf{Algorithm} \ref{alg.gs}, we could narrow down the radius based on the golden search. Initially, the search area is bounded by the height of the \ac{GS} $h_G$, and the maximum altitude the \ac{tUAV} could reach $ h_G + T_{\max}$. And, $v= \frac{1+\sqrt{5}}{2}$ is the golden ratio. We will find the possible upper $a_1$ or lower limit $a_2$, and calculate the corresponding EI at the particular altitude. Then iterate over and over to narrow the limits until the gap between them is smaller than the accuracy set $error$.

 \begin{algorithm}[t!] 
	\caption{Optimize the position of \ac{tUAV} through golden search }
	\label{alg.gs} 
	\begin{algorithmic} [1]
		
		     \STATE \textbf{Input}  %location of users $k \in \mathcal{K}_m$, location of the $m^\text{th}$ \ac{tUAV},
		     $\epsilon _{kj},j \in \mathcal{J}, k \in \mathcal{K}$, $\vartheta _{m{n}}, m \in \mathcal{M}, n \in \mathcal{N}$,
		     $r_{\min}$, $h_\mathrm{G}$, $T_{\max}$, $v$, $h_{\min}$
		  \FOR{m=1:M}
%		\STATE \textbf{Initialization:} 
          \STATE  generate initial lower bound $h_{\mathrm{l}}=h_\mathrm{G}$ and upper bound  $h_{\mathrm{u} }=h_\mathrm{G} + T_{\max} $
		\STATE  generate initial golden section altitudes  $h_{\mathrm{u}'} = h_{\mathrm{l}} + v\,(h_{\mathrm{u}}-h_{\mathrm{l}}) $ and $h_{\mathrm{l}'} = h_{\mathrm{u} }-v\,(h_{\mathrm{u}}-h_{\mathrm{l}})$. 
%		\STATE $h_{\min}$ is accuracy
		
		%\STATE  Calculate the radiation $\text{EI}_{a1}$ and $\text{EI}_{a2}$ these two altitudes, respectively
		
		\WHILE{$\left | h_\mathrm{u}-h_\mathrm{l} \right |> h_{\min}$}
		\STATE find barycenters (or nearest points of them in hovering area) $\mathbf{O}_{\mathrm{l}'}$ and $\mathbf{O}_{\mathrm{u}'}$ at golden section altitudes, then calculate the radiation at this two points $\mathrm{EI}_{\textrm{GS}}(\mathbf{O}_{\mathrm{l}'})$ and  $\mathrm{EI}_{\textrm{GS}}(\mathbf{O}_{\mathrm{u}'})$, respectively
		\IF {$\mathrm{EI}_{\textrm{GS}}(\mathbf{O}_{\mathrm{l}'}) < \mathrm{EI}_{\textrm{GS}}(\mathbf{O}_{\mathrm{u}'})$}
		\STATE  change the upper bound $h_{\mathrm{u}}= h_{\mathrm{u}'}$
		%$a=a_1, a_1=a_2$
		%\STATE $a_2=a+v*(b-a)$
		\ELSE 
		\STATE  change the lower bound $h_{\mathrm{l}}= h_{\mathrm{l}'}$
		%$b=a_2, a_2=a_1$
		%\STATE $a_1=b-v*(b-a)$
		\ENDIF 
		\STATE generate golden section altitudes $h_{\mathrm{u}'} = h_{\mathrm{l}} + v\,(h_{\mathrm{u}}-h_{\mathrm{l}}) $ and $h_{\mathrm{l}'} = h_{\mathrm{u} }-v\,(h_{\mathrm{u}}-h_{\mathrm{l}})$		
		%\STATE Calculate the radiation $\text{EI}_{a1}$ and $\text{EI}_{a2}$ these two altitudes, respectively
		\ENDWHILE	
		\STATE $h_m= (h_\mathrm{u}+h_\mathrm{l})/2$, and find the barycenter (or nearest points of it in hovering area) point $\mathbf{O}_{m}$ at this altitude, then compute corresponding $\theta _{m},T _{m},\varphi _{m}$ in spherical coordinates
		%as the final location of $j^\text{th}$ \ac{tUAV}
		\ENDFOR
		\STATE \textbf{Output} $\theta _{m},T _{m},\varphi _{m}, m \in \mathcal{M}$, i.e., the location of \acp{tUAV}  %$\mathbf{O}_{m}$, the location of $m^\text{th}$ \ac{tUAV} 

	\end{algorithmic}
\end{algorithm}

\subsubsection{Algorithm Based on 3D \acs{SR} Process}
The proposed algorithm is based on \ac{SR} process to optimize each \ac{tUAV} location individually by choosing the proper location among $T^{'}$ candidates around the current location of the \ac{tUAV}, denoted by $q_m^i$, at iteration $i$. Let us define $q_{m,t}^{i}$ as the  $t^{\text{th}}$ potential position for the location of the $m^{\text{th}}$ \ac{tUAV} on polyhedron around $q_m^i$ with radius $r^i$. 
%We consider We refer to those potential position with , t\in \mathcal{T^{'}}$, where $T^{'}$ is the number of possible locations.
%Initially, the \ac{tUAV} is located ${T_{\max}}/{2}$ meters directly above the \ac{GS}, $q_m^i$. Then, we start generating initial candidates for next positions $q_{m,t}^{i}, t\in \mathcal{T^{'}} \triangleq \{1,2, \cdots, T^{'}\}$ as a polyhedron with radius $r^i$ around the \ac{tUAV}.
%
Then, we calculate radiation $\mathrm{EI}_{q}(q_{m,t}^i) \triangleq \sum\limits_{k\in {{\mathcal{K}_m}}}e_{km}$ when \ac{tUAV} at those candidate locations. Then, we select the location, among the candidates, that minimizes the radiation $\mathrm{EI}_{q}$. After that, by \ac{SR} process, we generating $T^{'}$ new candidates on a polyhedron with radius  $r^{i+1}=r^{i}/2$ around  the \ac{tUAV}.
%Next, we find the best next position and compare it to where the \ac{tUAV} is located now, i.e., select the location that minimizes the radiation. After that, by \ac{SR} process, we generating $T^{'}$ new candidates on a polyhedron with radius  $r^{i+1}=r^{i}/2$ around where the \ac{tUAV} is now.
We repeat this process until the radius is less than the precision $r_{\min}^{'}$, as described in \textbf{Algorithm}~\ref{alg.SR}.

%$\mathrm{EI} \left(h \right) \triangleq \sum\limits_{k\in {{\mathcal{K}_j}}}e_{k}$

%The proposed algorithm has 2 main steps: $\mathit{(i)}$ select altitudes, $\mathit{(ii)}$  pick points on the horizontal planes.

%We purpose a heuristic algorithm  based on  \ac{SR} process, to find the 3D optimal location of the \ac{tUAV} with fixed user association. 
%We start \textbf{Algorithm} \ref{alg.SR} by generating initial candidates for next positions $q_{t}^{i}, t\in T^{'}$ as a polyhedron with radius $r(i)$ around \ac{tUAV}. Next, we find the best next position and compare it to where the \ac{tUAV} is located. After that, by \ac{SR} process, we generating $T^{'}$ new candidates on a polyhedron with radius  $r(i+1)=r(i)/2$ around \ac{tUAV}'s location $q^{i}$, we repeat this process until the search radius is less than precision $r_{\min}^{'}$.

%Therefore, its total complexity is an order of $I_{\max}^{'} \times T \times W_j$.
%For example, if the search accuracy is 0.5m and the radius is around 50m, $I_{\max}^{'}$ will be 6 or 7 iterations. 
	
   \begin{algorithm}[t!] 
	\caption{Optimize the position of \ac{tUAV} through \ac{SR} process}
	\label{alg.SR} 
	\begin{algorithmic} [1]
		
		 \STATE \textbf{Input}  $\epsilon _{kj},j \in \mathcal{J}, k \in \mathcal{K}$, $\vartheta _{m{n}}, m \in \mathcal{M}, n \in \mathcal{N}$, $r_{\min^{'}}$
		
%		\STATE \textbf{Initialization:} 
        \FOR{m=1:M}
 		\STATE i=1
%, precision is $r_{\min}^{'}$
		\STATE generate \ac{tUAV}'s location $q_m^i$
		\STATE generate initial candidates $\mathcal{T^{'}}$ in a polyhedron of radius $r(i)$ around \ac{tUAV}, $q_{m,t}^{i}, t\in \mathcal{T^{'}}$
		\WHILE{$r^i>r_{\min}^{'}$}
		\STATE find $q_{m,{t_0}}^{i}$ s.t.  $q_{m,{t_0}}^{i}=\arg {\min\limits _{\;t\in \mathcal{T^{'}}} \,\mathrm {EI}_{q} (q_{m,{t_0}}^i)}$ (i.e., choose the best candidate position which is in hovering area)
		\STATE $q^{i+1}_m=\arg\min{\left \{ \mathrm {EI}_{q} (q_{m}^i), \mathrm {EI}_{q} (q_{m,t}^i) \right \}}$

		\STATE $r^{i+1}=r^{i}/2$	
		\STATE $i=i+1$
		\ENDWHILE
		\STATE compute corresponding $\theta _{m},T _{m},\varphi _{m}$ in spherical coordinates for location $q_{m}^i$
		%as the final location of $j^\text{th}$ \ac{tUAV}
		\ENDFOR
		\STATE \textbf{Output} $\theta _{m},T _{m},\varphi _{m}, m \in \mathcal{M}$, i.e., the location of \acp{tUAV}  %location of $m^\text{th}$ \ac{tUAV} $q_{m,t}^i$
		
	\end{algorithmic}
\end{algorithm}

\subsubsection{Complexity Analysis}

Next, we analyze the complexity of \textbf{Algorithm} \ref{alg.gs} and \textbf{Algorithm} \ref{alg.SR}. The complexity is on the order of $M$, as the iteration of golden search or \Ac{SR} are fixed, does not depend on the system parameters, such as number of users or possible location of GS. Overall, the complexity of both algorithms can be expressed as $\mathcal{O}(M)$.
%Note that both of these algorithms will consider a single \ac{tUAV} and the users connected to that \ac{tUAV}, so it can only find the optimal location of one \ac{tUAV} at a time. So, the complexity of finding the optimal location for \acp{tUAV} depends on $M$, which is the number of \acp{tUAV}. Overall, the complexity of both algorithms can be expressed as $\mathcal{O}(M)$.

%\textbf{Algorithm} \ref{alg.gs} 
%
%
% \textbf{Algorithm} \ref{alg.SR} mainly consists of two parts, the initialization phase and \ac{SR} phase. We can get the number of iteration $I_{\max}^{'}$ by taking the log base 2 of a number, which is the ratio of the radius of \acp{UAV}' hovering area to the search accuracy. Also, the constant $T^{'}$ is the maximum number of candidate locations that need to be tried per iteration. Then for each candidate location, the gain $g_{kj}$ is calculated up to $W_j$ times, which is the maximum number of users connected to \ac{tUAV} $j$. Therefore, the number of iterations required to find the optimal location of one \ac{tUAV} in the hovering area can be regarded as a constant, which is $I_{\max}^{'} \times T \times W_j$. Overall, the complexity of finding the optimal location for \acp{tUAV} depends on $M$, which is the number of \acp{tUAV}.

\subsection{Overall Complexity} 

Generally speaking, there are two options for determining the deployment of \ac{tUAV}, modified $K$-mean and algorithm based on 2D \ac{SR} process, and the complexity of them is $\mathcal{O}(J \times K)$ and $\mathcal{O}(J^2 \times K)$, respectively. Therefore, only considering the complexity of algorithms, $K$-mean is a better choice at this step. After the deployment of the \ac{tUAV} is determined, the position of the \ac{tUAV} in the hovering area will be adjusted only slightly in the hovering area. And there are also two algorithms for this step, the algorithm based on 3D \ac{SR} process or golden search, both of which have complexity  $\mathcal{O}(M)$. 
%This also means that the performance of the two needs to be further compared to choose the better algorithm.
Therefore, the complexity of whole process is a polynomial order of $\mathcal{O}((M+1) \times K+M)$ or $\mathcal{O}((M+1)^2 \times K+1)$, which is overwhelmed by $(M \times K)$ or $(M^2 \times K)$.
%and $(M)$, where $J$, $K$, $M$ are number of \acp{gNB}, users and \ac{tUAV}, respectively. 
And in general, the number of users $K$ is usually much larger than other parameters, so the maximum complexity of the entire process should be determined mainly by the number of users.  In this regard, we propose the use of $K$-mean for connection between \acp{tUAV} and \acp{GS},  and \ac{SR} process to adjust \acp{tUAV} position in a small range, as they have the lowest complexity and best performance in terms of minimizing radiation as shown in numerical results.

\section{Dual problem: Cellular System Design with EMF Constraints} \label{sec:dual}
In previous sections, we consider the problem of minimizing the exposure, while achieving a target rate. Another architecture of interest for cellular operators is the dual problem, where we aim to maximize of the rate under a constraint on the users' exposure to \ac{EMF}.
%In this section, we are interested in the benefits of the network from the carrier's perspective.  
More precisely, our objective is to optimally deploy \acp{tUAV} in the space and associate the users in order to maximize the sum \ac{UL} rate. Such a problem can be formulated as
\begin{subequations} \label{eq:dualproblem}
	\begin{alignat}{2}
		\text{(P4): }&\underset{ \boldsymbol{\gamma}}{\text{minimize}}        &\qquad& \sum_{k=1}^{K} R_{k}^{\text{UL}}(\boldsymbol{\gamma}) \label{eq:optrate}\\
		&\textrm{subject to:}    &      & {\mathrm{SAR}_{k}^{\text{UL}}\, P_{k}^{\text{UL}} (\boldsymbol{\gamma}) \leq \mathrm{SAR}_\mathrm{limit}, \; \forall k \in \mathcal{K}, \label{eq:constraint9}} \\
		&                  &      & %\eqref{eq:constraint6},
		{\eqref{eq:constraint7},} \eqref{eq:constraint1}, \eqref{eq:constraint2}, \eqref{eq:constraint3}, \eqref{eq:constraint4},\eqref{eq:constraint5}, \eqref{eq:constraint8},
	\end{alignat}
\end{subequations}
where $\mathrm{SAR}_\mathrm{limit}$ is the whole body or local \ac{SAR} threshold. The constraint \eqref{eq:constraint9} guarantees that the  \ac{EMF} exposure of each user is less than a pre-specified limit, e.g., the \ac{ICNIRP} limit, to avoid the health risks, while other restrictions remain the same. In the following, we propose an algorithm to solve the problem. 

In order to improve the sum-rate, the \ac{UE} can send the maximum allowable power that satisfy constraint \eqref{eq:constraint7}. Nevertheless, the induced exposure to \ac{EMF} can be escalated above the required level, violating the constraint \eqref{eq:constraint9}. Hence,   an additional limit on transmit power is added to satisfy both of the constraints, i.e.,
%. Simultaneously, constraint \eqref{eq:constraint7} indicates that the mobile device cannot transfer more power than the permitted one. These two constraints, one is the legal provisions, the other is the hardware restrictions, are very crucial. So we have to satisfy both of them, and the actual transmit power can be expressed as 
\begin{equation} \label{eq:po}
	\PN=\text{min}\left \{ P_{{\max}},  \frac{\mathrm{SAR}_\mathrm{limit}}{\mathrm{SAR}_{k}^{\text{UL}}} \right \},	
\end{equation}
where transmit power $\PN$ is the minimum between the maximum transmit power $P_{{\max}}$ and the power to ensure the \ac{SAR} limitation ${\mathrm{SAR}_\mathrm{limit}}/{\mathrm{SAR}_{k}^{\text{UL}}}$. In fact, each user $k$ can have his own reference \ac{SAR}, $\mathrm{SAR}_{k}^{\text{UL}}$,  depending on several factors such as postures (standing or sitting), usages (voice or data), and device type (wearable, handheld, or in-body).

%\subsection{Algorithms for Solving Optimization}

Although the objective function in \textbf{P4} is different from the original one \textbf{P0}, the dual problem is also a \ac{MINLP}. 
%the difference of constraints is tiny. 
Therefore, the proposed algorithms in section~\ref{sec:alg} are still applicable and we only need to make some minor adjustments as follows{\color{black}:}
\begin{itemize}
    \item We consider a new power allocation strategy as in \eqref{eq:po} rather than \eqref{equ:rate}.
    \item For \textbf{Algorithm}~\ref{alg.limited_1}, the radiation $e_{kj}$, is replaced with the minus of the  rate $-R_{kj}$, and   the gain is redefined as $g_{kj} \triangleq R_{kj}-R_{k0}$. 
    
    \item For \textbf{Algorithm} \ref{alg.kmean}, we change the function from \ac{SAR}-weighted to Rate-weighted, i.e., using rate $-R_{kj}$ to replace $e_{kj}$ in the formula for calculating weighted barycenters.
    
    \item For \textbf{Algorithm} \ref{alg.SR_d}, the exposure metric $\mathrm{EI}_{\mathrm{SR}} (q_{j,t}^{i})$ is replaced with the rate $\mathrm{R}_{\mathrm{SR}} (q_{j,t}^{i}) \triangleq  \sum\limits_{k \in \mathcal{K}} \sum\limits_{j \in \mathcal{J}} (-R_{kj})\, \epsilon_{kj} $.
  \item  For \textbf{Algorithm} \ref{alg.gs}, we need to substitute $\mathrm{EI}_{\textrm{GS}} \left(\mathbf{O}_m \right)$ with $ \mathrm{R}_{\textrm{GS}}\left(\mathbf{O}_m \right) \triangleq \sum\limits_{k \in \mathcal{K}_m}  (-R_{km})  $.
  \item For \textbf{Algorithm} \ref{alg.SR}, the \ac{EMF} exposure $\mathrm{EI}_{q}(q_{j,t}^i)$  is replaced by $\mathrm{R}_{q}(q_{j,t}^i) \triangleq \sum\limits_{k\in {{\mathcal{K}_m}}}(-R_{km})$.
\end{itemize}

%$\mathrm{EI} (q_{j,t}^{i}) \triangleq \sum\limits_{k \in \mathcal{K}} \sum\limits_{j \in \mathcal{J}} e_{kj} \, \epsilon_{kj}$

%In particular, the previous objective is to minimize the radiation, but now you need to maximize the rate. Also, . 
%Adding a minus sign and turning it into a minimization can better apply the proposed algorithms.
%The first step is to find the association matrix under the condition of fixed  \acp{tUAV} position.
%The second step is to find the optimal connection between the \acp{tUAV} and the \acp{GS}. \textbf{Algorithm} \ref{alg.kmean} itself does not need any additional changes, only the output of calling \textbf{Algorithm} \ref{alg.limited_1} will change. Similarly,  The final step is to fine-tune the position of \acp{tUAV}. 

%In section \ref{sec:num} we will show correlative simulation results.

\section{Numerical Results} \label{sec:num}

In this section, we present some selected simulation results to demonstrate the advantages of the proposed architecture. The evaluation scenario is composed of a \ac{BS} and several \acp{tUAV} and \acp{GS} within an area of $1000$~m $\times$ $1000$~m, i.e., $A=1$~${\text{km}}^{2}$. 
The \ac{BS} is fixed in the middle of the area, and the \acp{GS} are evenly distributed in the area.\footnote{Note that if we have prior information about the users' locations, we can set more \acp{GS} in the areas where users are likely to gather.} Let us recall that $\bar{K}$ is the average number of  residents in the area (including users and non-users), their locations follow a superposition of two random processes: \textit{i)} \ac{PPP} with density $\bar{K}/(3 A)$;  \textit{ii)} \ac{PCP}, were each cluster represents one hotspot with a radius $100$~m, with average number of users $ \bar{K}/6$
%$2 \bar{K}/(3{ \color{black}???})$
, and the average number of clusters in the area is four. %{\color{black}?????}. 
In simulation, all the metrics  used are averaged over 1000 iterations. We consider that active users fall within two categories according to their usage, i.e., 
\emph{voice} and \emph{data} users. Each category has distinct \ac{UL} rate requirements and \ac{SAR} reference for the \ac{EMF} exposure. In fact,  since the mobile phone is closer to the person when making a call through the handset directly, the induced whole-body \ac{SAR} of voice usage is larger than that of data usage. For instance, we consider a reference \ac{SAR}, $\mathrm{SAR}_{k}^{\text{UL}} \in \{\mathrm{SAR}_{\text{v}},{\mathrm{SAR}}_{\text{d}}\}$ with ${\mathrm{SAR}}_{\text{v}}= 0.0047$ and ${\mathrm{SAR}}_{\text{d}}=0.0037$~W/kg per unit transmit power in the sub-$5$~GHz band  for \emph{voice} and \emph{data} users, respectively \cite[Table 27]{Tes2014LEXNET}.
%
%is ~W/kg, which is
%Here, we use the whole body \ac{SAR} which corresponds to a frequency of . After that,
%$2.6$
\begin{table}[t!]	
\renewcommand{\arraystretch}{1.3}
\caption{SIMULATION PARAMETERS}
\label{table_par}
\centering
\begin{tabular}{|c|c||c|c|}
\hline
\textbf{Constant} & \textbf{Value} & \textbf{Constant} & \textbf{Value}\\
\hline 
\hline
Area & $1000$~m $\times$ $1000$~m & ${\eta _{{\rm{LoS}}}}$ & $1.6$~dB\\
\hline
$a$ & 9.61 & ${\eta _{{\rm{NLoS}}}}$ &  $23$~dB  \\
\hline
$b$ & 0.16 &  $P_{\max}$ &  $26$~dBm\\
\hline
$c$ & $3\times 10^{8}$~m/s & $T_{\max}$ & $100$~m\\
\hline
$f_c$ & $3.5$~Ghz  & $\theta _{\min }$   &  $31^{\circ}$\\
\hline
$B$ & $10$~Mhz  &    $\alpha_{\rm{l}}$  &  $2$ \\
\hline
$h_\mathrm{B}$  &  $25$~m  &   $\alpha_{\rm{n}}$  &  $2$ \\
\hline
$W_{m,\max}$ & 6 &   $h_\mathrm{G}$  &  $30$~m\\
\hline
$\mathrm{SAR}_{\text{v}}$ & $0.0047$~W/kg & $\mathrm{SAR}_{\text{d}}$ & $0.0037$~W/kg \\
 \hline
\end{tabular}
\end{table}

% \begin{table}[t!]	
% \renewcommand{\arraystretch}{1.3}
% \caption{SIMULATION PARAMETERS}
% \label{table_par}
% \centering
% \begin{tabular}{|c|c||c|c|}
% \hline
% \textbf{Constant} & \textbf{Value} & \textbf{Constant} & \textbf{Value}\\
% \hline 
% \hline
% Area & $1000$~m $\times$ $1000$~m & ${\xi _{{\rm{LoS}}}}$ & $1.6$~dB\\
% \hline
% $a$ & 9.61 & ${\xi _{{\rm{NLoS}}}}$ &  $23$~dB  \\
% \hline
% $b$ & 0.16 &  $P_{\max}$ &  $26$~dBm\\
% \hline
% $c$ & $3\, 10^{8}$~m/s & $T_{\max}$ & $100$~m\\
% \hline
% $f_c$ & $3.5$~Ghz  & $\theta _{\min }$   &  $31^{\circ}$\\
% \hline
% $B$ & $10$~Mhz  &   $h_\mathrm{B}$  &  $25$~m\\
% \hline
% $W_{m,\max}$ & 6 &   $h_\mathrm{G}$  &  $30$~m\\
% \hline
% $\mathrm{SAR}_v$ & $0.0047$~W/kg & $\mathrm{SAR}_d$ & $0.0037$~W/kg \\
%  \hline
% \end{tabular}
% \end{table}
For convenience, the simulation parameters are summarized in Table \ref{table_par}, unless specified otherwise. In the following figures, for the scenario named \emph{\ac{BS} only}, we consider a single \ac{BS} in the center of the area, and all the users are connected to it for both \ac{UL} and \ac{DL}.
 %There are two types of users that can be activated at the same time, voice and speed, and they have different absorption rates, 1 and 2, respectively.
%There are two types of activated users, voice and data, and they have different rate requirements and absorption

%In almost all of the following figures, the scenario named \ac{BS} only would be used for comparison, so we describe it separately at the beginning. In this scenario, there is a single \ac{BS} in the center of the area, and users are connected to the \ac{BS}, which has enough \acp{RB} to provide services for all users, both \ac{DL} and \ac{UL}. Therefore, all users connect to the \ac{BS} by default, and there is no need to run any algorithm to adjust the connection.

% \textbf{\ac{BS} only}: 

\subsection{Performance of the Proposed Algorithms on Small Scale}

We move our attention to the performance of the proposed algorithms for each sub-optimization process. In section \ref{sec:alg}, it can be seen that we solved the optimization problem in three steps. Firstly, we determine the association between the user and \acp{gNB} for the given position of \acp{tUAV}, then ascertain the deployment of \acp{tUAV} in a 2D plane, and finally, slightly adjust the tether length and angle of each \ac{tUAV}. In the process of comparison, we consider the results of brute force search as a baseline. Therefore, in this simulation, we only consider two fixed \acp{tUAV} evenly distributed, each of which has two \acp{RB}. In addition, there are a \ac{BS} in the center of the region, $N=9$ evenly distributed \acp{GS}, and $\bar{K}=60$ residents. The number of active users $K$ ranges from 4 to 20, and all active users are \emph{data} users whose required \ac{UL} rate is $50$~Mbps. 

\subsubsection{Optimize the Association between Users and \acsp{gNB}}

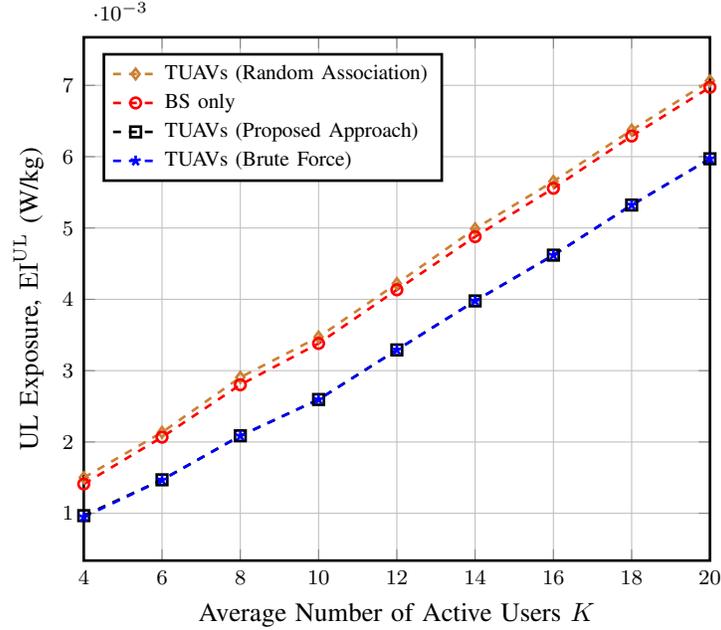
\begin{figure}[t!]
	\centering
	\pgfplotsset{every axis/.append style={
		font=\footnotesize,
		line width=1pt,
		legend style={font=\footnotesize, at={(0.98,0.31)}},legend cell align=left},
} %
\pgfplotsset{compat=1.13}
	\begin{tikzpicture}%[trim axis right]%[trim axis left, trim axis right]
\begin{axis}[
%axis
%loglog
%title= users' satisfaction ratio vs. data rate,
%axis x line=bottom,
%yticklabel style = {font=\footnotesize,xshift=0.5ex},
%xticklabel style = {font=\footnotesize,yshift=0.5ex},
%ylabel style = {font=\huge},
%xlabel style = {font=\huge},
legend pos=north west,
xlabel near ticks,
ylabel near ticks,
grid=major,
xlabel={Average Number of Active Users $K$},
ylabel={\acs{UL} Exposure, $\mathrm{EI}^\mathrm{UL}$ (W/kg) },
%width=0.79\linewidth,%3.5in,
%	ytick={10,0,-10,-20,-30,-40},
%	x label style={at={(axis description cs:0.5,-0.07)},anchor=north},
%   y label style={at={(axis description cs:-0.11,.5)},rotate=0,anchor=south},
%yticklabel style={/pgf/number format},
%axis y line=left,
width=0.6\linewidth,
%height=0.6\textwidth,
%	legend columns=1,	
%legend entries={firstmethos \cite{ElzGioChi:19}
%	,second $\x$,
%	third},
% 	log ticks with fixed point, 
%	scale only axis,
%	restrict
	xmin= 4, xmax=20,
%	ymin=0.5e-5, ymax=1e1,
ylabel style={font=\normalsize},
xlabel style={font=\normalsize},
%minor y tick num=10,
]

\addplot[orange!20!brown,mark=diamond,dashed,mark options=solid] table {figures/Association/ra.dat} ;

\addplot[red,mark=o,dashed,mark options=solid] table {figures/Association/bs.dat} ;

\addplot[black,mark=square,dashed,mark options=solid] table {figures/Association/pr.dat} ;

\addplot[blue,dashed,mark=star] table {figures/Association/bf.dat} ;

\legend{\acp{tUAV} (Random Association),BS only,\acp{tUAV} (Proposed Approach),\acp{tUAV} (Brute Force) }

%\addplot[black,mark=o,dashed,mark options=solid] table {figures/satisfiction/UAV_sat.dat} ;
%\addplot[orange!20!brown,mark=diamond,dashed] table {figures/satisfiction/SC_sat.dat} ;
%\addplot[blue,dashed,mark=star] table {figures/satisfiction/BS_sat.dat} ;
%
%\legend{BS with UAVs, BS with SCs, Only BS }

%\legend{$Only BS$,$BS with SCs$,$BS with UAVs$}
%	\addplot[red,dashed,mark=square] table {crbNoLISa4.dat};

%		\addplot[blue,mark=star] table {crbLISa1.dat};
%	\addplot[orange!20!brown ,mark=diamond,mark repeat=1] table {crbLISa2.dat};
%	\addplot[black,mark=o,mark options=solid] table {crbLISa3.dat};
% 	\addplot[red,mark=square] table {crbLISa4.dat};
%	\addplot[red,mark=square,mark repeat=4,dashed,mark options=solid] table {TSBCIM4fadingM4.dat};
%	\addplot[orange,mark=triangle,mark options=solid,mark repeat=2] table {UniformSD5fadingM4.dat}; 
%\addplot[orange,mark=star,mark size=1.5,mark repeat=6] table {UniformBCIM6fadingM4.dat};
%\addplot[brown] table {CapacityLBfadingM4.dat};
%\addplot[gray] table {PSDCEXPSF7.dat};
%\addplot[black,thick,mark repeat=4,mark options=solid] table {TxRatefadingM4.dat};
%\addplot[gray!50!black,thick,mark=o,mark options=solid, mark size=3] coordinates {
%	(6.65,1.13)
%};

%\addplot[gray,thin] table {Figures/FigPSDContMatlab/FFTFigPSDContSF12.txt};
%%/tikz/thin (no value)
%/tikz/ultra thin (no value)
%/tikz/very thin (no value)
%/tikz/semithick (no value)
%/tikz/thick (no value)
%/tikz/very thick (no value)
%/tikz/ultra thick
\end{axis}
\end{tikzpicture}
	\caption{Users' \ac{UL} EMF exposure for various users' association strategies, for $M=2$ and $W_{m,\max}=2$.}
	\label{fig:ass}
\end{figure}

In the beginning, we compare the effectiveness of different methods for determining the association matrix. In addition to the results of our proposed algorithm, we also show those of \ac{BS} only, randomly associating the user with \acp{gNB}, and brute force approach. As can be seen from Fig.~\ref{fig:ass}, randomly associating users with \acp{gNB} can even be counterproductive. The result of our proposed algorithm almost overlaps with that of brute force. Therefore, the association policy in the following simulations 
is determined via our proposed algorithm.
\subsubsection{Optimize the Deployment of \acsp{tUAV} to \acsp{GS} }

\begin{figure}[t!]
	\centering
	\pgfplotsset{every axis/.append style={
		font=\footnotesize,
		line width=1pt,
		legend style={font=\footnotesize, at={(0.98,0.31)}},legend cell align=left},
} %
\pgfplotsset{compat=1.13}
	\begin{tikzpicture}%[trim axis right]%[trim axis left, trim axis right]
\begin{axis}[
%axis
%loglog
%title= users' satisfaction ratio vs. data rate,
%axis x line=bottom,
%yticklabel style = {font=\footnotesize,xshift=0.5ex},
%xticklabel style = {font=\footnotesize,yshift=0.5ex},
%ylabel style = {font=\huge},
%xlabel style = {font=\huge},
legend pos=north west,
xlabel near ticks,
ylabel near ticks,
grid=major,
xlabel={Average Number of Active Users $K$},
ylabel={UL Exposure, $\mathrm{EI}^\mathrm{UL}$ (W/kg) },
%width=0.79\linewidth,%3.5in,
%	ytick={10,0,-10,-20,-30,-40},
%	x label style={at={(axis description cs:0.5,-0.07)},anchor=north},
%   y label style={at={(axis description cs:-0.11,.5)},rotate=0,anchor=south},
%yticklabel style={/pgf/number format},
%axis y line=left,
width=0.6\linewidth,
%height=0.6\textwidth,
%	legend columns=1,	
%legend entries={firstmethos \cite{ElzGioChi:19}
%	,second $\x$,
%	third},
% 	log ticks with fixed point, 
%	scale only axis,
%	restrict
	xmin= 4, xmax=20,
%	ymin=0.5e-5, ymax=1e1,
ylabel style={font=\normalsize},
xlabel style={font=\normalsize},
%minor y tick num=10,
]

%orange!20!brown,mark=diamond,dashed

\addplot[black,mark=o,dashed,mark options=solid] table {figures/deployment/bs_2.dat} ;

\addplot[orange!20!brown,mark=triangle,dashed,mark options=solid] table {figures/deployment/ra_2.dat} ;

%\addplot[blue,mark=triangle,dashed,mark options=solid] table {figures/deployment/ra_4.dat} ;

\addplot[blue,mark=square,dashed,mark options=solid,mark size=2.5pt] table {figures/deployment/km_2.dat} ;

\addplot[gray!50!black,mark=star,dashed,mark options=solid] table {figures/deployment/sr_2.dat} ;

\addplot[red,mark=diamond,dashed,mark repeat=2,mark options=solid] table {figures/deployment/bf_2.dat} ;

%\addplot[blue,mark=diamond,dashed,mark options=solid] table {figures/deployment/km_4.dat};
%
%\addplot[blue,mark=o,dashed,mark options=solid] table {figures/deployment/sr_4.dat} ;
%
%\addplot[blue,mark=star,dashed,mark options=solid] table {figures/deployment/bf_4.dat} ;

\legend{BS only,\acp{tUAV} (Random Deployment),\acp{tUAV} (Modified $K$-mean),\acp{tUAV} (2D SR Process),\acp{tUAV} (Brute Force) }

%\addplot[black,mark=o,dashed,mark options=solid] table {figures/satisfiction/UAV_sat.dat} ;
%\addplot[orange!20!brown,mark=diamond,dashed] table {figures/satisfiction/SC_sat.dat} ;
%\addplot[blue,dashed,mark=star] table {figures/satisfiction/BS_sat.dat} ;
%
%\legend{BS with UAVs, BS with SCs, Only BS }

%\legend{$Only BS$,$BS with SCs$,$BS with UAVs$}
%	\addplot[red,dashed,mark=square] table {crbNoLISa4.dat};

%		\addplot[blue,mark=star] table {crbLISa1.dat};
%	\addplot[orange!20!brown ,mark=diamond,mark repeat=1] table {crbLISa2.dat};
%	\addplot[black,mark=o,mark options=solid] table {crbLISa3.dat};
% 	\addplot[red,mark=square] table {crbLISa4.dat};
%	\addplot[red,mark=square,mark repeat=4,dashed,mark options=solid] table {TSBCIM4fadingM4.dat};
%	\addplot[orange,mark=triangle,mark options=solid,mark repeat=2] table {UniformSD5fadingM4.dat}; 
%\addplot[orange,mark=star,mark size=1.5,mark repeat=6] table {UniformBCIM6fadingM4.dat};
%\addplot[brown] table {CapacityLBfadingM4.dat};
%\addplot[gray] table {PSDCEXPSF7.dat};
%\addplot[black,thick,mark repeat=4,mark options=solid] table {TxRatefadingM4.dat};
%\addplot[gray!50!black,thick,mark=o,mark options=solid, mark size=3] coordinates {
%	(6.65,1.13)
%};

%\addplot[gray,thin] table {Figures/FigPSDContMatlab/FFTFigPSDContSF12.txt};
%%/tikz/thin (no value)
%/tikz/ultra thin (no value)
%/tikz/very thin (no value)
%/tikz/semithick (no value)
%/tikz/thick (no value)
%/tikz/very thick (no value)
%/tikz/ultra thick
\end{axis}
\end{tikzpicture}
	\caption{The \ac{UL} EMF exposure for different  \acp{tUAV} association methods with \acp{GS}, for $M=2$, $W_{m,\max}=2$, and $N=9$.}
	\label{fig:dep}
\end{figure}
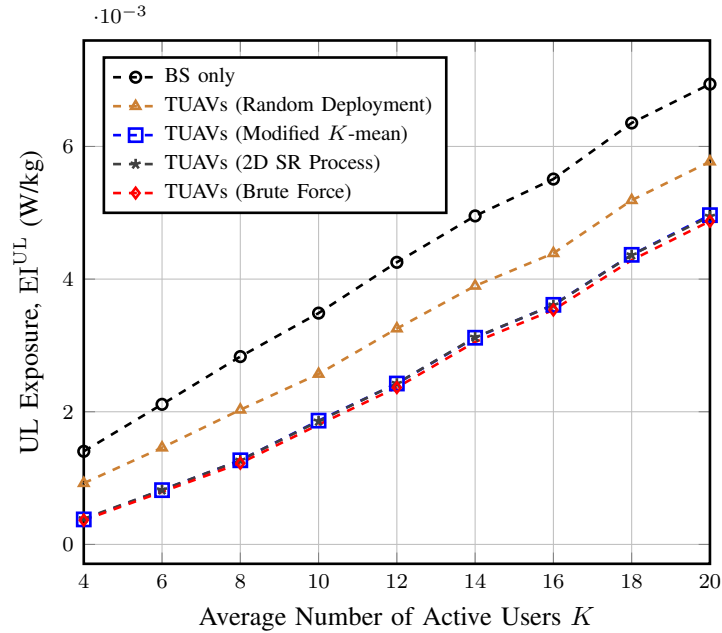

Fig.~\ref{fig:dep} reports the numerical evaluation of radiation, considering associating the \acp{tUAV} with proper \acp{GS}. We compare the results of \ac{BS} only, random deployment of \acp{tUAV}, and traversing all combinations (brute force search) with the results of the two proposed algorithms. As shown in Fig.~\ref{fig:dep}, the results of the modified $K$-mean and \ac{SR} algorithms almost coincide, and there is not much difference between them and that of brute force. Furthermore, these two algorithms can provide an additional $15\mbox{}\%$ gain compared to the \acp{tUAV}’ random deployment. Combined with the previous complexity analysis, we chose the modified $K$-mean with less complexity to determine the \acp{tUAV}’ deployment in the following results.

\subsubsection{Optimize \acs{tUAV}’s Position within the Hovering}

\begin{figure}[t!]
	\centering
	\pgfplotsset{every axis/.append style={
		font=\footnotesize,
		line width=1pt,
		legend style={font=\footnotesize, at={(0.98,0.31)}},legend cell align=left},
} %
\pgfplotsset{compat=1.13}
	\begin{tikzpicture}%[trim axis right]%[trim axis left, trim axis right]
\begin{axis}[
%axis
%loglog
%title= users' satisfaction ratio vs. data rate,
%axis x line=bottom,
%yticklabel style = {font=\footnotesize,xshift=0.5ex},
%xticklabel style = {font=\footnotesize,yshift=0.5ex},
%ylabel style = {font=\huge},
%xlabel style = {font=\huge},
legend pos=north west,
xlabel near ticks,
ylabel near ticks,
grid=major,
xlabel={Average Number of Active Users $K$},
ylabel={UL Exposure, $\mathrm{EI}^\mathrm{UL}$ (W/kg) },
%width=0.79\linewidth,%3.5in,
%	ytick={10,0,-10,-20,-30,-40},
%	x label style={at={(axis description cs:0.5,-0.07)},anchor=north},
%   y label style={at={(axis description cs:-0.11,.5)},rotate=0,anchor=south},
%yticklabel style={/pgf/number format},
%axis y line=left,
width=0.6\linewidth,
%height=0.6\textwidth,
%	legend columns=1,	
%legend entries={firstmethos \cite{ElzGioChi:19}
%	,second $\x$,
%	third},
% 	log ticks with fixed point, 
%	scale only axis,
%	restrict
	xmin= 4, xmax=20,
%	ymin=0.5e-5, ymax=1e1,
ylabel style={font=\normalsize},
xlabel style={font=\normalsize},
%minor y tick num=10,
]

%orange!20!brown,mark=diamond,dashed

\addplot[black,mark=o,dashed,mark options=solid] table {figures/length_angle/bs.dat} ;

\addplot[orange!20!brown,mark=triangle,dashed,mark options=solid] table {figures/length_angle/ra.dat} ;

%\addplot[blue,mark=triangle,dashed,mark options=solid] table {figures/deployment/ra_4.dat} ;

\addplot[blue,mark=square,dashed,mark options=solid] table {figures/length_angle/km.dat} ;

\addplot[gray!50!black,mark=star,dashed,mark options=solid] table {figures/length_angle/sr.dat} ;

\addplot[red,mark=diamond,dashed,mark options=solid] table {figures/length_angle/bf.dat} ;

%\addplot[blue,mark=diamond,dashed,mark options=solid] table {figures/deployment/km_4.dat};
%
%\addplot[blue,mark=o,dashed,mark options=solid] table {figures/deployment/sr_4.dat} ;
%
%\addplot[blue,mark=star,dashed,mark options=solid] table {figures/deployment/bf_4.dat} ;

\legend{BS only, \acp{tUAV} (Random Hovering),\acp{tUAV} (Modified Golden Search),\acp{tUAV} (3D SR Process),\acp{tUAV} (Brute Force) }

%\addplot[black,mark=o,dashed,mark options=solid] table {figures/satisfiction/UAV_sat.dat} ;
%\addplot[orange!20!brown,mark=diamond,dashed] table {figures/satisfiction/SC_sat.dat} ;
%\addplot[blue,dashed,mark=star] table {figures/satisfiction/BS_sat.dat} ;
%
%\legend{BS with UAVs, BS with SCs, Only BS }

%\legend{$Only BS$,$BS with SCs$,$BS with UAVs$}
%	\addplot[red,dashed,mark=square] table {crbNoLISa4.dat};

%		\addplot[blue,mark=star] table {crbLISa1.dat};
%	\addplot[orange!20!brown ,mark=diamond,mark repeat=1] table {crbLISa2.dat};
%	\addplot[black,mark=o,mark options=solid] table {crbLISa3.dat};
% 	\addplot[red,mark=square] table {crbLISa4.dat};
%	\addplot[red,mark=square,mark repeat=4,dashed,mark options=solid] table {TSBCIM4fadingM4.dat};
%	\addplot[orange,mark=triangle,mark options=solid,mark repeat=2] table {UniformSD5fadingM4.dat}; 
%\addplot[orange,mark=star,mark size=1.5,mark repeat=6] table {UniformBCIM6fadingM4.dat};
%\addplot[brown] table {CapacityLBfadingM4.dat};
%\addplot[gray] table {PSDCEXPSF7.dat};
%\addplot[black,thick,mark repeat=4,mark options=solid] table {TxRatefadingM4.dat};
%\addplot[gray!50!black,thick,mark=o,mark options=solid, mark size=3] coordinates {
%	(6.65,1.13)
%};

%\addplot[gray,thin] table {Figures/FigPSDContMatlab/FFTFigPSDContSF12.txt};
%%/tikz/thin (no value)
%/tikz/ultra thin (no value)
%/tikz/very thin (no value)
%/tikz/semithick (no value)
%/tikz/thick (no value)
%/tikz/very thick (no value)
%/tikz/ultra thick
\end{axis}
\end{tikzpicture}
	\caption{The \ac{UL} EMF exposure  for various positioning techniques for the  \acp{tUAV} within their hovering zone, for $M=2$, $W_{m,\max}=2$, and $N=9$.}
	\label{fig:angle}
\end{figure}
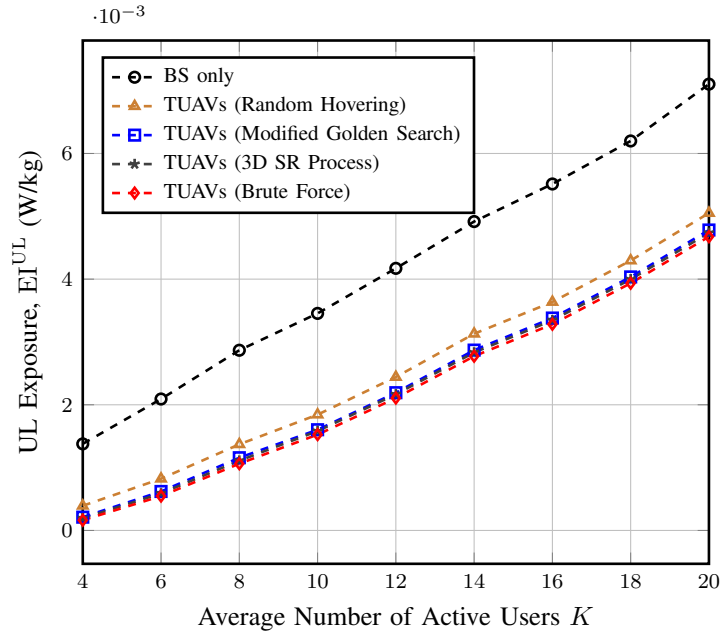

In the final step, we consider each \ac{tUAV} separately and only move them in a small area around the \ac{GS}. Furthermore, we also compare the results of the two proposed algorithms with the other three baselines, \ac{BS}, random selection, and brute force search over a gridded hovering zone. As can be seen from Fig.~\ref{fig:angle}, the optimization over the fine location of the \acp{tUAV} in the hovering area can result in a gain, albeit not as large as the previous two. Moreover, although the results of the two methods are relatively close, the gain brought by the adopted golden search is slightly smaller than that of the 3D \ac{SR} process. So the latter approach will be used in subsequent simulations.

{\color{black} Another alternate optimization iterations can be considered after solving the problem by three-step sub-algorithms. In this iteration, the optimized variables of the last iteration are provided as the initial values of the optimization variables.  Such alternate optimization should further enhance the performance, at the expense of additional complexity.}

%Moreover, when we use the proposed scheme to optimize the network in real life, most of the parameters are fixed or can be found in some tables. What might change is the number of \acp{tUAV}, the location and number of \acp{GS}, and number of \acp{RB} for each \ac{gNB}. \

%We can optimize these parameters according to users' location information, which can be found by wireless localization or \ac{GPS} 

%\subsection{Comparison of Different Types \acp{tUAV}}
%\subsection{Comparison of \acp{tUAV} Used to Densify the \ac{UL} and/or \ac{DL}}
%\subsection{\Acs{UL} and \Acs{DL} Exposure with \acsp{tUAV}}
\subsection{\acs{UL} and \acs{DL} Exposure with \acsp{tUAV} }

In the following, we analyze the effects of different types of \acp{tUAV} (i.e., various \ac{UL} and \ac{DL} decoupling methods) on reducing \ac{EMF} exposure. In each scenario, \acp{tUAV} have $N=25$ \acp{GS} to choose from. We have $\bar{K}= 120$ residents in the area, $K$ out of them are randomly selected as active users, who have two different ways of using the phone, voice ($20\%$ of users) or data ($80\%$). We consider that the maximum available \acp{RB} for each \ac{tUAV} is $W_{m,\max}=6$, while \ac{BS} has sufficient resources to allocate all users in the area. If \acp{tUAV} are used to densify(i.e., assist) both \ac{UL} and \ac{DL}, then each user needs to occupy two \acp{RB}, and the maximum number of users connected to each  \ac{tUAV} is ${W_{m,\max}}/{2}$. There are four scenarios, of which \emph{\ac{BS} only} is previously described, and the other three involve $M=4$ \acp{tUAV}, and %with $W_{m,\max}=6$ \Acp{RB}
they are described as follows{\color{black}:}
\begin{itemize}
%	\item \textbf{\ac{BS} only}: In this scenario, there is  1 \ac{BS} in the center of the area, and users are connected to the \ac{BS}, which has enough \acp{RB} to provide services for all users, both \ac{DL} and \ac{UL}. Therefore, all users connect to \ac{BS} by default, and there is no need to run any algorithm to adjust the connection.
	
	\begin{figure}[t!]
	\centering
	\pgfplotsset{every axis/.append style={
		font=\footnotesize,
		line width=1pt,
		legend style={font=\footnotesize, at={(0.98,0.31)}},legend cell align=left},
} %
\pgfplotsset{compat=1.13}
	\begin{tikzpicture}%[trim axis right]%[trim axis left, trim axis right]
\begin{axis}[
%axis
%loglog
%title= users' satisfaction ratio vs. data rate,
%axis x line=bottom,
%yticklabel style = {font=\footnotesize,xshift=0.5ex},
%xticklabel style = {font=\footnotesize,yshift=0.5ex},
%ylabel style = {font=\huge},
%xlabel style = {font=\huge},
legend pos= north west,
xlabel near ticks,
ylabel near ticks,
grid=major,
xlabel={Average Number of Active Users $K$},
ylabel={UL\&DL Exposure, $\mathrm{EI}$ (W/kg) },
%width=0.79\linewidth,%3.5in,
%	ytick={10,0,-10,-20,-30,-40},
%	x label style={at={(axis description cs:0.5,-0.07)},anchor=north},
%   y label style={at={(axis description cs:-0.11,.5)},rotate=0,anchor=south},
%yticklabel style={/pgf/number format},
%axis y line=left,
width=0.6\linewidth,
%height=0.6\textwidth,
%	legend columns=1,	
%legend entries={firstmethos \cite{ElzGioChi:19}
%	,second $\x$,
%	third},
% 	log ticks with fixed point, 
%	scale only axis,
%	restrict
	xmin= 6, xmax=48,
%	ymin=0.5e-5, ymax=1e1,
ylabel style={font=\normalsize},
xlabel style={font=\normalsize},
%minor y tick num=10,
]

\addplot[blue,dashed,mark=star] table {figures/UL_DL_LAM/bs.dat} ;

\addplot[black,dashed,mark=o,mark options=solid, mark size=2.9pt] table {figures/UL_DL_LAM/dd.dat} ;

\addplot[orange,dashed,mark=diamond,mark options=solid] table {figures/UL_DL_LAM/ud.dat} ;

\addplot[red,mark=square,dashed,mark options=solid] table {figures/UL_DL_LAM/uu.dat};

\legend{BS only,Special \acsp{tUAV} (Assisting DL) ,Regular \acsp{tUAV} (Assisting DL and UL),Green \acsp{tUAV} (Assisting UL)}

%\addplot[black,mark=o,dashed,mark options=solid] table {figures/satisfiction/UAV_sat.dat} ;
%\addplot[orange!20!brown,mark=diamond,dashed] table {figures/satisfiction/SC_sat.dat} ;
%\addplot[blue,dashed,mark=star] table {figures/satisfiction/BS_sat.dat} ;
%
%\legend{BS with UAVs, BS with SCs, Only BS }

%\legend{$Only BS$,$BS with SCs$,$BS with UAVs$}
%	\addplot[red,dashed,mark=square] table {crbNoLISa4.dat};

%		\addplot[blue,mark=star] table {crbLISa1.dat};
%	\addplot[orange!20!brown ,mark=diamond,mark repeat=1] table {crbLISa2.dat};
%	\addplot[black,mark=o,mark options=solid] table {crbLISa3.dat};
% 	\addplot[red,mark=square] table {crbLISa4.dat};
%	\addplot[red,mark=square,mark repeat=4,dashed,mark options=solid] table {TSBCIM4fadingM4.dat};
%	\addplot[orange,mark=triangle,mark options=solid,mark repeat=2] table {UniformSD5fadingM4.dat}; 
%\addplot[orange,mark=star,mark size=1.5,mark repeat=6] table {UniformBCIM6fadingM4.dat};
%\addplot[brown] table {CapacityLBfadingM4.dat};
%\addplot[gray] table {PSDCEXPSF7.dat};
%\addplot[black,thick,mark repeat=4,mark options=solid] table {TxRatefadingM4.dat};
%\addplot[gray!50!black,thick,mark=o,mark options=solid, mark size=3] coordinates {
%	(6.65,1.13)
%};

%\addplot[gray,thin] table {Figures/FigPSDContMatlab/FFTFigPSDContSF12.txt};
%%/tikz/thin (no value)
%/tikz/ultra thin (no value)
%/tikz/very thin (no value)
%/tikz/semithick (no value)
%/tikz/thick (no value)
%/tikz/very thick (no value)
%/tikz/ultra thick
\end{axis}
\end{tikzpicture}
	\caption{Total \ac{UL} and \ac{DL} EMF exposure for various decoupling techniques with \acp{tUAV}, for $M=4$ and $N=25$.}
	\label{fig:total}
\end{figure}
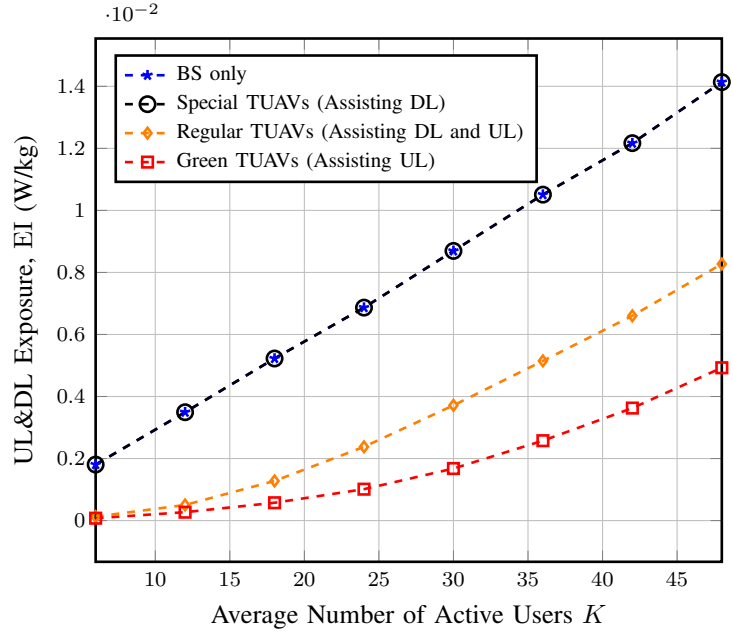

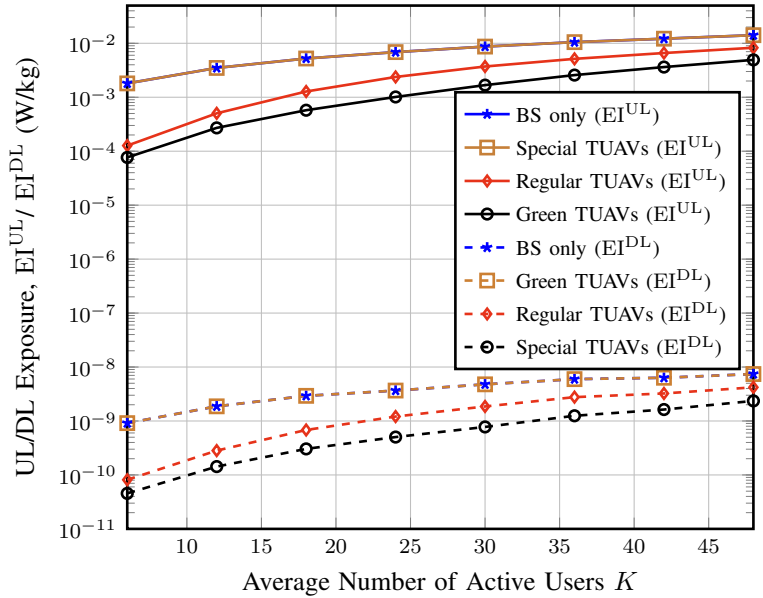
\begin{figure}[t!]
	\centering
	\pgfplotsset{every axis/.append style={
		font=\footnotesize,
		line width=1pt,
		legend style={font=\footnotesize, at={(0.98,0.58)}},legend cell align=left},
} %
\pgfplotsset{compat=1.13}
	\begin{tikzpicture}%[trim axis right]%[trim axis left, trim axis right]
\begin{semilogyaxis}[
%axis
%loglog
%title= users' satisfaction ratio vs. data rate,
%axis x line=bottom,
%yticklabel style = {font=\footnotesize,xshift=0.5ex},
%xticklabel style = {font=\footnotesize,yshift=0.5ex},
%ylabel style = {font=\huge},
%xlabel style = {font=\huge},
legend style={at={(0.98,0.57)},anchor=east},
xlabel near ticks,
ylabel near ticks,
grid=major,
xlabel={Average Number of Active Users $K$},
ylabel={UL/DL Exposure, $\mathrm{EI}^\mathrm{UL}$/ $\mathrm{EI}^\mathrm{DL}$ (W/kg) },
%width=0.79\linewidth,%3.5in,
%	ytick={10,0,-10,-20,-30,-40},
%	x label style={at={(axis description cs:0.5,-0.07)},anchor=north},
%   y label style={at={(axis description cs:-0.11,.5)},rotate=0,anchor=south},
%yticklabel style={/pgf/number format},
%axis y line=left,
width=0.6\linewidth,
%height=0.6\textwidth,
%	legend columns=1,	
%legend entries={firstmethos \cite{ElzGioChi:19}
%	,second $\x$,
%	third},
% 	log ticks with fixed point, 
%	scale only axis,
%	restrict
	xmin= 6, xmax=48,
	ymin=1e-11, ymax=0.5e-1,
ylabel style={font=\normalsize},
xlabel style={font=\normalsize},
%minor y tick num=10,
]

%dashed
\addplot[blue,mark=star] table {figures/UL_DL_LAM/ul_bs.dat} ;

\addplot[orange!10!brown,mark=square,mark size=2.5pt] table {figures/UL_DL_LAM/ul_bs.dat} ;

\addplot[red!60!brown,mark=diamond] table {figures/UL_DL_LAM/ul_ud.dat} ;

\addplot[black,mark=o,mark options=solid] table {figures/UL_DL_LAM/ul_uo.dat} ;

%DL

\addplot[blue,mark=star,dashed,mark options=solid] table {figures/UL_DL_LAM/dl_bs.dat} ;

\addplot[orange!20!brown,dashed,mark=square,mark size=2.5pt,mark options=solid] table {figures/UL_DL_LAM/dl_bs.dat} ;

\addplot[red!60!brown,dashed,mark=diamond,mark options=solid] table {figures/UL_DL_LAM/dl_ud.dat} ;

\addplot[black,mark=o,dashed,mark options=solid] table {figures/UL_DL_LAM/dl_do.dat} ;

%\legend{UL(BS only), UL(BS with 4 regular tUAVs(both DL and UL)),UL(BS with 4 green tUAVs(UL only)),DL(BS only), DL(BS with 4 regular tUAVs(both DL and UL)), DL(BS with 4 special tUAVs(DL only))}
\legend{BS only ($\mathrm{EI}^\mathrm{UL}$), Special \acsp{tUAV} ($\mathrm{EI}^\mathrm{UL}$), Regular \acsp{tUAV} ($\mathrm{EI}^\mathrm{UL}$), Green \acsp{tUAV} ($\mathrm{EI}^\mathrm{UL}$), BS only ($\mathrm{EI}^\mathrm{DL}$),  Green \acsp{tUAV} ($\mathrm{EI}^\mathrm{DL}$), Regular \acsp{tUAV} ($\mathrm{EI}^\mathrm{DL}$), Special \acsp{tUAV} ($\mathrm{EI}^\mathrm{DL}$) }

%\addplot[black,mark=o,dashed,mark options=solid] table {figures/satisfiction/UAV_sat.dat} ;
%\addplot[orange!20!brown,mark=diamond,dashed] table {figures/satisfiction/SC_sat.dat} ;
%\addplot[blue,dashed,mark=star] table {figures/satisfiction/BS_sat.dat} ;
%
%\legend{BS with UAVs, BS with SCs, Only BS }

%\legend{$Only BS$,$BS with SCs$,$BS with UAVs$}
%	\addplot[red,dashed,mark=square] table {crbNoLISa4.dat};

%		\addplot[blue,mark=star] table {crbLISa1.dat};
%	\addplot[orange!20!brown ,mark=diamond,mark repeat=1] table {crbLISa2.dat};
%	\addplot[black,mark=o,mark options=solid] table {crbLISa3.dat};
% 	\addplot[red,mark=square] table {crbLISa4.dat};
%	\addplot[red,mark=square,mark repeat=4,dashed,mark options=solid] table {TSBCIM4fadingM4.dat};
%	\addplot[orange,mark=triangle,mark options=solid,mark repeat=2] table {UniformSD5fadingM4.dat}; 
%\addplot[orange,mark=star,mark size=1.5,mark repeat=6] table {UniformBCIM6fadingM4.dat};
%\addplot[brown] table {CapacityLBfadingM4.dat};
%\addplot[gray] table {PSDCEXPSF7.dat};
%\addplot[black,thick,mark repeat=4,mark options=solid] table {TxRatefadingM4.dat};
%\addplot[gray!50!black,thick,mark=o,mark options=solid, mark size=3] coordinates {
%	(6.65,1.13)
%};

%\addplot[gray,thin] table {Figures/FigPSDContMatlab/FFTFigPSDContSF12.txt};
%%/tikz/thin (no value)
%/tikz/ultra thin (no value)
%/tikz/very thin (no value)
%/tikz/semithick (no value)
%/tikz/thick (no value)
%/tikz/very thick (no value)
%/tikz/ultra thick
\end{semilogyaxis}
\end{tikzpicture}
	\caption{\ac{UL} and \ac{DL} EMF exposure for various decoupling techniques with \acp{tUAV},  for $M=4$ and $N=25$.}
	\label{fig:ul_dl}
\end{figure}
	
	\item {\emph{Green \acp{tUAV} (Assisting \ac{UL})}}: This scenario is  our proposed network architecture, where
	the location of \ac{BS} is fixed like last scene, and we additionally add green \acp{tUAV} which can only receive  information from users and forward it to \ac{BS} through cable. Moreover, all users receive signals from the \ac{BS}, and in \ac{UL} up to 6 users are connected to each \ac{tUAV}. 
	%Here, we just focus on \ac{UL} and the \ac{DL} connection is no need to adjust because all the users receive signals from  \ac{BS}. 
	Therefore, we  run \textbf{Algorithm} \ref{alg.limited_1}  to adjust association matrix, while operate  \textbf{Algorithm} \ref{alg.kmean} to regulate the position of \acp{tUAV} in 2D space. After the above two steps converge, we will adjust the tether length and angles of each \ac{tUAV} through \textbf{Algorithm} \ref{alg.SR}. 
	
	\item {\emph{Regular \acp{tUAV} (Assisting \ac{UL} and \ac{DL})}}: In this scenario, the location of \ac{BS} is fixed like last scene, and we also additionally add regular \acp{tUAV} to densify both \ac{UL} and \ac{DL}. As we mentioned before, only 3 users can be connected to each \ac{tUAV} in this scenario, because each user  occupies 2 \acp{RB}. After that, we determine the location of \acp{tUAV} and user's association matrix as described in the previous scenario. The exposure is computed from \eqref{eq:EIboth}, \eqref{eq:ul}, and \eqref{eq.dl}.
	
	\item {\emph{Special \acp{tUAV} (Assisting \ac{DL})}}: In this scenario, the only difference between last scene is that \acp{tUAV} can only receive messages from \ac{BS} through cables, then forward them to users via wireless \ac{DL}, so the \ac{tUAV} can connect up to 6 users. Moreover, we also run the three algorithms mentioned above to minimize the \ac{DL}  exposure.
	
	%can also run the algorithms mentioned before to minimize \ac{EMF} exposure as much as possible.   %Here, for \ac{UL}, the presence of \acp{tUAV} don't change the user's association, so we only need to focus on \ac{DL}. Just like \ac{UL}, we can also run algorithms mentioned before to minimize \ac{EMF} exposure as much as possible.  
\end{itemize}

The effectiveness of different types of \acp{tUAV} in reducing \ac{EMF} exposure is shown in Fig.~\ref{fig:total} and Fig.~\ref{fig:ul_dl}. In this scenario, the number of active users $K$ varies from $6$ to $48$.  Among these active users, $20\%$ of them use mobile phones mainly for making calls with required \ac{UL} voice rate $5$~Mbps, the remaining $80\%$  use mobile phones for data with a high data rate requirement, i.e., $50$~Mbps for \ac{UL}. For \ac{DL}, the rate required by the voice is $5$~Mbps, while that of data is $100$~Mbps. %Notice that everyone in the area is affected by the \ac{DL} radiation due to the high power emitted by \acp{gNB}.
Fig.~\ref{fig:total} reveals that using \ac{tUAV} to densify \ac{DL} are not an effective way to reduce \ac{EMF} exposure. This is because \ac{EMF} exposure from \ac{DL} is almost negligible compare to that of \ac{UL}. To show this more clearly, the \ac{EMF} exposure separately generated by \ac{DL} and \ac{UL} in different scenarios are shown in Fig.~\ref{fig:ul_dl}. As can be seen, there is a great difference in order of magnitude, i.e., \ac{UL} radiation is about $10^6$ times that of \ac{DL}. In contrast, using \acp{tUAV} that can densify \ac{UL} could effectively reduce the \ac{EMF} exposure, especially green \acp{tUAV}. Because we use regular \ac{tUAV} to densify both \ac{UL} and \ac{DL}, each user will occupy two \acp{RB} and the maximum number of users connected to \ac{tUAV} will be halved. Therefore, the effect of regular \ac{tUAV} is not as good as the green \ac{tUAV}, given that they have the same amount of \acp{RB} which are relatively limited. Moreover, 
green \acp{tUAV}, with receiving only \ac{RF} chains, have lower complexity than regular \acp{tUAV}.

\subsection{Green Fixed \acsp{SC} and Green \acsp{tUAV}} 
%By comparing different kinds of \acp{tUAV}, green \ac{tUAV} proved to be an effective way to reduce EMF exposure.
%
In the following, we analyze the performance of green \acp{tUAV} and green \acp{SC}, both of which are used to densify the \ac{UL}. The only difference between them is that \acp{tUAV} are much more flexible, while  \acp{SC} have a fixed deployment, and they are evenly distributed in the area.  Since the achieved \ac{UL} data rate can be less than the required one due to the limitation in the devices' transmit power, we define the \emph{satisfied-users ratio} as the ratio of the number of users who satisfy their rate constraint \eqref{eq:constraint6} to that of total active users. We study the enhancement of the satisfied-users ratio when assisted by $M=4$ green \acp{tUAV} or green \acp{SC}. The locations of \acp{SC} are fixed, while \acp{tUAV} have $N=36$ \ac{GS} to choose from. We also consider $\bar{K}=240$ residents distributed in the area, $25\%$ of them are active, and each of them has the same data rate requirement, i.e., \emph{data} users. 
%Next, we present simulation results for two different optimization objectives. One is to minimize the radiation while meeting the user's basic \ac{UL} rate requirements as much as possible. Another one is the dual problem, maximizing the total \ac{UL} rate while keeping \ac{EMF} exposure received by each user less than a threshold.

%\subsubsection{Minimize \ac{EMF} exposure}
\begin{figure}[t!]
	\centering
	\pgfplotsset{every axis/.append style={
		font=\footnotesize,
		line width=1pt,
		legend style={font=\footnotesize, at={(0.98,0.31)}},legend cell align=left},
} %
\pgfplotsset{compat=1.13}
	\begin{tikzpicture}%[trim axis right]%[trim axis left, trim axis right]
\begin{axis}[
%axis
%loglog
%title= users' satisfaction ratio vs. data rate,
%axis x line=bottom,
%yticklabel style = {font=\footnotesize,xshift=0.5ex},
%xticklabel style = {font=\footnotesize,yshift=0.5ex},
%ylabel style = {font=\huge},
%xlabel style = {font=\huge},
legend pos=north east,
xlabel near ticks,
ylabel near ticks,
grid=major,
xlabel={Required \ac{UL} Data  Rate $R_{k,\min}$ (Mbps)},
ylabel={Satisfied-users Ratio (\%) },
%width=0.79\linewidth,%3.5in,
%	ytick={10,0,-10,-20,-30,-40},
%	x label style={at={(axis description cs:0.5,-0.07)},anchor=north},
%   y label style={at={(axis description cs:-0.11,.5)},rotate=0,anchor=south},
yticklabel style={/pgf/number format},
%axis y line=left,
width=0.6\linewidth,
%height=0.6\textwidth,
%	legend columns=1,	
%legend entries={firstmethos \cite{ElzGioChi:19}
%	,second $\x$,
%	third},
% 	log ticks with fixed point, 
%	scale only axis,
%	restrict
	xmin= 10, xmax=180,
%	ymin=0.5e-5, ymax=1e1,
ylabel style={font=\normalsize},
xlabel style={font=\normalsize},
%minor y tick num=10,
]

\addplot[black,mark=o,dashed,mark options=solid] table {figures/satisfiction/UAV_sat.dat} ;
\addplot[orange!20!brown,mark=diamond,dashed,mark options=solid] table {figures/satisfiction/SC_sat.dat} ;
\addplot[blue,dashed,mark=star] table {figures/satisfiction/BS_sat.dat} ;

\legend{Green \acsp{tUAV}, Green SCs, BS only }

%\legend{$Only BS$,$BS with SCs$,$BS with UAVs$}
%	\addplot[red,dashed,mark=square] table {crbNoLISa4.dat};

%		\addplot[blue,mark=star] table {crbLISa1.dat};
%	\addplot[orange!20!brown ,mark=diamond,mark repeat=1] table {crbLISa2.dat};
%	\addplot[black,mark=o,mark options=solid] table {crbLISa3.dat};
% 	\addplot[red,mark=square] table {crbLISa4.dat};
%	\addplot[red,mark=square,mark repeat=4,dashed,mark options=solid] table {TSBCIM4fadingM4.dat};
%	\addplot[orange,mark=triangle,mark options=solid,mark repeat=2] table {UniformSD5fadingM4.dat}; 
%\addplot[orange,mark=star,mark size=1.5,mark repeat=6] table {UniformBCIM6fadingM4.dat};
%\addplot[brown] table {CapacityLBfadingM4.dat};
%\addplot[gray] table {PSDCEXPSF7.dat};
%\addplot[black,thick,mark repeat=4,mark options=solid] table {TxRatefadingM4.dat};
%\addplot[gray!50!black,thick,mark=o,mark options=solid, mark size=3] coordinates {
%	(6.65,1.13)
%};

%\addplot[gray,thin] table {Figures/FigPSDContMatlab/FFTFigPSDContSF12.txt};
%%/tikz/thin (no value)
%/tikz/ultra thin (no value)
%/tikz/very thin (no value)
%/tikz/semithick (no value)
%/tikz/thick (no value)
%/tikz/very thick (no value)
%/tikz/ultra thick
\end{axis}
\end{tikzpicture}
	\caption{Satisfied-users ratio versus required \ac{UL} data rate, for $M=4$, $N=36$, and $K=60$.}
	\label{fig:satisfiction}
\end{figure}
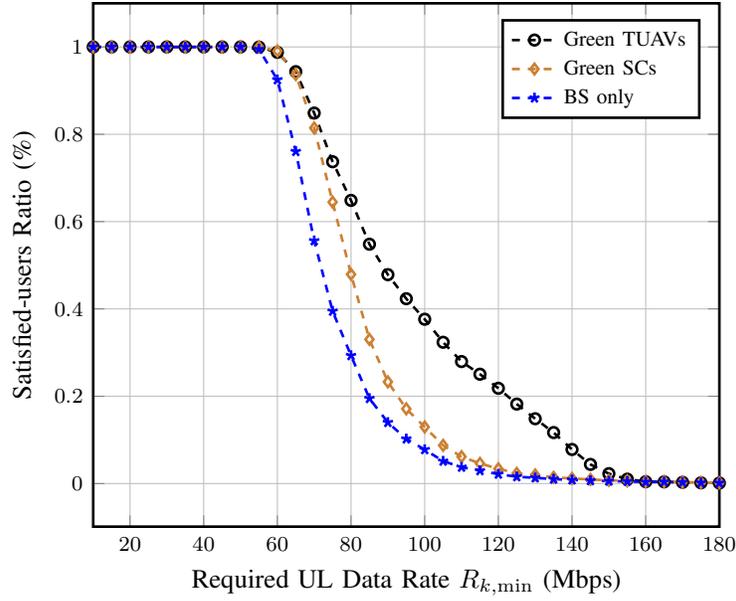
 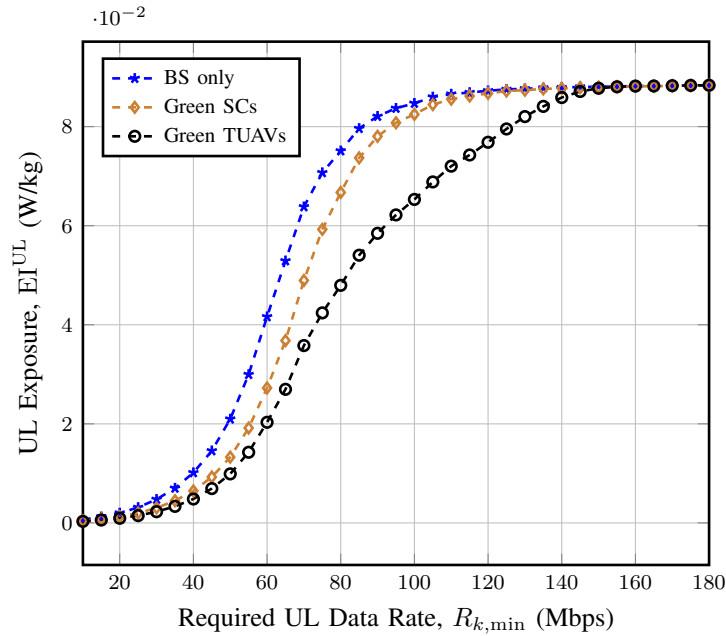
\begin{figure}[t!]
	\centering
	\pgfplotsset{every axis/.append style={
		font=\footnotesize,
		line width=1pt,
		legend style={font=\footnotesize, at={(0.98,0.31)}},legend cell align=left},
} %
\pgfplotsset{compat=1.13}
	\begin{tikzpicture}%[trim axis right]%[trim axis left, trim axis right]
\begin{axis}[
%axis
%loglog
%title= users' satisfaction ratio vs. data rate,
%axis x line=bottom,
%yticklabel style = {font=\footnotesize,xshift=0.5ex},
%xticklabel style = {font=\footnotesize,yshift=0.5ex},
%ylabel style = {font=\huge},
%xlabel style = {font=\huge},
legend pos=north west,
xlabel near ticks,
ylabel near ticks,
grid=major,
xlabel={Required \ac{UL} Data Rate, $R_{k,\min}$ (Mbps)},
ylabel={UL Exposure, $\mathrm{EI}^\mathrm{UL}$ (W/kg) },
%width=0.79\linewidth,%3.5in,
%	ytick={10,0,-10,-20,-30,-40},
%	x label style={at={(axis description cs:0.5,-0.07)},anchor=north},
%   y label style={at={(axis description cs:-0.11,.5)},rotate=0,anchor=south},
%yticklabel style={/pgf/number format},
%axis y line=left,
width=0.6\linewidth,
%height=0.6\textwidth,
%	legend columns=1,	
%legend entries={firstmethos \cite{ElzGioChi:19}
%	,second $\x$,
%	third},
% 	log ticks with fixed point, 
%	scale only axis,
%	restrict
	xmin= 10, xmax=180,
%	ymin=0.5e-5, ymax=1e1,
ylabel style={font=\normalsize},
xlabel style={font=\normalsize},
%minor y tick num=10,
]

\addplot[blue,dashed,mark=star] table {figures/satisfiction/BS_sat_emf.dat} ;
\addplot[orange!20!brown,mark=diamond,dashed,mark options=solid] table {figures/satisfiction/SC_sat_emf.dat} ;
\addplot[black,mark=o,dashed,mark options=solid] table {figures/satisfiction/UAV_sat_emf.dat} ;

\legend{BS only,Green SCs,Green \acsp{tUAV} }

%\addplot[black,mark=o,dashed,mark options=solid] table {figures/satisfiction/UAV_sat.dat} ;
%\addplot[orange!20!brown,mark=diamond,dashed] table {figures/satisfiction/SC_sat.dat} ;
%\addplot[blue,dashed,mark=star] table {figures/satisfiction/BS_sat.dat} ;
%
%\legend{BS with UAVs, BS with SCs, Only BS }

%\legend{$Only BS$,$BS with SCs$,$BS with UAVs$}
%	\addplot[red,dashed,mark=square] table {crbNoLISa4.dat};

%		\addplot[blue,mark=star] table {crbLISa1.dat};
%	\addplot[orange!20!brown ,mark=diamond,mark repeat=1] table {crbLISa2.dat};
%	\addplot[black,mark=o,mark options=solid] table {crbLISa3.dat};
% 	\addplot[red,mark=square] table {crbLISa4.dat};
%	\addplot[red,mark=square,mark repeat=4,dashed,mark options=solid] table {TSBCIM4fadingM4.dat};
%	\addplot[orange,mark=triangle,mark options=solid,mark repeat=2] table {UniformSD5fadingM4.dat}; 
%\addplot[orange,mark=star,mark size=1.5,mark repeat=6] table {UniformBCIM6fadingM4.dat};
%\addplot[brown] table {CapacityLBfadingM4.dat};
%\addplot[gray] table {PSDCEXPSF7.dat};
%\addplot[black,thick,mark repeat=4,mark options=solid] table {TxRatefadingM4.dat};
%\addplot[gray!50!black,thick,mark=o,mark options=solid, mark size=3] coordinates {
%	(6.65,1.13)
%};

%\addplot[gray,thin] table {Figures/FigPSDContMatlab/FFTFigPSDContSF12.txt};
%%/tikz/thin (no value)
%/tikz/ultra thin (no value)
%/tikz/very thin (no value)
%/tikz/semithick (no value)
%/tikz/thick (no value)
%/tikz/very thick (no value)
%/tikz/ultra thick
\end{axis}
\end{tikzpicture}
	\caption{\ac{UL} EMF exposure versus required \ac{UL} data rate, $M=4$, $N=36$, and $K=60$.}
	\label{fig:sat_emf}
\end{figure}

 {\color{black}Fig.~\ref{fig:satisfiction} plots the satisfied-users ratio versus required \ac{UL} data rate. As can be seen, when the required data rate is less than $50$~Mbps, each network can meet the needs of almost all users. However, as the data rate continues to increase, there will be significant gaps between each scenario.  For instance, considering a $100$~Mbps required data rate, the fixed \acp{SC} can achieve only a gain in the satisfied-users ratio of $70\%$ compared to \ac{BS} only approach. On the other hand,  the proposed \acp{tUAV} scheme boosts it with up to $400\%$ compared to the \ac{BS} only scenario. 
 
%  When the  reaches , compared to fixed \acp{SC} that can only increase the satisfied-users ratio by $70\%$ compared to \ac{BS} only approach,  boost it by almost  $400\%$.

 Such a significant gain is mostly due to the reduced area that each \ac{gNB} has to cover. Therefore, the power that the user should transmit to achieve a required data rate decreases significantly below the maximum allowable transmit power, escalating the satisfied-users ratio.
 }
 
 %   Most of the gains is due to using 4 \acp{tUAV} along with the \ac{BS}, each one has to serve only a small area. 
 %make sure that the gain ratios are correct OK???
 %Compared to small cell？the gain is 300%
 %no BS
%gain of tuav/gain of small cell?
% no gain tuab/BS only (boost it main the improve with respect to BS ) 
%you can detail it like 200% and 300% with respect to SC and BS, respectively. something like thi

Furthermore, Fig.~\ref{fig:sat_emf} depicts the \ac{UL} \ac{EMF} exposure versus the required \ac{UL} data rate. This figure shows that as the required \ac{UL} data rate increases, the total \ac{EMF} exposure increases up to a certain value then saturates. This can be attributed to the fact that each user can not accommodate transmit power higher than his maximum level $P_{\max}$, leading to a drop in his satisfied rate, as in Fig.~\ref{fig:satisfiction}, and saturation to the exposure. When the required data rate reaches $100$~Mbps, the proposed scheme reduces the exposure by $23\%$ and $21\%$ compared to \ac{BS} only and green \acp{SC}, respectively. Combined with Fig.~\ref{fig:satisfiction}, we can see that the proposed scheme can improve the satisfied-users ratio, equivalent to the \ac{UL} data rate while maintaining relatively low \ac{EMF} exposure.

 \begin{figure}[t!]
	\centering
	\pgfplotsset{every axis/.append style={
		font=\footnotesize,
		line width=1pt,
		legend style={font=\footnotesize, at={(0.98,0.31)}},legend cell align=left},
} %
\pgfplotsset{compat=1.13}
	\begin{tikzpicture}%[trim axis right]%[trim axis left, trim axis right]
\begin{axis}[
%axis
%loglog
%title= users' satisfaction ratio vs. data rate,
%axis x line=bottom,
%yticklabel style = {font=\footnotesize,xshift=0.5ex},
%xticklabel style = {font=\footnotesize,yshift=0.5ex},
%ylabel style = {font=\huge},
%xlabel style = {font=\huge},
legend pos=north west,
xlabel near ticks,
ylabel near ticks,
grid=major,
xlabel={Average Number of Active Users $K$},
ylabel={UL Exposure, $\mathrm{EI}^\mathrm{UL}$ (W/kg) },
%width=0.79\linewidth,%3.5in,
%	ytick={10,0,-10,-20,-30,-40},
%	x label style={at={(axis description cs:0.5,-0.07)},anchor=north},
%   y label style={at={(axis description cs:-0.11,.5)},rotate=0,anchor=south},
%yticklabel style={/pgf/number format},
%axis y line=left,
width=0.6\linewidth,
%height=0.6\textwidth,
%	legend columns=1,	
%legend entries={firstmethos \cite{ElzGioChi:19}
%	,second $\x$,
%	third},
% 	log ticks with fixed point, 
%	scale only axis,
%	restrict
	xmin= 5, xmax=90,
%	ymin=0.5e-5, ymax=1e1,
ylabel style={font=\normalsize},
xlabel style={font=\normalsize},
%minor y tick num=10,
]

\addplot[blue,dashed,mark=star] table {figures/EMF/BS_EMF_EXP.dat} ;
\addplot[orange!20!brown,mark=diamond,dashed,mark options=solid] table {figures/EMF/SC_EMF_EXP.dat} ;
\addplot[black,mark=o,dashed,mark options=solid] table {figures/EMF/UAV_EMF_EXP.dat} ;

\legend{BS only,Green SCs,Green \acsp{tUAV} }

%\addplot[black,mark=o,dashed,mark options=solid] table {figures/satisfiction/UAV_sat.dat} ;
%\addplot[orange!20!brown,mark=diamond,dashed] table {figures/satisfiction/SC_sat.dat} ;
%\addplot[blue,dashed,mark=star] table {figures/satisfiction/BS_sat.dat} ;
%
%\legend{BS with UAVs, BS with SCs, Only BS }

%\legend{$Only BS$,$BS with SCs$,$BS with UAVs$}
%	\addplot[red,dashed,mark=square] table {crbNoLISa4.dat};

%		\addplot[blue,mark=star] table {crbLISa1.dat};
%	\addplot[orange!20!brown ,mark=diamond,mark repeat=1] table {crbLISa2.dat};
%	\addplot[black,mark=o,mark options=solid] table {crbLISa3.dat};
% 	\addplot[red,mark=square] table {crbLISa4.dat};
%	\addplot[red,mark=square,mark repeat=4,dashed,mark options=solid] table {TSBCIM4fadingM4.dat};
%	\addplot[orange,mark=triangle,mark options=solid,mark repeat=2] table {UniformSD5fadingM4.dat}; 
%\addplot[orange,mark=star,mark size=1.5,mark repeat=6] table {UniformBCIM6fadingM4.dat};
%\addplot[brown] table {CapacityLBfadingM4.dat};
%\addplot[gray] table {PSDCEXPSF7.dat};
%\addplot[black,thick,mark repeat=4,mark options=solid] table {TxRatefadingM4.dat};
%\addplot[gray!50!black,thick,mark=o,mark options=solid, mark size=3] coordinates {
%	(6.65,1.13)
%};

%\addplot[gray,thin] table {Figures/FigPSDContMatlab/FFTFigPSDContSF12.txt};
%%/tikz/thin (no value)
%/tikz/ultra thin (no value)
%/tikz/very thin (no value)
%/tikz/semithick (no value)
%/tikz/thick (no value)
%/tikz/very thick (no value)
%/tikz/ultra thick
\end{axis}
\end{tikzpicture}
	\caption{\ac{UL} \ac{EMF} exposure versus the average number of active users, for $M=4$, $W_{m,\max}=6$, and $N=36$.}
	\label{fig:emf}
\end{figure}
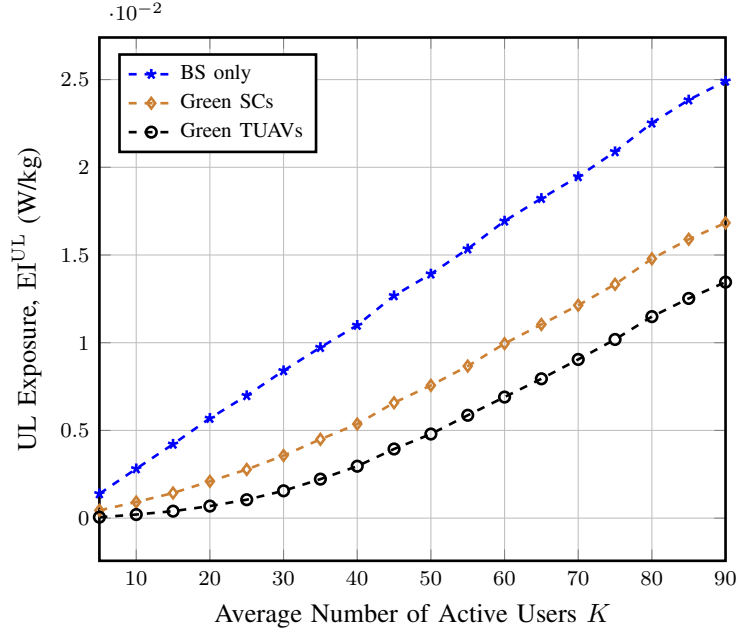
 
In Fig.~\ref{fig:emf}, we plot the  \ac{UL} \ac{EMF} exposure versus the average number of active users, which consider a scenario of $N=36$ \acp{GS} and $\bar{K}=240$ residents. The average number of active users varies between $5$ and $90$. Among these active users, $80\%$ are \emph{data} users with a target \ac{UL} rate of $50$~Mbps, and the remaining $20\%$ are \emph{voice} users with required \ac{UL} rate of $5$~Mbps. In the case of \ac{BS} only, the curve increases almost linearly from the start with the average number of active users. While the other two curves of \ac{UL} \ac{EMF} exposure exhibit the same basic shape, the slopes are small at first, then increase to certain values and remain stable due to the limited available \acs{RB}. In general, using \acp{tUAV} can adapt to changes in user distribution. For example, the proposed scheme can reduce the exposure, for $K=50$, by $65\%$ and $30\%$ compared to fixed BS only and \acp{SC}, respectively.

%Notice that there is also a slight difference in the final stable slope of these three curves and the gap between them is growing as expectation rise. In general, using \acp{tUAV} that can adapt to changes in user distribution can reduce \ac{EMF} exposure by an additional 40\% compared to fixed \acp{SC}.

%When the expectation is XX, the reduced \ac{EMF} exposure are XX and XX respectively, compared to the \ac{BS} only. But the gaps grows to XX and XX, when the expectation is XX. 

%the gap is 1 and 2, and when the number of users is 80, the gap increases to 3 and 4

	% In this figure, as the number of users increases,  
 %we can see that each line rises linearly at a different slope, and the gap between lines also increases a little.  

\begin{figure}[t!]
	\centering
	\pgfplotsset{every axis/.append style={
		font=\footnotesize,
		line width=1pt,
		legend style={font=\footnotesize, at={(0.98,0.90)}},legend cell align=left},
} %
\pgfplotsset{compat=1.13}
	\begin{tikzpicture}%[trim axis right]%[trim axis left, trim axis right]
\begin{semilogxaxis}[
%axis
%loglog
%title= users' satisfaction ratio vs. data rate,
%axis x line=bottom,
%yticklabel style = {font=\footnotesize,xshift=0.5ex},
%xticklabel style = {font=\footnotesize,yshift=0.5ex},
%ylabel style = {font=\huge},
%xlabel style = {font=\huge},
%legend pos=north east,
xlabel near ticks,
ylabel near ticks,
grid=major,
xlabel={Number of \acp{GS}, $N$},
ylabel={UL Exposure, $\mathrm{EI}^\mathrm{UL}$ (W/kg) },
%width=0.79\linewidth,%3.5in,
%	ytick={10,0,-10,-20,-30,-40},
%	x label style={at={(axis description cs:0.5,-0.07)},anchor=north},
%   y label style={at={(axis description cs:-0.11,.5)},rotate=0,anchor=south},
%yticklabel style={/pgf/number format},
%axis y line=left,
width=0.6\linewidth,
%height=0.6\textwidth,
%	legend columns=1,	
%legend entries={firstmethos \cite{ElzGioChi:19}
%	,second $\x$,
%	third},
% 	log ticks with fixed point, 
%	scale only axis,
%	restrict
	xmin= 2, xmax=128,
	log ticks with fixed point,
	xtick = {2,4,8,16,32,64,128},
%	extra x ticks={2},
	ymin=0.8e-2, ymax=2.5e-2,
ylabel style={font=\normalsize},
xlabel style={font=\normalsize},
%minor y tick num=10,
]

\addplot[blue,dashed] table {figures/num_possible_location_uavs/SC_2.dat} ;

\addplot[blue,mark=star] table {figures/num_possible_location_uavs/UAV_2.dat} ; \addlegendentry{2 green tUAVs}

\addplot[orange!20!brown,dashed] table {figures/num_possible_location_uavs/SC_4.dat} ;

\addplot[orange!20!brown,mark=diamond,mark options=solid] table {figures/num_possible_location_uavs/UAV_4.dat} ;

\addplot[black,dashed] table {figures/num_possible_location_uavs/SC_8.dat} ;
%\addlegend{4 green tUAVs}

\addplot[black,mark=o,mark options=solid] table {figures/num_possible_location_uavs/UAV_8.dat} ;

\legend{,$2$ Green \acsp{tUAV},,$4$ Green \acsp{tUAV},,$8$ Green \acsp{tUAV}}

%\legend{BS with 2 green SCs, BS with 2 green tUAVs ,BS with 4 green SCs,BS with 4 green tUAVs,BS with 8 green SCs,  BS with 8 green tUAVs}

%\addplot[black,mark=o,dashed,mark options=solid] table {figures/satisfiction/UAV_sat.dat} ;
%\addplot[orange!20!brown,mark=diamond,dashed] table {figures/satisfiction/SC_sat.dat} ;
%\addplot[blue,dashed,mark=star] table {figures/satisfiction/BS_sat.dat} ;
%
%\legend{BS with UAVs, BS with SCs, Only BS }

%\legend{$Only BS$,$BS with SCs$,$BS with UAVs$}
%	\addplot[red,dashed,mark=square] table {crbNoLISa4.dat};

%		\addplot[blue,mark=star] table {crbLISa1.dat};
%	\addplot[orange!20!brown ,mark=diamond,mark repeat=1] table {crbLISa2.dat};
%	\addplot[black,mark=o,mark options=solid] table {crbLISa3.dat};
% 	\addplot[red,mark=square] table {crbLISa4.dat};
%	\addplot[red,mark=square,mark repeat=4,dashed,mark options=solid] table {TSBCIM4fadingM4.dat};
%	\addplot[orange,mark=triangle,mark options=solid,mark repeat=2] table {UniformSD5fadingM4.dat}; 
%\addplot[orange,mark=star,mark size=1.5,mark repeat=6] table {UniformBCIM6fadingM4.dat};
%\addplot[brown] table {CapacityLBfadingM4.dat};
%\addplot[gray] table {PSDCEXPSF7.dat};
%\addplot[black,thick,mark repeat=4,mark options=solid] table {TxRatefadingM4.dat};
%\addplot[gray!50!black,thick,mark=o,mark options=solid, mark size=3] coordinates {
%	(6.65,1.13)
%};

%\addplot[gray,thin] table {Figures/FigPSDContMatlab/FFTFigPSDContSF12.txt};
%%/tikz/thin (no value)
%/tikz/ultra thin (no value)
%/tikz/very thin (no value)
%/tikz/semithick (no value)
%/tikz/thick (no value)
%/tikz/very thick (no value)
%/tikz/ultra thick
\end{semilogxaxis}
\end{tikzpicture}
	\caption{\ac{UL} EMF exposure versus the number of \acp{GS}, for $K=96$.}
	\label{fig:num}
\end{figure}
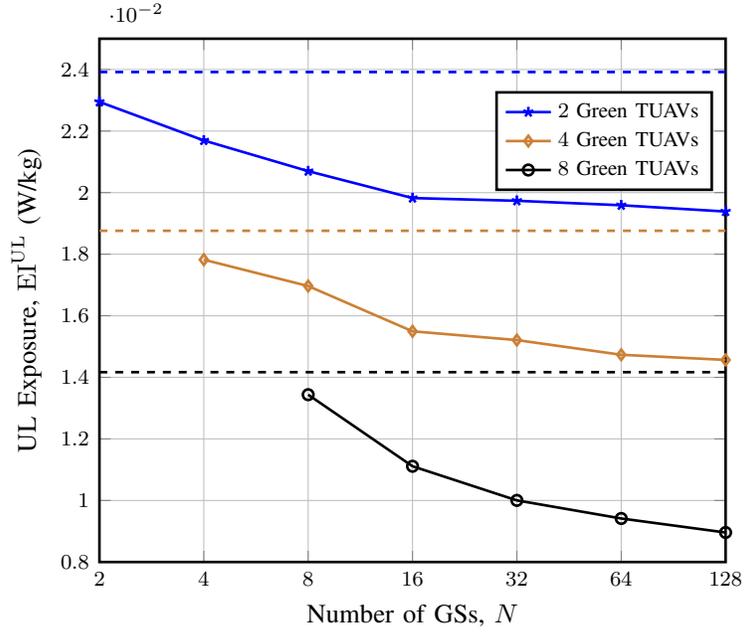

 In Fig.~\ref{fig:num}, we consider a scenario of $\bar{K}=240$ residents, and 40\% of them are active users and divided into two categories as scenario of Fig.~\ref{fig:emf}. After that, the solid lines represent radiation with green \acp{tUAV} assistance, and the dashed line parallel to the $x$-axis represents that of fixed deployed \acp{SC}. Moreover, the number of \acp{tUAV} and \acp{SC} $M \in \{2,4,8\}$, while that of \acp{GS} $N$ spans from $2$ to $128$. As can be seen from Fig.~\ref{fig:num}, increasing the number of \acp{tUAV} or \acp{SC} is a useful way to reduce \ac{EMF} exposure, but it's also costly. With the same number of \acp{tUAV} or \acp{SC}, \acp{tUAV} provide greater performance than \acp{SC}, even with $M=N$, i.e., also a fixed deployment for \acp{tUAV}. The proposed scheme of \acp{tUAV} can still get lower radiation by adjusting tether length and angles. Clearly, increasing the number of \ac{GS} can improve the performance of \ac{tUAV}, where for $N$ tends to infinity the \acp{tUAV} are equivalent to traditional un-tethered \acp{UAV}. But the improvements get smaller and smaller as the number of \acp{GS} increases.

% \subsection{Comparison of green \ac{SC} and \ac{tUAV} - dual problem}
 \subsection{Dual Problem: Cellular Design with \acs{EMF} Constraint}
 Here, we  present the result of the dual problem, which is to maximize the sum \ac{UL} data rate while keeping the \ac{EMF} exposure received by each user less than a threshold. The considered simulation scenario  is quite similar to the previous subsection, with $M=4$ green \acp{tUAV} (or \acp{SC}) assisting the communication. In addition, there are also $N=36$ evenly distributed \acp{GS} and $\bar{K}=240$ residents throughout the region, $25\%$ of
them are active, i.e., $K=60$ users. We consider the exposure  metric in terms of whole body \ac{SAR} with two levels: \textit{i)} the \ac{ICNIRP} limit, i.e., ${\mathrm{SAR}}_{\mathrm{limit}}=0.08$~W/kg; \textit{ii)} a stricter limit, i.e.,   $\mathrm{SAR}_\mathrm{limit}=0.0016$~W/kg, with a further safety factor of $50$ compared to the \ac{ICNIRP} limit.
 
 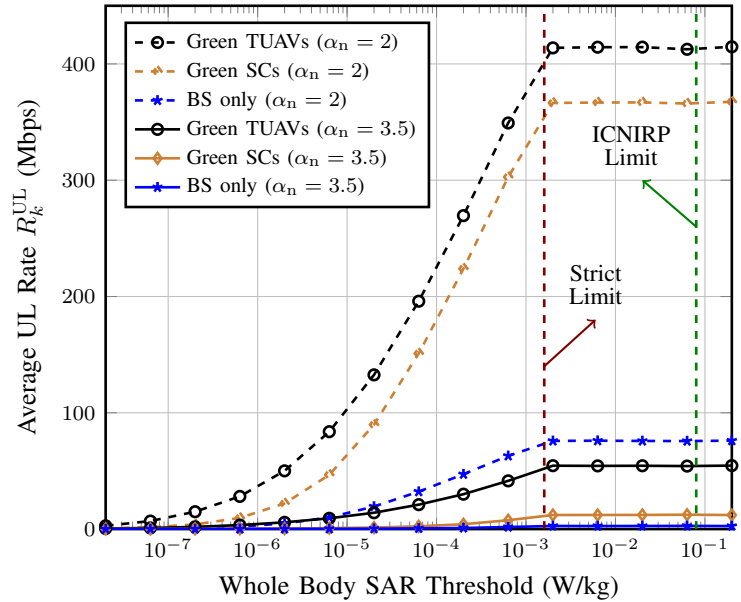
\begin{figure}[t!]
 	\centering
 	\pgfplotsset{every axis/.append style={
		font=\footnotesize,
		line width=1pt,
		legend style={font=\footnotesize, at={(0.98,0.31)}},legend cell align=left},
} %
\pgfplotsset{compat=1.13}
	\begin{tikzpicture}%[trim axis right]%[trim axis left, trim axis right]
\begin{semilogxaxis}[
%axis
%loglog
%title= users' satisfaction ratio vs. data rate,
%axis x line=bottom,
%yticklabel style = {font=\footnotesize,xshift=0.5ex},
%xticklabel style = {font=\footnotesize,yshift=0.5ex},
%ylabel style = {font=\huge},
%xlabel style = {font=\huge},
legend pos=north west,
xlabel near ticks,
ylabel near ticks,
grid=major,
ylabel={Average UL Rate $R_k^{\mathrm{UL}}$ (Mbps)},
xlabel={Whole Body SAR Threshold (W/kg)},
%width=0.79\linewidth,%3.5in,
%	ytick={10,0,-10,-20,-30,-40},
%	x label style={at={(axis description cs:0.5,-0.07)},anchor=north},
%   y label style={at={(axis description cs:-0.11,.5)},rotate=0,anchor=south},
%yticklabel style={/pgf/number format},
%axis y line=left,
width=0.6\linewidth,
%height=0.6\textwidth,
%	legend columns=1,	
%legend entries={firstmethos \cite{ElzGioChi:19}
%	,second $\x$,
%	third},
% 	log ticks with fixed point, 
%	scale only axis,
%	restrict
	xmin= 2e-8, xmax=2e-1,
	%extra x ticks={8e-2,1.6e-3},
	ymin=0, ymax=450,
ylabel style={font=\normalsize},
xlabel style={font=\normalsize},
%minor y tick num=10,
]

%/Users/lou/Dropbox/LOU/Lou_EMF_Aware_UAVs/figures/rateobjective

\addplot[black,mark=o,dashed,mark options=solid] table {figures/rateobjective/UAV.dat} ;
\addplot[orange!20!brown,mark=diamond,dashed] table {figures/rateobjective/SC.dat} ;
\addplot[blue,dashed,mark=star] table {figures/rateobjective/BS.dat} ;
\addplot[black,mark=o,mark options=solid] table {figures/rateobjective/UAV_c.dat} ;
\addplot[orange!20!brown,mark=diamond] table {figures/rateobjective/SC_c.dat} ;
\addplot[blue,mark=star] table {figures/rateobjective/BS_c.dat};
% \legend{Green \acp{tUAV} ($\alpha_l=\alpha_n = 2$), Green SCs ($\alpha_l=\alpha_n = 2$),BS only ($\alpha_l=\alpha_n = 2$), Green \acp{tUAV} ($\alpha_l=2.1, \alpha_n=3.5$),Green SCs ($\alpha_l=2.1, \alpha_n=3.5$),BS only ($\alpha_l=2.1, \alpha_n=3.5$) }
\legend{Green \acp{tUAV} ($\alpha_{\rm{n}} = 2$), Green SCs ($\alpha_{\rm{n}} = 2$),BS only ($\alpha_{\rm{n}} = 2$), Green \acp{tUAV} ($\alpha_{\rm{n}}=3.5$),Green SCs ($\alpha_{\rm{n}}=3.5$),BS only ($\alpha_{\rm{n}}=3.5$) }

\draw[line width=1pt, green!50!black, dashed] (0.08, 0) -- (0.08, 450);
\draw[line width=1pt, red!50!black, dashed] (0.0016, 0) -- (0.0016, 450);

\draw [green!50!black,thick,->] (0.08, 260) -- (0.02, 300);
\node at (0.015,340) {\small ICNIRP};
\node at (0.015,320) {\small Limit};

\draw [red!50!black,thick,->] (0.0016, 140) -- (0.006, 180);
\node at (0.006,220) {\small Strict};
\node at (0.006,200) {\small Limit};

\end{semilogxaxis}

\end{tikzpicture}
 	\caption{Average \ac{UL} rate versus whole body \ac{SAR} threshold that represents the \ac{EMF} constraint, for $M=4$, $N=36$, and $K=60$.}
 	\label{fig:rate}
 \end{figure}
 
     	 In Fig.~\ref{fig:rate}, we show the average \ac{UL} rate when the whole body \ac{SAR} limit spans from $2 \times 10^{-8}$~W/kg to $2\times 10^{-1}$~W/kg. 
     	 {\color{black} Moreover, we change the value of $\alpha_{\rm{n}}$, from $2$ to $3.5$, while keeping $\alpha_{\rm{l}}=2$, to see the impact of the path loss exponent for \ac{NLOS} on the performance. As shown in Fig.~\ref{fig:rate}, the average \ac{UL} rate would go down if the path loss exponent $\alpha_{\rm{n}}$ goes up, but the change of $\alpha_{\rm{n}}$ does not affect the trend of the curves. }
     	 %{\color{black} Moreover, we change the value of $\alpha_{\rm{n}}$, from $2$ to $3.5$, to see the differences. As shown in Fig.~\ref{fig:rate}, the average \ac{UL} rate would go down if the path loss exponent $\alpha_{\rm{n}}$ goes up, but the change of $\alpha_{\rm{n}}$ does not affect the trend of the curves.}
     	 Besides, it can be seen from the figure that the average rate increases as the threshold are relaxed. However, when the limit is increased to around $2\times 10^{-3}$~W/kg, further relaxation of the threshold does not improve the average rate. This can be explained by the hardware limitation of the user's transmit power. Moreover, the use of \ac{tUAV} improve the average rate by about $50$~Mbps compared to fixed \acp{SC} at \ac{ICNIRP} and stricter \ac{SAR} limits.
 
%  In Fig.~\ref{fig:rate}, we show the average \ac{UL} rate when the whole body \ac{SAR} limit spans from $2\, 10^{-8}$~W/kg to $2\, 10^{-1}$~W/kg. It can be seen from the figure that the average rate increases as the threshold are relaxed. However, when the limit is increased to around $2\, 10^{-3}$~W/kg, further relaxation of the threshold does not improve the average rate. This can be explained by the hardware limitation of the user's transmit power. Moreover, the use of \ac{tUAV} improve the average rate by about $50$~Mbps compared to fixed \acp{SC} at \ac{ICNIRP} and stricter \ac{SAR} limits.

  \begin{figure}[t!]
 	\centering
 	\pgfplotsset{every axis/.append style={
		font=\footnotesize,
		line width=1pt,
		legend style={font=\footnotesize, at={(0.98,0.31)}},legend cell align=left},
} %
\pgfplotsset{compat=1.13}
	\begin{tikzpicture}%[trim axis right]%[trim axis left, trim axis right]
\begin{axis}[
%axis
%loglog
%title= users' satisfaction ratio vs. data rate,
%axis x line=bottom,
%yticklabel style = {font=\footnotesize,xshift=0.5ex},
%xticklabel style = {font=\footnotesize,yshift=0.5ex},
%ylabel style = {font=\huge},
%xlabel style = {font=\huge},
legend pos=north west,
xlabel near ticks,
ylabel near ticks,
grid=major,
ylabel={Average UL Rate $R_k^{\mathrm{UL}}$ (Mbps)},
xlabel={Power Threshold $P_{\max}$ (dBm)},
%width=0.79\linewidth,%3.5in,
%	ytick={10,0,-10,-20,-30,-40},
%	x label style={at={(axis description cs:0.5,-0.07)},anchor=north},
%   y label style={at={(axis description cs:-0.11,.5)},rotate=0,anchor=south},
%yticklabel style={/pgf/number format},
%axis y line=left,
width=0.6\linewidth,
%height=0.6\textwidth,
%	legend columns=1,	
%legend entries={firstmethos \cite{ElzGioChi:19}
%	,second $\x$,
%	third},
% 	log ticks with fixed point, 
%	scale only axis,
%	restrict
	xmin= 20, xmax=36,
	ymin=0, ymax=700,
ylabel style={font=\normalsize},
xlabel style={font=\normalsize},
%minor y tick num=10,
]

%/Users/lou/Dropbox/LOU/Lou_EMF_Aware_UAVs/figures/rate_objective
\addplot[black,mark=o,dashed,mark options=solid] table {figures/rate_power/UAV8.dat} ;
\addplot[black,mark=o,mark options=solid] table {figures/rate_power/UAV16.dat} ;

\addplot[orange!20!brown,mark=diamond,dashed] table {figures/rate_power/SC8.dat} ;
\addplot[orange!20!brown,mark=diamond] table {figures/rate_power/SC16.dat} ;

\addplot[blue,dashed,mark=star] table {figures/rate_power/BS8.dat} ;
\addplot[blue,mark=star] table {figures/rate_power/BS16.dat} ;

\legend{Green \acsp{tUAV} (\ac{ICNIRP}), Green \acsp{tUAV} (Strict), Green SCs (\ac{ICNIRP}),Green SCs (Strict),BS only (\ac{ICNIRP}), BS only (Strict)}

\end{axis}

\end{tikzpicture}
 	\caption{Average \ac{UL} rate versus power threshold, for $M=4$, $N=36$, and $K=60$.}
 	\label{fig:power}
 \end{figure}
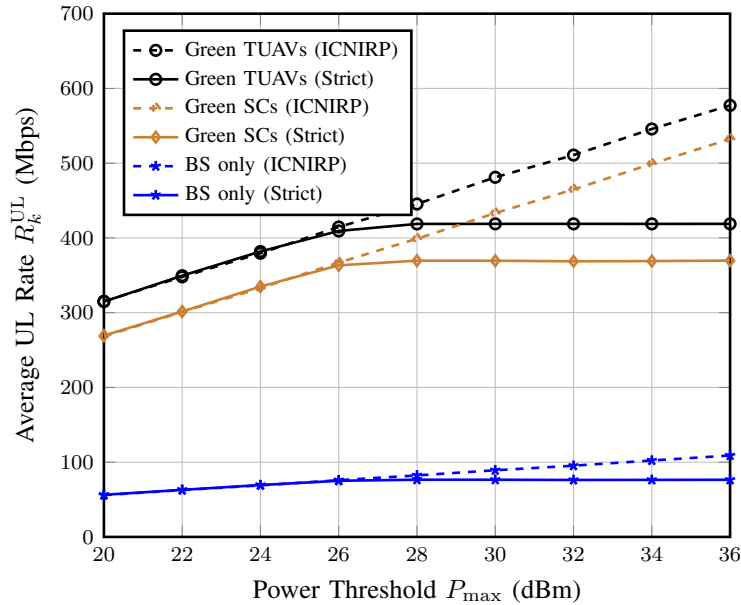

In Fig.~\ref{fig:power}, the maximum transmit power changes from $20$~dBm to $36$~dBm, while the \ac{SAR} limit remains the same at \ac{ICNIRP} limit $0.08$~W/kg or strict limit $0.0016$~W/kg. As can be seen that the addition of auxiliary green \acp{SC} or \acp{tUAV}  can effectively increase the average rate. Besides, when the \ac{ICNIRP} limit is considered, the average rate is positively correlated with the logarithm of the power threshold. When the \ac{SAR} limit is relatively strict, it can be seen that the curve saturates after rising, which can be explained by \eqref{eq:po}, i.e., the power allocation policy is limited by the exposure rather than the hardware constraint in this region. Moreover, under the same case, the \acp{tUAV} can adapt to the distribution of users, thus improving by an additional $50$~Mbps compared to fixed \acp{SC}.

%\section{Conclusions and Further Discussion} \label{sec:con}
\section{Conclusions} \label{sec:con}
A prevailing theory in the non-scientific community claims that \ac{EMF} exposure is uncontrolled and exponentially increasing due to \ac{gNB} densification. Also, there is a debate about the adverse health impacts due to the long-term non-thermal exposure to \ac{RF} radiations from \acp{BS}. In order to reduce the population exposure to \ac{EMF}, we proposed a novel architecture with \acp{tUAV} carrying green antennas to assist the communication system.  According to relevant work and confirmed by our numerical results, the \ac{UL} exposure is the dominant factor, e.g., $6$ orders of magnitudes higher than the \ac{DL} exposure for the considered simulation settings. In particular, we formulated two optimization problems either to minimize the \ac{UL} \ac{EMF} exposure and ensure a target data rate or to maximize the sum rate with an \ac{EMF} constraint. Moreover, we proposed several low-complexity algorithms to solve the optimization problems. The simulation results illustrated that using green \ac{tUAV} to densify \ac{UL} can provide good results when the number of \acp{RB} is limited. 
{\color{black} 
		On the other hand, comparing fixed green \ac{SC} to mobile green \ac{tUAV}, the latter tend to provide a better performance, both \ac{EMF} exposure and satisfied-users ratio, reducing by $23\%$ and boosting by $400\%$ respectively when required \ac{UL} data rate is $100$~Mbps.  The proposed scheme for cellular system design with \ac{EMF} constraint can significantly improve the average \ac{UL} rate by almost  $15\%$ and $350\%$ compared to \acp{SC} and \ac{BS} only architectures when considering the \ac{ICNIRP} exposure limit.}
%On the other hand, comparing fixed green \ac{SC} to mobile green \ac{tUAV}, the latter tend to provide a better performance, both \ac{EMF} exposure and satisfied-users ratio, reducing by $23\%$ and boosting by $400\%$ respectively.  The proposed scheme for cellular system design with \ac{EMF} constraint can significantly improve the average \ac{UL} rate by almost  $15\%$ and $350\%$ compared to \acp{SC} and \ac{BS} only architectures, when considering the \ac{ICNIRP} exposure limit. 
Therefore, we believe that the proposed architecture can be of interest for both the general public,  cellular operators, and  the research community, as reducing the exposure can alleviate the adverse impacts due to the \ac{EMF} exposure. 

%in the context of intense debate about the health risks of long-term exposure to \ac{EMF}.

%Moreover, it's an effective way to ease the fierce debate about the health risks of long-term exposure to \ac{EMF}.

%Therefore, we believe that the proposed architecture can be of interest for both the general public,  cellular operators, and  the research community.  

%Finally, we believe that this work can be useful for future research into the next generation of wireless communication systems. To this aim, maximizing the satisfied-users ratio in the constrain of ensuring that the \ac{EMF} exposure is below the threshold could be an interesting direction. 
%In addition, adopting a more practical propagation model is another attractive research direction. 
%In particular, the influence between users is to increase the noise power of other users, and the transmit power can also cause non-negligible exposure to nearby residents.

\bibliographystyle{IEEEtran}
\bibliography{references.bib}
\end{document}